\pdfoutput=1
\documentclass[11pt,a4paper]{article}

\usepackage{MySty}
\usepackage{amsmath,amsfonts,amssymb,graphics}
\usepackage[normalem]{ulem}
\usepackage{graphicx}
\usepackage{mathrsfs}
\usepackage{bbm}
\usepackage{pifont}
\usepackage{braket}
\usepackage{footmisc}
\usepackage[dvipsnames]{xcolor}
\usepackage{mathtools}
\usepackage{xspace}
\usepackage{booktabs}
\usepackage{siunitx}
\usepackage{slashed}
\usepackage{tikz}
\usetikzlibrary{positioning,shapes,arrows,calc}
\usetikzlibrary{decorations.pathmorphing}
\usetikzlibrary{decorations.markings}
\usepackage{standalone}
\usepackage{csquotes}
\usepackage{multirow}
 
\graphicspath{{figs/}}

\title{Type II Seesaw Leptogenesis in a Majoron background}
\author{Maximilian Berbig}
\affiliation{Departament de Física Teòrica, Universitat de València, 46100 Burjassot, Spain,\\
Instituto de Física Corpuscular (CSIC-Universitat de València), Parc Científic UV,\\
v
C/Catedrático José Beltrán, 2, E-46980 Paterna, Spain}
\emailAdd{berbig@ific.uv.es}

\abstract{
We discuss spontaneous Leptogenesis in the Type II Seesaw model of neutrino masses featuring an electroweak triplet scalar $T$ in a coherent pseudo Nambu-Goldstone boson (pNGB) background. In the \enquote{wash-in} scenario the inverse decays of Higgs bosons to $T$ generate a chemical potential for the triplet, that is then transmitted to the lepton sector via the leptonic decays of $T$. Our mechanism works with a single triplet, that can be as light as 1 TeV, and has a vacuum expectation value $v_T$ in the window $\mathcal{O}(\SI{1}{\kilo\electronvolt})<v_T<\mathcal{O}(\SI{1}{\mega\electronvolt})$. This range of $v_T$ can lead to appreciable decays of the triplet's doubly charged component into both same sign di-leptons and same sign pairs of $W$-bosons, which could potentially  allow for an experimental distinction from a recently proposed inflationary Type II Seesaw Affleck-Dine scenario  preferring the leptonic mode. 
In the \enquote{singlet-doublet-triplet Majoron} UV-completion of the Type II Seesaw model, the required pNGB is automatically included in the form of the Majoron, that originates from the phase of the lepton number breaking singlet scalar. The coherent motion of the Majoron can furthermore explain the dark matter relic abundance via the kinetic misalignment mechanism. Cogenesis of dark matter and the baryon asymmetry can work for a lepton number breaking scale of  $\mathcal{O}(\SI{e5}{\giga\electronvolt})<v_\sigma< \mathcal{O}(\SI{e8}{\giga\electronvolt})$ and a Majoron mass of $\mathcal{O}(\SI{1}{\electronvolt}) > m_j >\mathcal{O}(\SI{1}{\micro\electronvolt})$.}

\begin{document}
\maketitle

\section{Introduction}
\subsection{Overview}

\begin{figure}[t!]
    \centering
    \includegraphics[width=0.9\textwidth]{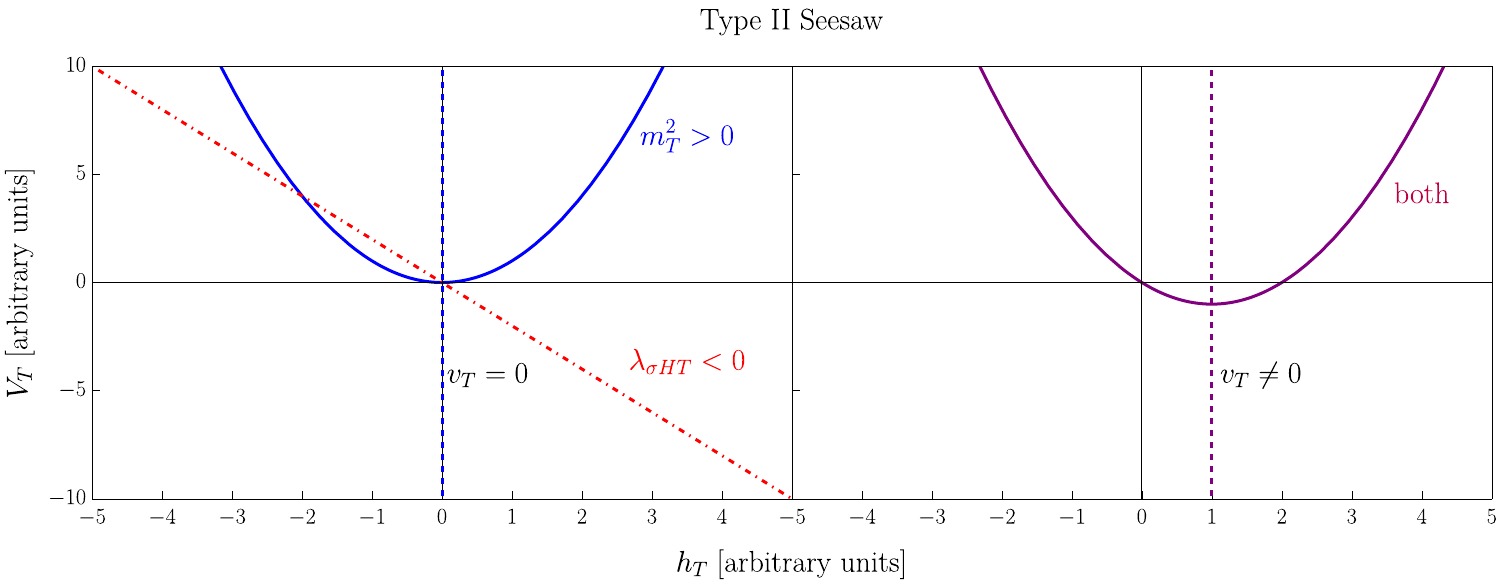}
    \caption{Basic overview for how a parametrically small triplet vev $v_T$ can be generated in the Type II Seesaw mechanism. The triplet has a positive mass squared \textit{(blue, left)} that by itself would lead to a trivial minimum $v_T=0$. After lepton number and the SM gauge symmetry are spontaneously broken by the vevs $v_\sigma$, $v_H$ of $\sigma$ and $H$ respectively, there exists a linear term in the triplet potential proportional to $\kappa=\lambda_{\sigma H T} v_\sigma/\sqrt{2}$ \textit{(red, left)} from Eq.~\eqref{eq:coupl}, which tilts the potential to generate a small $v_T\neq0$ \textit{(purple, right)}. Here $h_T$ denotes the field value of the electrically neutral, $CP$ even component of $T$ and $V_T$ is the potential energy. The plot is not up to scale and all parameters were chosen for illustration only.}
    \label{fig:TypeII}
\end{figure}

Non-vanishing neutrino masses have been robustly established from numerous observations of neutrino flavor oscillations at different length scales (see Refs.\cite{deSalas:2020pgw,Esteban:2020cvm} for an overview), and a precise measurement of the $\beta$-decay spectrum by the \verb|KATRIN| collaboration \cite{KATRIN:2024cdt}. Consequently the massive nature of neutrinos provides unambiguous evidence for physics beyond the Standard Model (SM).  
The Type II Seesaw mechanism \cite{Lazarides:1980nt,Schechter:1980gr,Mohapatra:1980yp,PhysRevD.22.2860,Wetterich:1981bx} stands out among the three known tree level UV completions of the Weinberg operator \cite{Weinberg:1979sa} for parametrically small Majorana neutrino masses, because it is the only one that relies on a scalar mediator in the form of electroweak triplet $T$ carrying hypercharge. 
Another distinction is given by the fact that it is the only tree level scheme for Majorana neutrino masses that predicts the existence of electrically double charged particles\footnote{For Majorona neutrinos doubly charged scalars can arise at two-loops in the Zee-Babu construction \cite{Zee:1985id,Babu:1988ki}. For Dirac neutrinos there exists one model at tree level that predicts doubly charged scalars from a bidoublet \cite{Berbig:2022hsm}.}, which might be accessible at collider experiments due to their couplings to electroweak gauge bosons. Similar to the right handed gauge singlet neutrino of the Type I Seesaw \cite{Minkowski:1977sc,Yanagida:1979as,Gell-Mann:1979vob,PhysRevLett.44.912}  the scalar triplet can be accommodated in the grand unified theory $\text{SO}(10)$, where it usually appears together with the Type I Seesaw. 

At its core the Type II Seesaw  mechanism can also be understood as  sourcing a parametrically small vacuum expectation value (vev) $v_T$  for the neutral component of the triplet, that can be closer to the observed scale of neutrino masse than the vev of the SM Higgs. This occurs because the positive quadratic potential for the triplet $m_T^2 |T|^2$ is biased in one direction by the impact of the term $\kappa H^t \varepsilon T H\;(\varepsilon=i\sigma_2)$ after the neutral component of the Higgs doublet condenses with a vev $v_H$. Here $\kappa$ is a coupling of mass dimension one, that is a priori unrelated to any other energy scale and it can be understood as a spurion of the global lepton number symmetry $\text{U}(1)_L$ with charge minus two.  A schematic illustration of the origin of this minimum with $v_T\neq 0$ can be observed in Fig.~\ref{fig:TypeII}.

Where the previously described model breaks lepton number explicitly and \textit{induces} a vev for $T$, one could also try to use the triplet to spontaneously break $\text{U}(1)_L$ with e.g. an eV-scale vev. Such a model gives rise to a Nambu-Goldstone boson (NGB) known as the triplet Majoron \cite{Gelmini:1980re}. However since the massless Majoron arises from a hypercharged weak multiplet, it would be produced in decays of the $Z$ boson. In fact this decay mode would contribute more than one additional generations of neutrinos to the $Z$-width, which is robustly ruled out by data from \verb|LEP 1| \cite{Berezhiani:1992cd}. This serious caveat is absent in models that utilize a scalar gauge singlet to spontaneously break lepton number, and their NGBs are usually referred to as singlet Majorons \cite{Chikashige:1980qk,Chikashige:1980ui}.\footnote{From a historical point of view this mirrors the development of \enquote{invisible} QCD axion models quite closely: The original Weinberg-Wilczek axion \cite{PhysRevLett.40.223,PhysRevLett.40.279} embedded the axion $a$ into a two Higgs doublet model, which lead to too large decay widths for processes like $K\rightarrow \pi a$ \cite{Wilczek:1977zn,Hall:1981bc}, that were insufficiently suppressed by the too small axion decay constant $f_a=\mathcal{O}(v_H)$. The next generation of models known as KSVZ \cite{PhysRevLett.43.103,SHIFMAN1980493}  and DFSZ \cite{Zhitnitsky:1980tq,DINE1981199} invisible axion models avoided this problem by embedding the axion in a gauge singlet scalar, whose vev is allowed to be $f_a\gg v_H$.}

Nevertheless one can salvage the triplet Majoron via a synthesis with the Type II Seesaw: In the 
singlet-doublet-triplet model \cite{Schechter:1981cv,Choi:1989hj,Choi:1991aa}
the dimensionful spurion $\kappa$ is UV completed via the vev $v_\sigma$ of a singlet scalar $\sigma$ that predominantly houses the Majoron and one obtains $\kappa = \lambda_{\sigma H T} v_\sigma/\sqrt{2}$ in terms of the dimensionless scalar coupling $\lambda_{\sigma H T}$. The advantage of having a Majoron in the spectrum is that it can be a good dark matter (DM) candidate (see Refs.~\cite{Garcia-Cely:2017oco,Brune:2018sab,Heeck:2019guh,Reig:2019sok,Akita:2023qiz} for an overview), if further explicit breaking of $\text{U}(1)_L$ renders it to be a massive pseudo Nambu-Goldstone boson (pNGB). Previous analyses of dark matter in the singlet-doublet-triplet Majoron scenario can be found in Refs.~\cite{Chao:2022blc,Biggio:2023gtm}.

It has been known for a long time that the neutrino sector can not just help with the problem of dark matter, but it can also address the origin of the observed asymmetry between matter and anti-matter of $\Delta_B = 8.76\times 10^{-11}$ \cite{Planck:2018vyg,Schoneberg:2024ifp} via the mechanism known as Leptogenesis \cite{Fukugita:1986hr} from the out-of-equilibrium decays of the heavy Seesaw mediators.
Type II Seesaw Leptogenesis was first studied in Refs.~\cite{Ma:1998dx,Hambye:2000ui,Hambye:2003ka,Hambye:2005tk}, and whereas neutrino masses only need a single triplet, the $CP$-violation in the out-of-equilibrium decay scenario requires at least two such generations. For a hierarchical triplet spectrum one needs \cite{Hambye:2005tk}
\begin{align}
    m_T>\mathcal{O}(\SI{e10}{\giga\electronvolt}),    
\end{align}
to explain the observed baryon asymmetry, which is about one order of magnitude stronger than the limit on the right handed neutrino masses in Type I Seesaw Leptogenesis \cite{Davidson:2002qv}. This estimate illustrates, why it is almost hopeless to test the required $m_T$ with laboratory experiments.
A resonant enhancement of the self-energy graphs on the other hand could bring the needed triplet mass down to the TeV-scale at the cost of having at least two almost degenerate triplets \cite{Pilaftsis:2003gt,Bechinger:2009qk}.

An elegant resolution of the previously motivated problems can be found in non-thermal Leptogenesis scenarios of the Affleck-Dine archetype \cite{Affleck:1984fy,Dine:1995kz,Murayama:1993em}. Here the required $CP$ violation comes from the coherent motion of a scalar field in the complex plane, and a single  triplet suffices. The coherent rotation of the condensate is possible due to a phase-dependent scalar interactionl, that explicitly violates baryon- or lepton-number. At the end of the scalar field's evolution its baryonic or leptonic charge is dumped into the plasma via decays, which lends itself
naturally to post-inflationary reheating, especially in models that identify the Affleck-Dine field with the inflaton.
The idea of inflation from a non-minimal coupling to gravity was merged with the Type II Seesaw and Affleck-Dine Leptogenesis in 
Refs.~\cite{Barrie:2021mwi,Barrie:2022cub}, and this set-up was further discussed in Refs.~\cite{Han:2023kjg,Kaladharan:2024bop}. Such an approach can also work with  TeV-scale triplets, and avoiding the washout of the leptonic asymmetry into the Higgs sector limits the triplet vev to be \cite{Barrie:2021mwi,Barrie:2022cub}
\begin{align}\label{eq:AD}
    v_T < \SI{8.5}{\kilo\electronvolt} \sqrt{\frac{\SI{1}{\tera\electronvolt}}{m_T}}.
\end{align}

\subsection{This work}
The second staple of non-thermal Baryogenesis scenarios from the coherent motion of a scalar field is known as 
Spontaneous Baryogenesis \cite{Cohen:1987vi,Cohen:1988kt} (see Ref.~\cite{DeSimone:2016ofp} for a review).
In its minimal form this scenario involves only a single $CP$ odd scalar, that eventually begins to oscillate around the minimum of its potential. The key idea is that the pseudoscalar's field velocity $\dot{\theta}$ evolves as an adiabatic background field, that spontaneously violates $CPT$ and can thus allow for asymmetry production in thermal equilibrium. The trade-off is that one needs to supply an additional baryon- or lepton-number violating interaction, that is active in the plasma, and decouples eventually. Here interactions that would usually \enquote{wash-out} the baryon- or lepton number asymmetry in the out-of-equilibrium decay scenario, instead \enquote{wash-in} the asymmetry \cite{Domcke:2020quw}. A natural realization of this mechanism is found in heavy\footnote{This statement will be elaborated upon in section \ref{sec:AD}.} Majoron models \cite{Kusenko:2014uta,Ibe:2015nfa} (see also Ref.~\cite{Mishra:2025twb} for a recent implementation in the singlet-doublet-triplet model). The needed Majorons typically overclose the universe, or are not long-lived enough to be the dark matter \cite{Chun:2023eqc}.

In recent years this idea has received renewed attention, because the authors of Refs.~\cite{Co:2019wyp,Co:2020jtv} realized that the pNGB velocity $\dot{\theta}$ , that drives spontaneous Baryogenesis, can be induced via the Affleck-Dine mechanism. The prerequisite explicit breaking of e.g. baryon- or lepton-number then automatically generates a mass for the NGB in question  \cite{Berbig:2023uzs,Wada:2024cbe} and the pNGBs can be much lighter than in the original spontanoues Baryogenesis scenario. It was shown in Ref.~\cite{Co:2019wyp} that the pNGB rotation experiences only a negligible back-reaction from driving Baryogenesis, which is why it can survive until late times, and source the dark matter relic abundance via the kinetic misalignment mechanism \cite{Co:2019jts,Chang:2019tvx,Co:2020dya} (see also \cite{Barman:2021rdr}). 
This idea stimulated a lot of studies 
\cite{Co:2020xlh,Harigaya:2021txz,Kawamura:2021xpu,Co:2021qgl,Domcke:2022wpb,Co:2022aav,Barnes:2022ren,Co:2023mhe,Barnes:2024jap,Co:2024oek,Chun:2024gvp,Chen:2025awt},
and was applied to both theories of Dirac neutrino masses in Ref.~\cite{Chakraborty:2021fkp,Berbig:2023uzs}, as well as to Majorana neutrino masses in Refs.~\cite{Chao:2023ojl,Chun:2023eqc,Wada:2024cbe}.

An investigation of these dynamics in a Type I Seesaw context \cite{Chun:2023eqc} revealed that \enquote{wash-in} Leptogenesis can proceed via the inverse decays $LH \rightarrow N$, where $N$ is a right handed neutrino. This approach has an additional benefit, that has not received a lot of attention so far: It can also significantly lower the right handed neutrino mass scale down to values above $\SI{2}{\giga\electronvolt}$, where the lower limit comes from the cosmological self-consistency of the analysis. In a similar vein Ref.~\cite{Barnes:2024jap} obtained an upper limit of $\SI{10}{\giga\electronvolt}$ in the context of asymmetry production from scattering in  a supersymmetric QCD axion model.
Apart from the attractive features of a built-in dark matter candidate and production mechanism, this makes Leptogenesis in a Majoron background a serious candidate for low-scale Leptogenesis, all without quasi-degenerate Seesaw mediators, or even multiple genrations of mediators.  In particular we expect the discussion of Ref.~\cite{Chun:2023eqc}  to also apply to the hypercharge-less triplet fermions of the Type III Seesaw \cite{Foot:1988aq}, where the lowest triplet mass is already constrained by collider searches at the LHC to be at or above the TeV-scale \cite{Ashanujjaman:2021jhi,Ashanujjaman:2021zrh}.

The present work follows this path and adapts it to the peculiarities of the Type II Seesaw. In particular our scenario works with a single generation of scalar triplets, whose mass is only limited by bounds from laboratory experiments and could well be close to the TeV-scale.
Here the inverse decay of two Higgses $HH\rightarrow T$ first generates a chemical potential for $T$ proportional to $\dot{\theta}$ that is then transmitted to the lepton sector via the decays $T\rightarrow L^\dagger L^\dagger$. We do not consider a rotation of the triplet, but rather employ it as an intermediate reservoir for the asymmetry, which works, because it is not self-conjugate. In the \enquote{strong wash-in} regime, where both of the aforementioned processes are fast on cosmological time scales, we predict that the triplet vev lies in the following window
\begin{align}
 \SI{8.5}{\kilo\electronvolt} \sqrt{\frac{\SI{1}{\tera\electronvolt}}{m_T}} <   v_T < \SI{277}{\kilo\electronvolt} \sqrt{\frac{\SI{1}{\tera\electronvolt}}{m_T}}\sqrt{\frac{\overline{m}_\nu^2}{\SI{2.65e-3}{\electronvolt\squared}}}.
\end{align}
However out set-up can also work outside of this window in the \enquote{weak wash-in} regime, so that a larger range of
\begin{align}
 \mathcal{O}(\SI{1}{\kilo\electronvolt}) <   v_T < \mathcal{O}(\SI{1}{\mega\electronvolt})
\end{align}
is viable. In principle these ranges for $v_T$ allow our model to be  experimentally distinguished  from the inflationary Affleck-Dine scenario of Refs.~\cite{Barrie:2021mwi,Barrie:2022cub,Han:2023kjg,Kaladharan:2024bop}, that predicts the different range of $v_T$ shown in Eq.~\eqref{eq:AD}. For completeness we also discuss the high-scale scenario with a super-heavy triplet $m_T\gtrsim \mathcal{O}(\SI{e14}{\giga\electronvolt})$, that is too heavy to be present in the plasma, and corresponds to the well studied regime of spontaneous Leptogenesis  from the Weinberg-operator. If the radial mode of $\sigma$ has a large field value, this high-scale regime only works for a narrow range of parameters. 

This manuscript is structured in the following way:
We begin with a review over the Type II Seesaw model and its associated phenomenology in section~\ref{sec:TIIpheno}. Then we provide an overview over the phenomenological aspects of the singlet-doublet-triplet Majoron in section~\ref{sec:Majoron}, where special emphasis is put on a review of the known mechanisms for generating a time dependent Majoron condensate, including the Affleck-Dine scheme in subsection~\ref{sec:AD} and a host of alternatives in subsection~\ref{sec:NOAD}. Majoron dark matter from the kinetic misalignment mechanism and its detection prospects are discussed in section~\ref{sec:DM}. We encourage the avid reader, that is already familiar with these subjects, to directly skip ahead to sections~\ref{sec:low}-\ref{sec:high}, where the key features of our Leptogenesis  scenario are discussed in detail.  The combined parameter space from all these ingredients is displayed in Figs.~\ref{fig:triplet}, \ref{fig:9-6} and \ref{fig:3-0}. Section~\ref{sec:high} discusses the high scale scenario, where the triplet is too heavy to be in the thermal bath, and contains our arguments, for why we do not focus on this specific region of parameter space.  The details of our predictions will be  elucidated in section~\ref{sec:discuss}, before we provide our conclusions in section~\ref{sec:conclusion}.
Appendix \ref{sec:KG} explicitly demonstrates that the effective chemical potential prescription in a Majoron background also holds for bosons coupled to the Majoron. Appendix \ref{sec:tripletRot} explains why we do not consider a rotation of the triplet scalar itself.

\section{Type II Seesaw}\label{sec:TIIpheno}
In this section we introduce the Type II Seesaw model \cite{Lazarides:1980nt,Schechter:1980gr,Mohapatra:1980yp,PhysRevD.22.2860,Wetterich:1981bx} in the \enquote{singlet-doublet-triplet} formulation  \cite{Schechter:1981cv,Choi:1989hj,Choi:1991aa}, and review its salient phenomenological aspects.

\subsection{The Model}
\begin{figure}[t]
 \centering
  \tikzset{
  blackline/.style={thin, draw=black, postaction={decorate},
    decoration={markings, mark=at position 0.6 with {\arrow[black]{triangle 45}}}},
     grayline/.style={thin, draw=gray, postaction={decorate},
    decoration={markings, mark=at position 0.6 with {\arrow[gray]{triangle 45}}}},
    blueline/.style={thin, draw=blue, postaction={decorate},
    decoration={markings, mark=at position 0.6 with {\arrow[blue]{triangle 45}}}},
    redline/.style={thin, draw=red, postaction={decorate},
    decoration={markings, mark=at position 0.6 with {\arrow[red]{triangle 45}}}},
    greenline/.style={thin, draw=green, postaction={decorate},
    decoration={markings, mark=at position 0.6 with {\arrow[green]{triangle 45}}}},    
    graydashed/.style={dashed, draw=gray, postaction={decorate},
    decoration={markings}},
   yellowdashed/.style={dashed, draw=orange, postaction={decorate},
    decoration={markings}},
    photon/.style={decorate, draw=red,
    decoration={coil,amplitude=12pt, aspect=0}},
  gluon/.style={dashed, decorate, draw=black,
    decoration={coil, segment length=5pt, amplitude=8pt}}
  line/.style={thick, draw=black, postaction={decorate},
    decoration={markings}}
}

\begin{tikzpicture}[node distance=1cm and 1cm]

\coordinate[label = left: $L$] (start);
\coordinate[right=4cm of start,label=right: $L$] (end);

\coordinate[right=2cm of start, label = below: $y_L$] (vertex);
\coordinate[above=1.7 cm of vertex] (vertex2);
\coordinate[above=1.5 cm of vertex2, label =  above: $\color{gray} \sigma$] (vertex3);
\coordinate[below right=1 cm of vertex2, label = right: $\color{gray} T$ ] (test2);
\coordinate[above left =2cm of vertex2, label =  above: $\color{gray} H$ ] (vev2);
\coordinate[above right =2cm of vertex2, label = above: $\color{gray} H$ ] (vev3);

\fill (vertex) circle (2pt);
\fill[gray] (vertex2) circle (2pt);

\draw[blackline] (start)   -- (vertex);
\draw[blackline] (end)   -- (vertex);
\draw[graydashed] (vertex2) -- (vev2);
\draw[graydashed] (vertex2) -- (vev3);
\draw[graydashed] (vertex) -- (vertex2);
\draw[graydashed] (vertex2) -- (vertex3);

\end{tikzpicture}
  \caption{Diagrammatic representation of the dimension six operator for the   Type II  Seesaw with spontaneously broken lepton number  giving rise to Majorana masses for the active neutrinos. The arrows of the fermion line correspond to their chirality following the two-component spinor notation of Ref.~\cite{Dreiner:2008tw}.}
  \label{fig:SeesawII}
\end{figure}
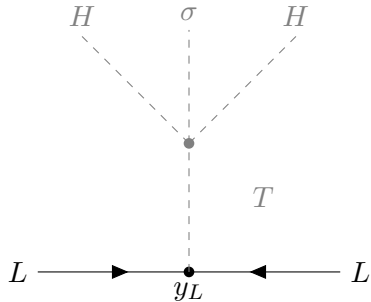

We introduce an electroweak scalar  triplet $T$ with hypercharge $Q_Y[T] =1$ and lepton number $Q_L[T]=-2$ that can be decomposed as 
\begin{align}
    T = \begin{pmatrix}
        \frac{T^+}{\sqrt{2}} & T^{++}\\
        T^0& -\frac{T^+}{\sqrt{2}}
    \end{pmatrix},
\end{align}
together with an SM singlet scalar $\sigma$ with $Q_L[\sigma]=2$
\begin{align}
    \sigma = \frac{h_\sigma + v_\sigma}{\sqrt{2}} e^{i \frac{j}{v_L}},
\end{align}
where $v_\sigma$ is the scale of lepton number breaking and $j$ the Majoron. 
The relevant  interactions with the lepton doublet $L$ and the Higgs doublet $H$ read
\begin{align}\label{eq:coupl}
    - \mathcal{L} &= y_L L^t \varepsilon T L + \text{h.c.},\\
    V(\sigma,H,T) &\supset m_T^2 \text{Tr}\left(T^\dagger T\right) +\lambda_{\sigma H T} \sigma H^\dagger \varepsilon T H^*  + \text{h.c.},
\end{align}
where $\varepsilon= i \sigma_2$  in terms of the second Pauli matrix is needed to symmetrize the $\text{SU}(2)$ contractions. Throughout this work we use the two-component spinor notation of Ref.~\cite{Dreiner:2008tw}.
The full scalar potential can be found in e.g. Ref.~\cite{Biggio:2023gtm} and the coupling $\lambda_{\sigma H T}$ can be made real by a field redefinition of either $\sigma$ or $T$. Consequently a complex vev $v_T$ can only arise from spontaneous $CP$ violation, which requires at least two triplets \cite{Ferreira:2021bdj}.
The necessary and sufficient conditions for vacuum stability were compiled in Ref.~\cite{Bonilla:2015eha}.

We take  $m_T^2>0$ and assume the following potential for the spontaneous breaking of lepton number 
\begin{align}\label{eq:Vsigma}
    V_\sigma= \lambda_\sigma (v_\sigma^2- |\sigma|^2)^2.
\end{align}
The role of the spontaneously broken $\text{U}(1)_L$ symmetry is to ultra-violet (UV) complete the commonly assumed explicit lepton number breaking term $\kappa H^t \varepsilon T H$ via the identification $\kappa= \lambda_{\sigma H T} v_\sigma/\sqrt{2}$.
At low energies one can integrate out the triplet to obtain the following dimension six operator depicted in Fig.~\ref{fig:SeesawII}
\begin{align}\label{eq:SeeswII}
    y_L \lambda_{\sigma H T} \frac{v_H^2}{m_T^2} \;\sigma^* (L H)^2,
\end{align}
which reduces to the well known dimension five Weinberg-operator $(L H)^2$ \cite{Weinberg:1979sa} once the lepton number breaking scalar $\sigma$ condenses.
Non-zero values for the vev $v_H$ of the neutral component of $H$ and  $v_\sigma$  induce  a vev for the  neutral component of the triplet (see Fig.~\ref{fig:TypeII} for an illustration) 
\begin{align}\label{eq:vT}
    v_T = \lambda_{\sigma H T} \frac{v_\sigma v_H^2}{m_T^2},
\end{align}
and the neutrino mass is found to be 
\begin{align}\label{eq:mNu}
    m _\nu = y_L \frac{v_T}{\sqrt{2}}.
\end{align}
Taking $m_T^2 \gg \lambda_{\sigma H T} v_\sigma v_H$ ensures that $v_T\ll v_H$. In section \ref{sec:discuss} we will motivate, why we focus on a TeV-scale triplet with $\mathcal{O}(\SI{1}{\kilo\electronvolt}) <   v_T < \mathcal{O}(\SI{1}{\mega\electronvolt})$ and $v_\sigma >\mathcal{O}(\SI{e5}{\giga\electronvolt})$, for which the 
neutrino mass scale is realized via a combination of  small $y_L$ and $\lambda_{\sigma H T}$. 
Parametrically we find
\begin{align}\label{eq:m1}
    m_\nu = \SI{0.05}{\electronvolt} \left(\frac{y_L}{7.1\times10^{-5}}\right)\left(\frac{v_T}{\SI{1}{\kilo\electronvolt}}\right),
\end{align}
and
\begin{align}\label{eq:m2}
    v_T = \SI{1}{\kilo\electronvolt} \left(\frac{\lambda_{\sigma H T}}{1.7\times 10^{-13}}\right) \left(\frac{v_\sigma}{\SI{e8}{\giga\electronvolt}}\right) \left(\frac{\SI{1}{\tera\electronvolt}}{m_T}\right)^2.
\end{align}

One could see such a low value for the scalar coupling of $\lambda_{\sigma H T}=1.7\times 10^{-13}$ as defeating the purpose of constructing a seesaw mechanism. However note that the required Yukawa coupling of the neutrinos of $y_L= 7.1\times10^{-5}$ is almost two orders of magnitude larger than the smallest Yuakwa coupling in the SM, namely the electron's Yukawa coupling. Therefore our model is still a Seesaw mechanism, in the sense that it allows one to increase the size of the neutrino Yukawa coupling. We further remark that small values of $\lambda_{\sigma H T}$ are technically natural as setting this coupling to zero increases the symmetry of the scalar potential, since this  would allows us to rephase each scalar multiplet independently. Of course this does not address why this coupling is small to begin with. The overall situation closely mirrors the case of the small lepton-number violating $\mu$-term in the Inverse Seesaw model \cite{Mohapatra:1986bd}, see Ref.~\cite{CentellesChulia:2020dfh} for a recent review. It is conceivable that some of the mechanisms that aim to address the origin of this small $\mu$-term such as Refs.~\cite{Ma:2009gu,Ahriche:2016acx,Berbig:2024uwm} or multiple iterations of the Type II Seesaw structure \cite{Gu:2019ird} could also explain the smallness of $\lambda_{\sigma H T}$.

\subsection{Phenomenology}\label{eq:pheno}
\begin{table}[t]
\centering
\begin{tabular}{ |c||c|c|c|c||c|c|c|c| } 
\hline
  process &  type  & $v_T>$ & experiment & reference \\
\hline
\hline
\multirow{2}{0.25\linewidth}{$\mu\rightarrow e \gamma $}  & limit & $ \SI{0.95}{\electronvolt} \left(\frac{\SI{1}{\tera\electronvolt}}{m_T}\right)$& MEG & \cite{Adam:2013vqa}\\
& projection & $\SI{1.85}{\electronvolt} \left(\frac{\SI{1}{\tera\electronvolt}}{m_T}\right)$ & \verb|MEGII| &  \cite{MEGII:2018kmf}\\
\hline
\hline
\multirow{2}{0.25\linewidth}{$\mu\rightarrow 3 e$} & limit & $ \left(\frac{\SI{1}{\tera\electronvolt}}{m_T}\right) \begin{cases}
        \SI{0.45}{\electronvolt} \; \text{(NO)}\\
        \SI{0.98}{\electronvolt} \; \text{(IO)}
    \end{cases}$   & \verb|SINDRUM| & \cite{SINDRUM:1987nra}\\
&projection & $ \left(\frac{\SI{1}{\tera\electronvolt}}{m_T}\right) \begin{cases}
        \SI{4.54}{\electronvolt} \; \text{(NO)}\\
        \SI{9.83}{\electronvolt} \; \text{(IO)}
    \end{cases}$& \verb|Mu3e| &  \cite{Mu3e:2020gyw}  \\
\hline 
\hline
\multirow{2}{0.25\linewidth}{$\mu-e\; \text{conversion}$}  &  limit (Ti) & $\SI{9.10}{\electronvolt}  \left(\frac{\SI{1}{\tera\electronvolt}}{m_T}\right) \sqrt{\frac{4v_T^2 |C_{\mu e}^{(II)}|}{\SI{1}{\electronvolt}}}$ & \verb|SINDRUMII| & \cite{SINDRUMII:1993gxf}\\
& limit (Al) & $\SI{71.86}{\electronvolt}  \left(\frac{\SI{1}{\tera\electronvolt}}{m_T}\right) \sqrt{\frac{4v_T^2 |C_{\mu e}^{(II)}|}{\SI{1}{\electronvolt}}}$ & \verb|Mu2e| & \cite{Mu2e:2014fns}\\
\hline
\end{tabular}
\caption{Bounds on the vev $v_T$ from lepton flavor violation. The limits were adapted from Refs.~\cite{Barrie:2022ake,Han:2025ifi}.}
\label{tab:LFV}
\end{table}

Most of the Type II Seesaw phenomenology is controlled by the two parameters $v_T$ and $m_T$, together with the sum of squared neutrino masses $\sum_i m_{\nu_i}^2$. For the numerical evaluation of this quantity  we take the mass splittings from the global fit in Ref.~\cite{Esteban:2020cvm}, and impose the constraint on the sum of neutrino masses $\sum_i m_{\nu_i} < \SI{0.12}{\electronvolt}$ \cite{Planck:2018vyg}\footnote{To be conservative here we quote the results without including the constraints arising from the latest DESI BAO datasets \cite{DESI:2024mwx,DESI:2025zgx}. The pull towards smaller values of $\sum_i m_{\nu_i}$ (see e.g. \cite{DESI:2024mwx,Naredo-Tuero:2024sgf}) and the recent hints in favor of dynamical dark energy \cite{DESI:2024mwx,DESI:2025wyn} are both consequences of a preference for a larger matter fraction $\Omega_m$ in the data \cite{Loverde:2024nfi,Tang:2024lmo,Jhaveri:2025neg}. This can be seen from the fact that analyses featuring dynamical dark energy typically obtain a relaxed bound on the neutrino mass of scale of $\sum_i m_{\nu_i}=0.19^{+0.15}_{-0.18}\;\text{eV}$ \cite{RoyChoudhury:2024wri,RoyChoudhury:2025dhe} in sharp contrast to analyses using pure $\Lambda$CDM that find $\sum_i m_{\nu_i}<\SI{0.072}{\electronvolt}$  \cite{DESI:2024mwx}.}
from Planck temperature and polarization data together with information from gravitational lensing and BAO 
\begin{align}\label{eq:overline}
\text{(NO)}\quad 
  \SI{2.65e-3}{\electronvolt\squared} &<  \overline{m}_\nu^2 \equiv \sum_i m_{\nu_i}^2< \SI{5.34e-3}{\electronvolt\squared},\\
\text{(IO)}\quad 
  \SI{4.92e-3}{\electronvolt\squared} &<  \overline{m}_\nu^2 \equiv \sum_i m_{\nu_i}^2< \SI{5.68e-3}{\electronvolt\squared},
\end{align}
where NO (IO) denotes normal (inverted) mass ordering. The lower value refers to the case of a massless lightest neutrino and the upper value comes from saturating the cosmological limit.
Throughout this work we limit ourselves to the NO mass hierarchy and fix 
\begin{align}
    \overline{m}_\nu^2 = \SI{2.65e-3}{\electronvolt\squared}.
\end{align}

The mass eigenstates of the triplet are given by the $CP$ even (odd) $h_T \;(a_T)$, the singly charged $h^\pm$ and the doubly charged $h^{\pm\pm}$ that are (mostly) aligned with the gauge eigenstates $T^0, T^\pm$ and  $T^{\pm\pm}$. Their mass spectrum reads in the limit $v_H \gg v_T$ 
\begin{align}
    m_{h_T}^2 \simeq m_{a_T}^2 \simeq m_{h^\pm}^2 + \frac{\lambda_{HT2}}{4} v_H^2 \simeq m_{h^{\pm\pm}}^2 + \frac{\lambda_{HT2}}{2} v_H^2 \simeq m_T^2,
\end{align}
where $\lambda_{HT2}$ is the coefficient of the term $\text{Tr}\left(H^\dagger T^\dagger T H\right)$, and a full overview over the masses and mixing can be found in Ref.~\cite{Biggio:2023gtm}. 
At one-loop we find following Ref.~\cite{Cirelli:2005uq} that for $m_T\gg m_Z, m_W$
\begin{align}\label{eq:loop}
    |m_{h^{\pm\pm}}-m_{h^\pm}|\simeq \frac{\alpha_2}{2} (m_W + (3 \sin{(\theta_W)}^2-1) m_Z)\simeq \SI{0.85}{\giga\electronvolt},
\end{align}
where $\theta_W$ denotes the weak mixing angle and $\alpha_2\simeq 1/30$ the $\text{SU}(2)_L$ fine structure constant.

Electroweak precision data further constrains the mass splitting between the doubly and singly-charged scalar \cite{Chun:2012jw,Aoki:2012jj,Das:2016bir,Primulando:2019evb}
\begin{align}
    |m_{h^{\pm\pm}}-m_{h^\pm}| < \SI{40}{\giga\electronvolt}.
\end{align}
In this work we will consider  both $m_T^2\gg v_H^2$ and  $\lambda_{HT2}\ll1$, which limits the discussion to a degenerate mass spectrum of common mass $m_T$ as the one loop correction in Eq.~\eqref{eq:loop} is negligible. We denote the mass eigenstate that predominantly arises from the radial mode of $\sigma$ as $h_\sigma$ and its mass reads
\begin{align}
    m_S^2 \simeq 2 \lambda_\sigma v_\sigma^2.
\end{align}
We introduce the $CP$ even boson, that predominantly arises from the neutral component of $H$, as $h_H$, and identify it with the $\SI{125}{\giga\electronvolt}$ resonance observed at the LHC.

The doubly charged scalar $h^{\pm\pm}$ is the smoking gun signature for models involving a hypercharged triplet, such as the Type II Seesaw for Majorana neutrino masses.
Present day collider searches already set limits on the doubly charged triplet mass and the strongest constraint comes from the pair production of $h^{\pm\pm}$ via the Drell-Yan process, followed by the decays $h^{\pm\pm}\rightarrow W^\pm W^\pm$ or $h^{\pm\pm}\rightarrow l^\pm l^\pm$.\footnote{The cascade decay into a charged scalar and off-shell $W$-boson  $h^{\pm\pm}\rightarrow h^{\pm}W^{\pm *}$ is only relevant for mass splittings $|m_{h^{\pm\pm}}-m_{h^\pm}|>\mathcal{O}(\SI{1}{\giga\electronvolt})$ \cite{Ashanujjaman:2021txz}, which is why we neglect it here, see e.g. Eq.~\eqref{eq:loop}.}  Their branching fraction reads \cite{Ashanujjaman:2021txz}
\begin{align}
\frac{\Gamma\left(h^{\pm\pm}\rightarrow \sum_i l_i^\pm l_i^\pm\right)}{\Gamma\left(h^{\pm\pm}\rightarrow W^\pm W^\pm\right)} = \frac{1}{4}  \frac{\sum_i m_{\nu_i}^2}{m_T^2} \left(\frac{v_H}{v_T}\right)^4,
\end{align}
and from this one finds that $h^{\pm\pm}\rightarrow W^\pm W^\pm$ dominates for 
\begin{align}\label{eq:vTW}
    v_T > \SI{40}{\kilo\electronvolt} \left(\frac{\overline{m}_\nu^2}{\SI{2.65e-3}{\electronvolt\squared}}\right)^\frac{1}{4}\sqrt{\frac{\SI{1}{\tera\electronvolt}}{m_T}}.
\end{align}
The negative result of a search for the $W^{\pm} W^{\pm}$ final state carried out by the \texttt{ATLAS} collaboration leads to the following bound for a degenerate mass spectrum \cite{ATLAS:2018ceg} 
\begin{align}\label{eq:LHC1}
  m_T > \SI{220}{\giga\electronvolt},
\end{align}
and the stronger bound of $m_T>\SI{350}{\giga\electronvolt}$ \cite{ATLAS:2021jol} for triplet mass splittings below $\SI{100}{\giga\electronvolt}$.

Associated production of a doubly-charged scalar together with a singly charged one  gives a limit of $m_T > \SI{230}{\giga\electronvolt}$ \cite{ATLAS:2021jol}.
For the case of decays to lepton dominates ($v_T<\SI{40}{\kilo\electronvolt}$) there exist dedicated searches by \verb|ATLAS| \cite{ATLAS:2017xqs,Novak:2024igo} and \verb|CMS| \cite{CMS:2017pet}. The strongest limit for a degenerate mass spectrum reads \cite{Novak:2024igo} 
\begin{align}\label{eq:LHC2} 
    m_T > \SI{1080}{\giga\electronvolt},
\end{align}
and this limit decreases to $\SI{1020}{\giga\electronvolt}$, if not all triplet components have the same mass.
Optimized multi-lepton searches at the LHC with more luminosity are expected to improve these limits for $v_T>\SI{40}{\kilo\electronvolt}$ to \cite{Ashanujjaman:2021txz}
\begin{align}
    m_T > \SI{640}{\giga\electronvolt}
\end{align}
and for $v_T<\SI{40}{\kilo\electronvolt}$ to \cite{Ashanujjaman:2021txz}
\begin{align}
    m_T > \SI{1.49}{\tera\electronvolt}.
\end{align}
Future colliders are also expected to be able to probe $m_T= \mathcal{O}(\text{TeV})$ \cite{Mandal:2022ysp} and for example a 100 TeV proton-proton collider could improve these bounds to $m_T>\SI{1}{\tera\electronvolt}$ for $v_T>\SI{40}{\kilo\electronvolt}$ as well as  $m_T>\SI{4.5}{\tera\electronvolt}$ for $v_T<\SI{40}{\kilo\electronvolt}$ respectively \cite{Du:2018eaw}.

The electroweak scale depends on the doublet and triplet vevs
\begin{align}
  v_\text{EW} \equiv   \sqrt{v_H^2 + 2 v_T^2} = \SI{246}{\giga\electronvolt},
\end{align}
and the dimensionless measure of custodial symmetry violation called the $\rho$-parameter \cite{ROSS1975135,Veltman:1976rt} turns out to be 
\begin{align}
    \rho = \frac{m_W^2}{m_Z^2 \cos{(\theta_W)}^2}= \frac{v_H^2 + 2v_T^2}{v_H^2 + 4 v_T^2}= \frac{v_\text{EW}^2}{v_\text{EW}^2 + 2v_T^2}.
\end{align}
The current experimental limit on this quantity reads $\rho=1.0002\pm0.0009$ \cite{ParticleDataGroup:2024cfk} and therefore the triplet vev is bounded to be \cite{Biggio:2023gtm}
\begin{align}\label{eq:EWPT}
    v_T < \SI{7}{\giga\electronvolt}
\end{align}
at $95\%$ confidence level.
A lower limit on $v_T$ follows from the demand, that the atmospheric mass splitting $m_\text{atm.}^2 =\SI{2.52e-3}{\electronvolt\squared}$ should be reproduced for perturbative values of $y_L< \sqrt{4\pi}$, and therefore
\begin{align}\label{eq:vTLow}
    v_T > \sqrt{\frac{2 m_\text{atm.}^2}{4 \pi }}= \SI{0.02}{\electronvolt}.
\end{align}

Neutrinoless double-beta decay can be mediated by the doubly-charged component of the triplet \cite{Mohapatra:1981pm}, and the corresponding amplitude is expected to be suppressed by $p_\nu^2/m_T^2$ \cite{Barry:2012ga} compared to the usual light neutrino mediated amplitude in Type I Seesaw models, where $p_\nu\simeq  \SI{100}{\mega\electronvolt}$ is the momentum exchange inside the nucleus. For our range of $m_T\gtrsim \mathcal{O}(\SI{1}{\tera\electronvolt})$ we expect the rate for this process to be well within existing bounds. 

Lepton flavor violation for our scenario was studied in e.g. Refs.~\cite{Barrie:2022ake,Han:2025ifi} and the authors obtained limits on $\kappa = \lambda_{\sigma H T} v_\sigma/\sqrt{2}$ by using the best fit values for the neutrino mass splittings, and numerically marginalizing over the PMNS mixing angles and phases. We recast these limits in terms of  $v_T$.
We compiled the relevant bounds in the table \ref{tab:LFV}, and they include 
the non-observation of the decay mode $\mu\rightarrow e \gamma$ at \verb|MEG| \cite{Adam:2013vqa}, as well as an improved limit expected from future \verb|MEGII| \cite{MEGII:2018kmf} data. 
The next decay mode is $\mu\rightarrow 3 e$  with a limit from \verb|SINDRUM| experiment  \cite{SINDRUM:1987nra} and a projected limit for  \verb|Mu3e| experiment \cite{Mu3e:2020gyw}.
Nuclear conversion of $\mu$ to $e$  was searched for by the \verb|SINDRUMII| experiment \cite{SINDRUMII:1993gxf} using titanium (Ti)  and in aluminum (Al) by the \verb|Mu2e| experiment \cite{Mu2e:2014fns}. 
Here the definition of the quantity $C_{\mu e}^{(II)}$ involves the PMNS matrix elements  and a loop factor that depends on $m_T$ (see Refs.~\cite{Barrie:2022ake,Han:2025ifi} for more details), which is why it can be suppressed for both NO and IO, or even cross zero for NO. The precise bounds for this process have to be found numerically by varying the PMNS phases and the lightest neutrino mass \cite{Barrie:2022ake,Han:2025ifi}, which is beyond the scope of this work. 
Future experiments such as \verb|COMET| \cite{Moritsu:2022lem} and \verb|Mu2e-II| \cite{Mu2e-II:2022blh} are expected to improve the aforementioned limits. 
The anomalous magnetic moment of the muon can not be explained by loops involving the triplet, because the resulting contribution has the wrong sign \cite{Lindner:2016bgg}.

We showcase the collected limits from this section in the $m_T$ versus $v_T$ plane in Fig.~\ref{fig:triplet} without the more complicated bounds from nuclear conversion.

\section{The Majoron}\label{sec:Majoron}
Here we discuss the singlet-doublet-triplet Majoron model \cite{Schechter:1981cv,Choi:1989hj,Choi:1991aa}, specify the origins of the Majoron mass and summarize the known mechanisms for generating the Majoron velocity.

\subsection{Phenomenology}
Global symmetries like $\text{U}(1)_L$ are expected to be only approximate in the context of quantum gravity \cite{Georgi:1981pu,Dine:1986bg,Coleman:1989zu,Abbott:1989jw,Holman:1992us,Kamionkowski:1992mf,Barr:1992qq,Ghigna:1992iv,Kallosh:1995hi,Alonso:2017avz}. 
We parameterize the explicit breaking of lepton number required for the Majoron mass as a dimension $d=m+n$ operator
\begin{align}\label{eq:highdim}
    V_{\slashed{L}} = c_{m+n} \frac{\sigma^m \sigma^{*\;n}}{M_\text{Pl.}^{m+n-4}} +\text{h.c.},
\end{align}
where we require $m\neq n$. The resulting Majoron mass reads
\begin{align}\label{eq:mj}
    m_j^2 = 2^{1-\frac{m+n}{2}} (m-n)^2 |c_{m+n}| \cos[\delta] v_\sigma^2 \left(\frac{v_\sigma}{M_\text{Pl.}}\right)^{m+n-4},
\end{align}
and we further defined $\delta \equiv \text{Arg}[c_{m+n}]$.
Note that a priori there is no reason for $V_\slashed{L}$ to be a $d>4$ operator, and that the coupling $c_{m+n}$ might be exponentially suppressed by some action similar to an instanton effect (see e.g. Ref.~\cite{Kallosh:1995hi}).

In section \ref{sec:discuss} we will demonstrate that our preferred parameter space involves the following range for the lepton number breaking scale   $\mathcal{O}(\SI{e5}{\giga\electronvolt})<v_\sigma< \mathcal{O}(\SI{e8}{\giga\electronvolt})$ and the Majoron mass $\mathcal{O}(\SI{1}{\electronvolt}) > m_j >\mathcal{O}(\SI{1}{\micro\electronvolt})$. For $v_\sigma =\mathcal{O}(\SI{e5}{\giga\electronvolt})$ we find that the right $m_j$ can be realized with a $d=6$ operator
\begin{align}
    m_j \simeq \SI{0.8}{\electronvolt} \sqrt{|c_6|\cos[\delta]} \left(\frac{v_\sigma}{\SI{e5}{\giga\electronvolt}}\right)^2,
\end{align}
and for $v_\sigma =\mathcal{O}(\SI{e8}{\giga\electronvolt})$ we would need a $d=8$ operator
\begin{align}
    m_j \simeq \SI{7}{\micro\electronvolt}\sqrt{|c_8|\cos[\delta]} \left(\frac{v_\sigma}{\SI{e8}{\giga\electronvolt}}\right)^3.
\end{align}

The mass eigenstate Majoron $\hat{j}$ is given by the following linear combination of the phases of the neutral component of the doublet (triplet) $a_H\;(a_T)$ \cite{Schechter:1981cv}
\begin{align}\label{eq:profile}
    \hat{j}\simeq j - \frac{2v_T^2}{v_\sigma v_H} a_H + \frac{v_T}{v_\sigma} a_T,
\end{align}
and since we take $v_\sigma \gg v_H \gg v_T$ we can take $\hat{j}\simeq j$ to simplify our notation. Consult Refs.~\cite{Choi:1991aa,Bonilla:2015jdf} for the full mixing matrix. 
A non-zero Majoron mass modifies the above estimate for the mixing between $j$ and $a_T$ only by a completely negligible contribution of 
$(m_j/m_T)^2 v_T/v_\sigma \ll1$ \cite{Biggio:2023gtm}. 
The admixture of the Majoron in the phase of the neutral component of $H$ that becomes the longitudinal mode of the $Z$-boson, is also given by $2v_T^2(v_\sigma v_H)$ \cite{Choi:1991aa} and consequently the flavor-diagonal Majoron couplings to SM fermions read \cite{Biggio:2023gtm}
\begin{align}\label{eq:cferm}
    c_{\nu_i}\simeq -\frac{m_{\nu_i}}{2 v_\sigma},\; c_{l_i} = -\frac{2 v_T^2}{v_H^2} \frac{m_{l_i}}{v_\sigma},\; c_{u_i} = \frac{2 v_T^2}{v_H^2} \frac{m_{u_i}}{v_\sigma},\; c_{d_i} = -\frac{2 v_T^2}{v_H^2} \frac{m_{d_i}}{v_\sigma}.
\end{align}
Since the singlet-doublet-triplet Majoron couplings are aligned with the charged lepton and quark mass matrices, they do not give rise to flavor changing processes like $\mu \rightarrow e j$, or $K\rightarrow \pi j$ at tree level. Lepton flavor violating decays can nevertheless happen at one loop via diagrams involving the singly and doubly charged components of the triplet, where the Majoron is emitted from the triplet or lepton lines. However since the Majoron coupling will always involve a small factor of $1/v_\sigma$, we expect the rates to be even more suppressed than $\mu\rightarrow e \gamma$ from section \ref{eq:pheno}, and therefore we do not discuss this mode in more detail.

The coupling of the Majoron to the $\nu_e$ flavor eigenstate is constrained by the non-observation of Majoron emission in neutrinoless double-beta decays, and the strongest limit  $|c_{\nu_e}|<(0.4-0.9)\times10^{-5}$ was obtained by the EXO-200 experiement using $^{136}$Xe nuclei \cite{Kharusi:2021jez} (consult also Ref.~\cite{CUPID-0:2022yws} for an overview over the limits from different neutrinoless double-beta decay experiments). Even without taking the  flavor composition of $\nu_e$ into account it is obvious, that $c_\nu$ in this model will in general far to suppressed to matter, as can be seen from a simple estimate of 
\begin{align}\label{eq:cnu}
    |c_{\nu}| \lesssim \frac{m_\text{atm.}}{2v_\sigma} = 2.5\times 10^{-10} \left(\frac{\SI{e8}{\giga\electronvolt}}{v_\sigma}\right).
\end{align}
When it comes to the Majoron electron coupling, stellar cooling arguments demand that $|c_e|<10^{-13}$  \cite{Choi:1989hi}, which in turn implies that 
\begin{align}\label{eq:j-e}
    v_T < \SI{24.33}{\giga\electronvolt} \sqrt{\frac{v_\sigma}{\SI{e8}{\electronvolt}}}.
\end{align}

The Majoron can couple to photons either via mixing with the neutral pion \cite{Bauer:2017ris}, or at one loop due via the above couplings to charged SM fermions. Since lepton number is not anomalous with respect to electromagnetism, the loop induced contribution scales with the masses of the fermions running in the loop, and it is dominated by the lightest such fermions \cite{Bazzocchi:2008fh,Garcia-Cely:2017oco}, which is why we limit ourselves to the electron, as well as the up and down quarks. Thus the effective coupling reads at leading order 
\begin{align}\label{eq:gamma}
g_{j\gamma\gamma} \simeq \frac{2\alpha}{\pi} \frac{v_T^2}{v_H^2 v_\sigma} \left(\frac{m_j^2}{m_j^2-m_{\pi_0}^2} -B_1\left(\frac{4 m_e^2}{m_j^2}\right) + \frac{4}{3}B_1\left(\frac{4 m_u^2}{m_j^2}\right)-\frac{1}{3}B_1\left(\frac{4 m_d^2}{m_j^2}\right)\right),
\end{align} 
in terms of the loop functions $B_1$ defined in Ref.~\cite{Bauer:2017ris}.

The phenomenology of the Type II Seesaw Majoron has been extensively studied in Refs.~\cite{Joshipura:1992hp,Diaz:1998zg,Bonilla:2015jdf}. We have to check that the branching ratio for  Higgs decays to   Majorons are within existing bounds. Working to lowest order in $v_T$ we obtain an irreducible contribution from the $\lambda_{\sigma H T}$ term of\footnote{We checked that this is indeed the dominant contribution. Our result disagrees with the leading order result obtained in Ref.~\cite{Chao:2022blc} when neglecting the mixing between $h_H$ and the radial mode of $T_0$. The authors of Ref.~\cite{Chao:2022blc} focused on the $v_\sigma$ term when expanding the $\lambda_{\sigma H T}$ interaction in the cartesian representation and then use the mixing of $a_H$ given by $V_{13} \simeq 2 i v_T^2/(v_\sigma v_H)$ and of $a_T$ of $V_{23}= i v_T/v_\sigma$ to get $j^2$. We on the other hand focus on the $v_H$ term and use the fact that $j\simeq \hat{j}$ so we only need the mixing of $a_T$ with $V_{23}= i v_T/v_\sigma$ to obtain the second $j$. Thus while our amplitude is smaller by $v_H/v_\sigma$ the lack of a second mixing term enhances it by $1/V_{13}= v_\sigma v_H/(2 i v_T^2)$, which combines to $v_H^2/v_T^2\gg1$ and the final decay width is larger by $v_H^4/v_T^4$.} 
\begin{align}\label{eq:inv1}
    \Gamma(h_H\rightarrow j j) \simeq \frac{1}{64\pi}  \left(\frac{m_T}{v_\sigma}\right)^4 \frac{v_T^4}{v_H^2 m_{h_H}},
\end{align}
where $m_{h_H}=\SI{125}{\giga\electronvolt}$ is the mass of the SM-like Higgs boson. The invisible branching fraction is constrained to be $\text{BR}\equiv\text{BR}(h_H \rightarrow \text{inv.})=0.18$ \cite{CMS:2022qva}, and the SM prediction for the  decay width of the Higgs to visible states reads $\Gamma(h_H \rightarrow \text{vis.})= \SI{4.1}{\mega\electronvolt}$ \cite{ATLAS:2023dnm}. From the condition $\Gamma(h_H\rightarrow j j) < \text{BR}/(1-\text{BR})\Gamma(h_H \rightarrow \text{vis.})$ we obtain a weak bound on $v_T$ of 
\begin{align}\label{eq:vTInv1}
    v_T < \SI{2.4e6}{\giga\electronvolt} \left(\frac{v_\sigma}{\SI{e8}{\giga\electronvolt}}\right) \left(\frac{\SI{1}{\tera\electronvolt}}{m_T}\right).
\end{align}
The decay width for $h_H \rightarrow Z j$ was parameterized in terms of the mixing matrix elements in Ref.~\cite{Biggio:2023gtm}, and the  admixture of $j$ in the $Z$-boson is given by $2 v_T^2 /(v_\sigma v_H)$ \cite{Choi:1991aa}, so that
\begin{align}\label{eq:inv2}
    \Gamma(h_H \rightarrow Zj) = \frac{g^2}{16\pi \cos{(\theta_W)}^2}\frac{v_T^4}{v_\sigma^2 v_H^2} \frac{m_{h_H}^3}{m_Z^2} \left(1-\frac{m_Z^2}{m_{h_H}^2}\right)^3,  
\end{align}
where $g$ denotes the $\text{SU}(2)_\text{L}$ gauge coupling and $\theta_W$ the weak mixing angle. 
The corresponding bound on $v_T$ from invisible Higgs decays reads 
\begin{align}\label{eq:vTInv2}
    v_T < \SI{3.8e5}{\giga\electronvolt} \sqrt{\frac{v_\sigma}{\SI{e8}{\giga\electronvolt}}}.
\end{align}
We do not consider the decay mode $h_H \rightarrow \nu^\dagger \nu j$, because the neutrinos only couple to the SM like Higgs via the potentially small mixing with the neutral component of the triplet, and the emission of the Majoron introduces another small factor of $m_\nu/v_\sigma$.

Invisible decays of the $Z$-boson can involve Majorons too, and the first process we consider is $Z\rightarrow \nu \nu j$.
Ref.~\cite{Dev:2024ygx} was the first work to properly account for the treatment of the infrared divergence that appears when taking $m_j\rightarrow 0$, and the authors found a limit of 
$|c_\nu|<1.4$ \cite{Dev:2024ygx},
which translates to 
\begin{align}
    v_\sigma > \frac{\overline{m}_\nu}{2.8},
\end{align}
and this bound is hardly constraining at all. 
The decays $Z\rightarrow h_i j$ are forbidden as all the $CP$-even scalars $h_i = h_\sigma, h_H, h_T$, are heavier than the $Z$-boson. This illustrates why the singlet Majoron extension of the Type II Seesaw is not ruled out by constraints from \verb|LEP 1| \cite{Berezhiani:1992cd}, unlike the original triplet Majoron model \cite{Gelmini:1980re}.

\subsection{Mechanisms for Majoron Rotation}\label{sec:rot}
We begin with the most well known approach for generating an angular rotation for a complex scalar, that is known as the Affleck-Dine mechanism \cite{Affleck:1984fy,Dine:1995kz}, which was applied to pNGBs in Ref.~\cite{Co:2019wyp}. Afterwards we provide an overview of numerous alternative schemes for realizing a non-zero angular velocity of the Majoron. 

\subsubsection{Affleck-Dine mechanism}\label{sec:AD}
We limit our discussion to the epoch of radiation domination, where the Hubble rate and entropy density are given by 
\begin{align}
    H(T) = \sqrt{\frac{8\pi^3 g_\rho(T)}{90}} \frac{T^2}{M_\text{Pl.}}, \quad s(T)= \frac{2\pi^2}{45} g_S(T) T^3,
\end{align}
in  terms of the number of relativistic degrees of freedom in energy (entropy) $g_\rho\;(g_S)$.

In models with an oscillating Majoron  \cite{Cohen:1987vi,Cohen:1988kt,Kusenko:2014uta,Ibe:2015nfa,Datta:2024xhg} the field velocity is given by
\begin{align}
    \dot{\theta} \simeq m_j
\end{align}
and successful Leptogenesis typically requires very large values of $m_j$ as can be seen from the following estimate: The baryon asymmetry scales as 
\begin{align}
    \Delta_B \simeq \frac{45}{2\pi^2}\frac{\dot{\theta}(T_\text{dec.})}{g_S(T_\text{dec.}) T_\text{dec.}}, 
\end{align}
where $T_\text{dec.}$ is the model dependent temperature at which the lepton or baryon number changing processes decouple. For a rough model independent estimate we take  $T_\text{dec.} \simeq T_\text{osc.}\simeq 0.78 \sqrt{m_j M_\text{Pl.}}/g_\rho(T_\text{osc.})^{1/4}$, where  $T_\text{osc.}$ is the temperature at which the Majoron oscillations commence.
This leads us to 
\begin{align}
    \Delta_B \simeq \frac{3 g_\rho(T_\text{osc.})^\frac{1}{4}}{g_S(T_\text{osc.})}\sqrt{\frac{m_j}{M_\text{Pl.}}},
\end{align}
and reproducing the observed baryon asymmetry of implies with $g_\rho(T_\text{osc.})\simeq  g_S(T_\text{osc.}) \simeq 106.75$ that
\begin{align}
   m_j \simeq \SI{12}{\giga\electronvolt}.
\end{align}
Such a heavy Majoron can lead to overclosure, which could be fixed via sufficient entropy dilution. However a large $m_j$ also implies a short life-time for the Majoron (see e.g. Eq.~\eqref{eq:lifetime} in one of the next sections) that can be below the age of the universe, making it hard to realize Majoron dark matter in these scenarios. 

The advantage of the Affleck-Dine mechanism for NGBs \cite{Co:2019wyp} is that it decouples $\dot{\theta}$ from its present day mass $m_j$, because the dynamics   of the radial mode with initial value $S_i$ lead to an attractor solution \cite{Gouttenoire:2021jhk} for the  velocity of \cite{Co:2020jtv} 
\begin{align}
    \braket{\dot{\theta}} = \varepsilon m_S (S_i) \gg m_j,
\end{align}
where an average over the radial oscillations is understood, and we will define $\varepsilon, m_S$  in the following.
During inflation the field value $S$ of the radial mode can be far larger than its minimum today of $v_\sigma$. If the effective mass of $S$ given by 
\begin{align}
   m_S(S)\simeq \sqrt{3\lambda_\sigma}S 
\end{align}
is below the approximately constant Hubble rate during inflation $H_I<\SI{6e13}{\giga\electronvolt}$ \cite{Planck:2018jri}, then diffusion due to stochastic quantum fluctuations can push the field value to  \cite{Starobinsky:1994bd}
\begin{align}
    S_i \simeq \left(\frac{1}{8\pi^2}\right)^\frac{1}{4} \frac{H_I}{\lambda_\sigma^\frac{1}{4}},
\end{align}
and due to inflation this field value will be almost homogeneous in our Hubble patch. 
Alternatively one may consider a coupling of the form $-R|\sigma|^2$ or finite energy effects in supersymmetric theories \cite{Dine:1995uk} to induce a negative Hubble dependent squared mass  for $S$ during inflation that leads to
\begin{align}
    S_i \simeq  \frac{H_I}{\sqrt{\lambda_\sigma}}.
\end{align}
The radial mode will begin to oscillate when $m_S(S_i)\simeq 3 H(T_\text{osc.})$, which implicitly defines 
\begin{align}\label{eq:Tosc}
    T_\text{osc.}  \simeq \SI{3e15}{\giga\electronvolt} \left(\frac{\lambda_\sigma}{10^{-12}}\right)^\frac{1}{4}  \sqrt{\frac{S_i}{M_\text{Pl.}}},
\end{align}
and throughout this work we assume that this happens during radiation domination. Furthermore we will use $T_\text{osc.}$ as a free parameter instead of $\lambda_\sigma$. 
From $H(T_\text{osc.})<H_I$ and the CMB bound on $H_I$ \cite{Planck:2018jri} from the non=observation of inflationary tensor modes one finds that \cite{Wada:2024cbe}
\begin{align}\label{eq:ToscLimit}
    T_\text{osc.} < \SI{6.5e15}{\giga\electronvolt} \sqrt{\frac{H_I}{\SI{6e13}{\giga\electronvolt}}},
\end{align}
which in turn implies 
\begin{align}\label{eq:lambdasig}
    \lambda_\sigma < 1.8\times 10^{-11} \left(\frac{H_I}{\SI{6e13}{\giga\electronvolt}}\right)^2 \left(\frac{M_\text{Pl.}}{S_i}\right)^2.
\end{align}

As long as $S_i <2 M_\text{Pl.}$ the energy density stored in these oscillations is subdominant to the energy density of the radiation bath \cite{Co:2020dya,Gouttenoire:2021jhk}. 
In the Affleck-Dine mechanism \cite{Affleck:1984fy,Dine:1995kz} the radial oscillations are converted into a coherent motion in the angular direction. The required angular curvature comes from the higher dimensional operator defined in Eq.~\eqref{eq:highdim} that is also responsible for the Majoron mass. 
The lepton number charge stored in the $\sigma$ condensate with $\theta \equiv j/S$ reads
\begin{align}
    n_\theta \equiv \dot{\theta} S^2,
\end{align}
and the Klein-Gordon equation in the expanding space-time can be cast as \cite{Co:2020jtv} 
\begin{align}
    \dot{n}_\theta + 3 H n_\theta =\frac{\partial V_\slashed{L}}{\partial \theta}.
\end{align}
For small couplings $|c_{m+n}|\ll1$ or $\sin{((m-n)\theta + \delta)}\simeq 0$ we can work perturbatively in the slow-roll approximation, and neglect the evolution of $\theta$ away from its initial value $\theta_i$ \cite{Co:2022qpr}, to find the dimensionless charge yield
\begin{align}
   Y_\theta\equiv \frac{n_\theta}{s} \simeq \frac{2^{1-\frac{m+n}{2}} (m-n) |c_{m+n}| M_\text{pl.}^4 }{3 H(T_\text{osc.}) s(T_\text{osc.})}  \left(\frac{S_i}{M_\text{Pl.}}\right)^{m+n}  \sin{((m-n)\theta_i + \delta)},
\end{align}
and for larger $|c_{m+n}|$ or $\sin{((m-n)\theta + \delta)}$ one has to solve the equations of motion numerically. Note that the right hand side is proportional to the Majoron mass squared from Eq.~\eqref{eq:mj} evaluated at $S_i$, which scales as $m_j^2 (S_i/v_\sigma)^{m+n-2}$.
A higher dimensional operator for $V_\slashed{L}$ is favored, because the explicit breaking of lepton number is switched of dynamically as the radial field value decreases.
It is useful  to define a dimensionless measure of the charge stored in the condensate \cite{Co:2019wyp}
\begin{align}\label{eq:eps}
    \varepsilon\equiv \frac{n_\theta}{V_\sigma(S_i)/m_s(S_i)},
\end{align}
where $V_\sigma$ was introduced in Eq.~\eqref{eq:Vsigma}, and this can be understood as the ratio of the minor semi-axis of the two-dimensional orbit in field space over the major semi-axis. The aforementioned motion will in general be elliptic $\varepsilon<1$ and a perfectly circular motion is encoded in $\varepsilon=1$. Once the radial mode is thermalized the rotation will be perfectly circular \cite{Gouttenoire:2021jhk}. The case of  $\varepsilon>1$ is not allowed, as the Majoron would roll to one of the minima of its potential before the radial oscillations begin, and since the angular gradient vanishes at the minimum, an angular rotation will not be excited \cite{Co:2022qpr}. Another way to interpret the danger of $\varepsilon>1$ is that the effective Majoron mass evaluated at $S_i$ becomes larger than the radial mass $m_S(S_i)$, so that the treatment of the Majoron in terms of a pNGB breaks down \cite{Berbig:2023uzs}. Note that as a consequence of the previously employed slow-roll approximation the above formula for $\varepsilon$ only holds for $\varepsilon\ll1$ \cite{Co:2022qpr}, or else numerics have to be employed. 

Owing to the constraint in Eq.~\eqref{eq:lambdasig} one typically finds that $m_S(S_i)<3H_I/2$, and from the previous discussion it is clear that the angular mode will be even lighter. Hence both fields will be subject to quantum fluctuations during inflation, that get imprinted on the Baryon asymmetry and dark matter abundance \cite{Co:2020dya,Co:2020jtv}.\footnote{See Ref.~\cite{Co:2023mhe} for an exception in the context of supergravity.} Note that the fluctuations in the angular direction can be tuned away for specific choices of the initial misalignment angle $\theta_i$ in the regime $\varepsilon\ll1$ \cite{Co:2020jtv}. This is not possible for the radial mode and for a quartic potential these fluctuations typically lead to a limit of \cite{Co:2020dya,Co:2020jtv}
\begin{align}\label{eq:lambdaAD}
    \lambda_\sigma < \mathcal{O}(10^{-10}).
\end{align}
We refrain from estimating the precise constraint, but note that for a given $S_i$ it will lead to an upper limit for $T_\text{osc.}$. 
The leading correction to $\lambda_\sigma$ arises at two loops and reads
\begin{align}
    \delta \lambda_\sigma \simeq \frac{\lambda_{\sigma H T}^2}{16 \pi^2} \left(2\lambda_{\sigma H} + \lambda_{\sigma T}\right),
\end{align}
where $\lambda_{\sigma H}\;(\lambda_{\sigma T})$ is the coefficient of the operator $|\sigma|^2 |H|^2\;(|\sigma|^2 |T|^2)$. Since these couplings are irrelevant for our phenomenology, we can make them arbitrarily small, and therefore do not expect isocurvature constraints to impact our preferred parameter space.

A related scenario in which the vev of the radial mode decreases due to thermal corrections, so that the height of the Majoron's potential barriers decrease in time allowing it to \enquote{slide} over them, was analyzed in Ref.~\cite{Chun:2024gvp}.  
The Affleck Dine mechanism can also be realized in models where the radial mode of $\sigma$ is the inflaton see e.g. Refs.~\cite{Lee:2023dtw,Lee:2024bij} (see also Refs.~\cite{Maleknejad:2020yys,Maleknejad:2020pec} for a related scenario where inflation is driven by the angular mode itself).

\subsubsection{Other mechanisms}\label{sec:NOAD}
Here we provide an overview over possible mechanisms apart from the previously discussed Affleck-Dine approach for sourcing a Majoron velocity. Due to the large number of such mechanisms we take the viewpoint that the Majoron rotation can be quite generic in the early universe.

Apart from directly applying the Affleck-Dine mechanism to the dynamics of the lepton number breaking scalar, there exists also the possibility that another rotating scalar transfers part of its charge to the Majoron \cite{Domcke:2022wpb}. For instance the Majoron and the phase of the Affleck-Dine field $\theta_\text{AD}$ could have a coupling to a   non-abelian gauge group with field strength tensor $G_{\mu\nu}$ of  $(\theta+\theta_\text{AD})G_{\mu\nu} \tilde{G}^{\mu\nu}$. The charge initially stored only in the Affleck-Dine field reads $m r_0^2$ for a quadratic potential with mass $m$. For $r_0\gg v_\sigma$ the charged is mostly stored in the Affleck Dine field, but since the vev $r_0$ decreases due to Hubble friction, eventually once $r_0 < v_\sigma$ almost all of the charge can be transferred into the Majoron rotation. Depending on whether the transfer rate $\gamma_S T^2/v_\sigma^2$ defined in terms of the sphaleron transition rate $\gamma_S$ is fast or not, one can estimate the Majoron velocity to be \cite{Domcke:2022wpb}
\begin{align}
    \dot{\theta} \simeq -m~\text{Min}\left[1,\frac{\gamma_S T^2}{3 v_\sigma^2 H(T)}\right]. 
\end{align}

Next we introduce some mechanisms that do not necessarily rely on the motion of a radial mode. 
Suppose the Majoron possesses a non-minimal coupling to the Ricci scalar $R$ \cite{Takahashi:2015waa,Bettoni:2018utf,Berbig:2024ufe,Chen:2025awt}
of the form $\xi M_\text{Pl.}^2 R (1-\cos{(\theta)})$.
During inflation $R=12 H_I^2$, and if inflation is followed by an epoch of kination \cite{Spokoiny:1993kt,Joyce:1996CP,Ferreira:1997hj,Co:2021lkc,Gouttenoire:2021wzu,Gouttenoire:2021jhk} the sign of $R$ changes to $R=-6 H^2$. This \enquote{flip} of the Majoron potential interchanges its maxima and minima thus forcing the Majoron to roll to a new minimum after inflation with an initial velocity given by \cite{Chen:2025awt}
\begin{align}
    \dot{\theta}\simeq \frac{\sqrt{12\xi}M_\text{Pl.} H_I}{v_\sigma}.
\end{align}
If Majoron domain walls are formed (see the next section \ref{sec:defects} for more details) the motion of the walls with velocity $v_W$ and Lorentz factor $\gamma_W$ through space during their collapsing phase due to a bias term  \cite{Kibble:1976sj,Vilenkin:1981zs,Sikivie:1982qv} causes a time dependent angular velocity of \cite{Daido:2015gqa,Mariotti:2024eoh} 
\begin{align}
    \dot{\theta} = \frac{2 m_j \gamma_W v_W}{\cosh{(m_j \gamma_W v_W (t-t_0))}},
\end{align}
where the wall is located at the origin of the chosen coordinate system at the time $t_0$.
For scenarios with a strong first order phase transition in the lepton number breaking sector, the evolution of the bubble walls in the radial direction of $\sigma$ can also drive the evolution of $\theta$ for certain choices of explicit symmetry breaking potentials just as in e.g. the thin wall regime of electroweak Baryogenesis \cite{Cohen:1991iu}.

Lepton number might be anomalous with respect to an additional $\text{U}(1)$ gauge group leading to a topological term $\alpha /(8\pi) \theta F_{\mu \nu} \tilde{F}^{\mu\nu}$, where $F_{\mu\nu}$ is the field strength tensor of the new interaction. In the presence of a parity violating gauge field background one finds that $\braket{F_{\mu \nu} \tilde{F}^{\mu \nu}}\neq 0$ and this can push the Majoron to a velocity of \cite{Kobayashi:2020bxq}
\begin{align}
    \dot{\theta} = \frac{1}{\frac{3(\omega+3)}{2}-n} \frac{\alpha}{8\pi } \frac{\braket{F_{\mu \nu} \tilde{F}^{\mu \nu}}}{H f_a^2},
\end{align}
where $\alpha$ is the dark fine structure constant and the scaling of $F_{\mu \nu} \tilde{F}^{\mu \nu} \sim 1/a^n$ was assumed. Here $\omega$ is the equation of state parameter of the dominant component that drives the expansion of the universe and we require $n\neq 5$ during radiation domination.
For the conditions that ensure the survival of such a background consult Refs.~\cite{Domcke:2019mnd,Kobayashi:2020bxq} and references therein. 

One of the main advantages of sourcing the Majoron velocity via the above mechanisms is that they can take place after inflation, allowing for a post-inflationary breaking of the lepton number symmetry, and hence avoiding the stringent constraints from isocurvature fluctuations altogether.
This comes at the price of allowing the formation of domain walls discussed in the next section.
In the \enquote{flipped} model of Refs.~\cite{Bettoni:2018utf,Chen:2025awt} the Majoron is already present during inflation, but isocurvature fluctuations can be avoided as the non-minimal coupling can source an effective Majoron mass larger than $H_I$ \cite{Berbig:2024ufe}.

\section{Majoron Dark Matter}\label{sec:DM}
\subsection{Kinetic Misalignment}\label{sec:KMM}
The general phenomenology of Majoron DM was discussed in Refs.~\cite{Garcia-Cely:2017oco,Brune:2018sab,Heeck:2019guh,Reig:2019sok}.  Previous studies about Majoron dark matter in Type II scenarios focused on production via misalignment \cite{Chao:2022blc} or Freeze-in \cite{Biggio:2023gtm}. Coherent oscillations \cite{Preskill:1982cy,Abbott:1982af,Dine:1982ah} of the Majoron begin when $3H(T_\text{osc.}^j)=m_j$, which implies during radiation domination
\begin{align}
    T_\text{osc.}^j  \simeq \SI{27}{\tera\electronvolt}   \sqrt{\frac{m_j}{1\;\text{eV}}},
\end{align}
and the corresponding relic abundance is 
\begin{align}\label{eq:mis}
    \Omega_j^\text{mis.}\;h^2 \simeq 0.12 \sqrt{\frac{m_j}{1\;\text{eV}}} \left(\frac{\braket{\theta_i^2}}{(2.1)^2}\right)  \left(\frac{v_\sigma }{\SI{4.3e11}{\giga\electronvolt}}\right)^2,
\end{align}
where $\theta_i$ denotes the initial misalignment angle of the Majoron after inflation. 
If $\text{U}(1)_L$ is broken before or during the visible part of inflation as e.g. in the Affleck-Dine scenario, $\theta_i$ is an unknown initial condition. On the contrary for post-inflationary lepton number breaking one can average over all Hubble patches to find that $\sqrt{\braket{\theta_i^2}}=2.1$ \cite{GrillidiCortona:2015jxo}, which we employ as a reference value 

In this work we will focus on  the kinetic misalignment mechanism introduced in Refs.~\cite{Co:2019jts,Chang:2019tvx,Co:2020dya} (see also Ref.~\cite{Barman:2021rdr}).
This scenario relies on the fact that the initially large kinetic energy of the axion $\dot{\theta}^2 v_\sigma^2/2$ redshifts due to the Noether charge conservation  $\dot{\theta}\sim 1/a^3 \sim s$.  Eventually the kinetic energy becomes comparable to the maximum value of the potential $2 m_j v_\sigma^2$ and the Majoron becomes trapped at a temperature  defined by $\dot{\theta}(T^j_\text{trap})= 2 m_j$
\begin{align}\label{eq:trap}
    T_\text{trap}^j \simeq  \SI{100}{\giga\electronvolt} \left(\frac{m_j}{\SI{1}{\electronvolt}}\right)^\frac{1}{3} \left(\frac{v_\sigma}{\SI{e8}{\electronvolt}}\right)^\frac{2}{3} 
    \left(\frac{0.22}{Y_\theta}\right)^\frac{1}{3}.
\end{align}
Kinetic misalignment 
dominates over the usual misalignment mechanism  if   $T_\text{trap}^j <T_\text{osc.}^j$, which can be expressed as a bound on the condensate yield  $Y_\theta = \dot{\theta}v_\sigma^2 / s$ \cite{Co:2019jts,Barman:2021rdr}
\begin{align}\label{eq:condKM}
    Y_\theta > 6\times 10^{-8} \sqrt{\frac{\SI{1}{\electronvolt}}{m_j}}\left(\frac{v_\sigma}{\SI{e8}{\giga\electronvolt}}\right)^2.
\end{align}
Since the trapping happens later than the onset of the conventional misalignment, the Majoron energy density has less time to dilute, and the observed relic abundance can be reproduced for smaller values of $v_\sigma$ than in Eq.~\eqref{eq:mis}. Because the Majoron is trapped close to the hilltop of its potential, it probes the anharmonic part of the cosine potential, and  the yield for the Majoron rotation $Y_j$ is actually larger by a factor of two
\begin{align}
    Y_j = 2 Y_\theta,
\end{align}
as  explained in Refs.~\cite{Co:2019jts,Gouttenoire:2021jhk}. The dark matter relic abundance then turns out to be 
\begin{align}\label{eq:abund}
    \frac{\rho_j}{s} = m_j Y_j = \SI{0.44}{\electronvolt} \left(\frac{m_j}{\SI{1}{\electronvolt}}\right) \left(\frac{Y_\theta}{0.22}\right).
\end{align}
In the following we assume that the dark matter from the kinetic misalignment mechanism makes up all of the relic abundance and trade $Y_\theta$ for  $m_j$  via the relation
\begin{align}\label{eq:theta-param}
    Y_\theta = \frac{\SI{0.22}{\electronvolt}}{m_j}.
\end{align}
Using the above relation we can re-express the condition for kinetic misalignment from Eq.~\eqref{eq:condKM} as
\begin{align}\label{eq:KMMoperative}
    v_\sigma < \SI{2e11}{\giga\electronvolt} \left(\frac{\SI{1}{\electronvolt}}{m_j}\right)^\frac{1}{4}.
\end{align}

So far we treated the Majoron as a homogeneous condensate also known as a zero mode. Higher momentum modes can be excited as the Majoron rolls over many minima of its potential, which leads to fast oscillations of the Majoron mass. This process is known as fragmentation and was investigated in Refs.~\cite{Jaeckel:2016qjp,Berges:2019dgr,Fonseca:2019ypl,Morgante:2021bks,Eroncel:2022vjg}.
Fragmentation starts to happen for \cite{Eroncel:2022vjg}
\begin{align}\label{eq:fragment}
    v_\sigma  <   \SI{5e9}{\giga\electronvolt}  \cdot \left(\frac{\SI{1}{\electronvolt}}{m_j}\right)^\frac{1}{4},
\end{align}
and for smaller $v_\sigma$ it can even occur before the trapping of the zero mode has taken place. 
However fragmentation is not expected to significantly change the relic abundance, as it occurs at the same time the Majoron becomes non-relativistic  \cite{Co:2021rhi,Harigaya:2021txz}, so the characteristic momentum scale is given by $m_j$. A recent lattice study in Ref.~\cite{Fasiello:2025ptb} confirms the expectation that the prediction for the relic abundance can change by $\mathcal{O}(1)$ numbers.

Recently another mechanism for dark matter called \enquote{acoustic misalignment} was put forward in Refs.~\cite{Bodas:2025eca,Eroncel:2025qlk}. Here curvature perturbations excite fluctuations of the rotating pNGB condensate, that initially redshift like radiation, before later becoming non-relativistic. In certain cases this contribution can be larger than the energy density of the zero mode from kinetic misalignment, but this intimately depends on the cosmological history, as well as on the curvature power spectrum $\mathcal{P}_{\mathcal{R}}(k)$ on scales $k$ not constrained by the CMB observations. Using the results of Ref.~\cite{Eroncel:2025qlk} we find that the ratio of the present day energy densities for acoustic misalignment (AMM) and kinetic misalignment obtained from the Affleck Dine mechanism read
\begin{align}\label{eq:AMM}
    \frac{\rho_{\text{AMM},0}}{\rho_{j,0}}\simeq \sqrt{\varepsilon} \left(\frac{S_i}{0.05 M_\text{Pl.}}\right) \left(\frac{\SI{e8}{\giga\electronvolt}}{v_\sigma}\right) \left(\frac{\mathcal{P}_{\mathcal{R}}(k_\text{kin})}{2\times 10^{-9}}\right),
\end{align}
where $k_\text{kin}$ is the wavenumber of the mode that reenters the horizon at the time of the onset of the Majoron's kination like behavior, which occurs once $S=v_\sigma$. For reasons that will be explained  below we take $\varepsilon\simeq 1$. One can see that acoustic misalignment dominates over kinetic misalignment  for e.g. $v_\sigma < \SI{e8}{\giga\electronvolt}$, if $S_i$ lies parametrically far below $M_\text{Pl.}$. Alternatively to suppress the acoustic contribution one needs smaller values of the curvature power-spectrum $\mathcal{P}_{\mathcal{R}}(k_\text{kin})$ on non-CMB scales, which is hard to reconcile with conventional single field inflation that predicts an almost scale-invariant spectrum.
For the other rotation mechanisms we can express the ratio of energy densities in terms of the Majoron yield as 
 \begin{align}\label{eq:acoustic}
   \frac{\rho_{\text{AMM},0}}{\rho_{j,0}}\simeq  0.2 Y_\theta \left(\frac{T_\text{kin}}{v_\sigma}\right) \left(\frac{\SI{e10}{\giga\electronvolt}}{v_\sigma}\right)  \left(\frac{\mathcal{P}_{\mathcal{R}}(k_\text{kin})}{2\times 10^{-9}}\right),
 \end{align}
 where $T_\text{kin}$ denotes the model-dependent temperature at which the kination behavior of the Majoron was initiated in a radiation dominated cosmology. For the Affleck Dine mechanism one has $T_\text{kin}=T_S$. We will always take $T_\text{kin}<v_\sigma$ to avoid thermal symmetry restoration. In principle one could try to suppress the acoustic misalignment contribution by taking $T_\text{kin}\ll v_\sigma$, but this has to be checked on a model-by-model basis, which we relegate to future work. For successful cogenesis of dark matter and the baryon asymmetry we require $T_\text{kin}\gtrsim \tilde{T}$, where $\tilde{T}$ defined in Eq.~\eqref{eq:Tfo} is the temperature at which the baryon asymmetry is frozen in.

The oscillations of the radial mode can also efficiently produce angular fluctuations via the parametric resonance effect \cite{Co:2017mop}, which can be the dominant source of non-thermal Majorons. However this production channel suffers from the drawback that the Majoron abundance has too large momenta, which can be in conflict with bounds on the warmness of dark matter. In Ref.~\cite{Co:2020dya} it was found that parametric resonance in a quartic potential is absent for $\varepsilon>0.8$ (see Eq.~\eqref{eq:eps}), or if the radial mode is thermalized before parametric can become efficient \cite{Co:2022qpr}. For a quartic potential this would need to occur before $S$ reaches $0.01 S_i$  \cite{Kasuya:1998td,Kasuya:1999hy,Kawasaki:2013iha}.

\subsection{Topological defects}\label{sec:defects}
If $\text{U}(1)_L$ is broken before or during inflation, as in the Afflek-Dine mechanism from section~\ref{sec:AD}, or the \enquote{flipped} mechanism of Refs.~\cite{Bettoni:2018utf,Chen:2025awt} from section~\ref{sec:NOAD}, the Majoron field will have a correlation length that is larger than the cosmological horizon. Therefore the Majoron is homogenized over our entire universe, so that no domain wall arise. 

On the contrary, if lepton number breaking occurs after inflation, the Majoron stochastically takes a different initial value in each Hubble patch. In this second scenario global cosmic strings are formed at the temperature $T_\text{string, form.}\simeq v_\sigma$ when $\text{U}(1)_L$ is spontaneously broken after inflation, so they are not exponentially diluted. The higher dimensional operator in Eq.~\eqref{eq:highdim} explicitly breaks $\text{U}(1)_L$ down a to $\mathcal{Z}_{|m-n|}$. Domain walls are formed
when the explicit symmetry breaking that gives rise to the Majoron's potential becomes cosmologically relevant at $T_\text{DW, form.}\simeq \sqrt{m_j M_\text{Pl.}}$ \cite{Reig:2019sok}. In our preferred  parameter space we have $m_j<\mathcal{O}(\SI{1}{\electronvolt})$ and $v_\sigma > \mathcal{O}(\SI{e6}{\giga\electronvolt}$) (see section \ref{sec:discuss}), so that $T_\text{string, form.}\gg T_\text{DW, form.}$, which implies that the walls are formed long after the strings.

The domain wall number reads
\begin{align}
    N_\text{DW}= |m-n|
\end{align}
and for $N_\text{DW}>1$ one can use bias terms from additional explicit symmetry breaking to make the walls collapse \cite{Kibble:1976sj,Vilenkin:1981zs,Sikivie:1982qv}, which will be fleshed out at the end of this section. Furthermore there exist other dedicated mechanisms to solve the domain wall problem  \cite{Lazarides:1982tw,Barr:1982bb,Barr:2014vva,Reig:2019vqh,Zhang:2023gfu}.

For $N_\text{DW}=1$ the strings \textit{might} chop the domain walls into smaller pieces similar to the case of  QCD axion  models \cite{Vilenkin:1982ks,Barr:1986hs}, but a careful dedicated analysis is required. For $N_\text{DW}=1$ the nucleation of closed string loops \cite{PhysRevD.26.435} can also destroy the walls, but we do not expect it to matter here because the nucleation rate is exponentially suppressed by $(v_\sigma/m_j)^2$.

In the Affleck-Dine mechanism quantum fluctuations of the radial mode occur for  $m_S(S_i)<3 H_I/2$, and induce fluctuations of the angular mode via the operator in Eq.~\eqref{eq:highdim}. Ref.~\cite{Co:2020dya} demonstrated that these fluctuations can cause domain walls without attached strings. However if parametric resonance \cite{Co:2017mop} occurs, the $\text{U}(1)_L$ symmetry is restored and strings are formed again during  the subsequent breaking of lepton number. The strings cause  the walls to collapse at the price of an abundance of warm Majorons from parametric resonance (see the discussion at the end of section~\ref{sec:KMM}). 

Irrespective of the dynamics underlying the walls' origins (fluctuations or $N_\text{DW}>1$) one can invoke an even higher dimensional operator, than the one in Eq.~\eqref{eq:highdim} with powers $m_2>m,\;n_2>n$.
If $|m_2-n_2|$ is not co-prime with $|m-n|$, then this operator can act \cite{Reig:2019sok} as a small bias term \cite{Kibble:1976sj,Vilenkin:1981zs,Sikivie:1982qv} that corresponds to a tiny correction of the Majoron mass $\Delta m_j\ll m_j$, and lifts one vacuum over the others, eventually causing the walls to collapse. Alternatively one could use the tiny contribution to the Majoron potential from the anomalous $\text{U}(1)_L$ breaking via instantons of the weak interaction, which requires a separate interaction that explicitly breaks $B+L$ \cite{Berbig:2025nrt}, to generate the bias term  \cite{Preskill:1991kd}. The abundance of Majorons released in the wall collapse is of the order of the usual misalignment contribution from Eq.~\eqref{eq:mis} multiplied by $m_j/\Delta m_j\gg1$ \cite{Reig:2019sok,Berbig:2023uzs}. While this can also enhance the Majoron relic abundance at smaller $v_\sigma$, depending in the ratio $m_j/\Delta m_j$, we do not consider this avenue further.

\subsection{Dark Matter stability and (in)direct detection}\label{sec:DMfeatures}
The ultralight Majoron should be stable on cosmological time-scales in order for it to constitute the dark matter. Due to its lightness the most important decay modes are to neutrinos and photons. For the decay width of the Majoron to neutrinos  one finds 
\begin{align}\label{eq:dec-nu}
    \Gamma\left(j\rightarrow \sum_i \nu_i  \nu_i\right) \simeq \frac{\sum_i m_{\nu_i}^2}{16\pi v_\sigma^2} m_j,
\end{align}
where we ignored the phase space suppression, and work in the single effective neutrino approximation with  $m_j > 2 \sqrt{\sum_i m_{\nu_i}^2}$. 
Normalized to the age of the universe $t_0 \simeq 13.79\;\text{Gyr}$ one obtains for the Majoron life-time \cite{Reig:2019sok}
\begin{align}\label{eq:lifetime}
    \tau_\nu= 1/ \Gamma\left(j\rightarrow \sum_i \nu_i  \nu_i\right) \simeq 3\times 10^5 \;t_0 \left(\frac{\SI{2.65e-3}{\electronvolt\squared}}{\overline{m}_\nu^2}\right) \left(\frac{1\;\text{eV}}{m_j}\right)  \left(\frac{v_\sigma}{10^{8}\;\text{GeV}}\right)^2.
\end{align}
The decay width of the Majoron to photons from the coupling in Eq.~\eqref{eq:gamma} is given by
\begin{align}\label{eq:dec-gamma}
    \Gamma\left(j\rightarrow \gamma \gamma \right) \simeq  \frac{\alpha^2}{64\pi^3} |g_{j\gamma\gamma}|^2 m_j^3.
\end{align}
By focusing on the coupling from the electron loop we can write down an analytical estimate for the Majoron lifetime 
\begin{align}
        \tau_\gamma= 1/ \Gamma\left(j\rightarrow \gamma \gamma \right) \simeq 1.5\times 10^4 \; t_0 \left(\frac{m_j}{\SI{1}{\electronvolt}}\right) \left(\frac{v_\sigma}{\SI{e8}{\electronvolt}}\right)^2 \left(\frac{\SI{1}{\mega\electronvolt}}{v_T}\right)^4,
\end{align}
which illustrates that the lifetime decreases for larger $v_T$.

The precise limit on the dark matter life-time depends on its mass and possible decay products, see Refs.~\cite{Audren:2014bca,Enqvist:2019tsa,Nygaard:2020sow,Alvi:2022aam}.
Here we impose the following limit on the Majoron lifetime derived from  CMB and Baryon Acoustic Oscillation (BAO) data  \cite{Alvi:2022aam}
\begin{align}\label{eq:tau}
    \tau_j > 250\;\text{Gyr}.
\end{align}

Direct detection experiments utilizing electron recoils could offer the opportunity to detect Majoron DM, and the first run of  \verb|XENONnT| already placed bounds on Majorons with $\SI{1}{\kilo\electronvolt}<m_j<\SI{130}{\kilo\electronvolt}$ with $|c_{e}|<10^{-13}\;(10^{-14})$ for $m_j=\SI{100}{\kilo\electronvolt}\;(\SI{1}{\kilo\electronvolt})$ \cite{XENON:2022ltv}.
By comparing with the limit from stellar cooling in Eq.~\eqref{eq:j-e} via the majoron-electron coupling, we can deduce that direct detection constraints only matter for $v_T\geq \mathcal{O}(\SI{1}{\giga\electronvolt})$.

In section \ref{sec:discuss} we will see that our preferred parameter space involves sub-eV Majorons, and hence we focus on the constraints for such light masses: Apart from the dark matter lifetime there exists a multitude of limits on dark matter decaying to X-ray and gamma ray
photons from various satellites that were compiled for $m_j<\SI{1}{\mega\electronvolt}$ in Ref.~\cite{Panci:2022wlc}. Numerically we find that these limits are only relevant for $v_T>\mathcal{O}(\SI{1}{\mega\electronvolt})$. Due to the smallness of our predicted majoron masses the decay $j\rightarrow e^+ e^-$ is kinematically forbidden.

Majoron dark matter decaying to neutrinos can also lead to signals \cite{Palomares-Ruiz:2007egs} in neutrino detectors such as \verb|Super-Kamiokande|  \cite{Super-Kamiokande:2002exp,Super-Kamiokande:2013ufi}, \verb|Borexino| \cite{Borexino:2010zht} or \verb|KamLAND| \cite{KamLAND:2011bnd}. As a consequence of the Majoron's coupling to neutrino mass eigenstates, the flavor composition of the emitted neutrino lines does not change between their production and their arrival on earth \cite{Garcia-Cely:2017oco}.
The authors of Ref.~\cite{Garcia-Cely:2017oco} find that the aforementioned neutrino experiments are only sensitive to $m_J>\mathcal{O}(\SI{1}{\mega\electronvolt})$, which is why we do not include these bounds. 

One potential probe of ultralight Majorons decaying to neutrinos \cite{McKeen:2018xyz,Chacko:2018uke}  are proposed experiments aiming at a direct detection of the cosmic neutrino background such as \verb|PTOLEMY| \cite{Betts:2013uya,PTOLEMY:2018jst}, and an overview over the field can be found in Ref.~\cite{Bauer:2022lri}.

The authors of Ref.~\cite{Reig:2019sok} concluded that this could work for  $m_j=\mathcal{O}(\SI{0.1}{\electronvolt}-\SI{1}{\electronvolt})$ and lifetimes of $\tau_\nu=(10-100)\;t_0$. While we note that this is an attractive possibility, we do not analyze this idea further, because the \verb|PTOLEMY| collaboration is still investigating the potential complications of actually reaching the required energy resolution required for neutrino detection \cite{PTOLEMY:2022ldz}, that were first pointed out by Ref.~\cite{Cheipesh:2021fmg}.

\subsection{Dark Radiation and Hot Dark Matter}

\begin{table}[t]
    \centering
    \begin{tabular}{|c|c|c|}
    \hline
       $\Delta N_\text{eff.}<$ & experiment & reference\\
     \hline
     \hline
        0.285 & \verb|Planck|+lensing+BAO & \cite{Planck:2018vyg,Abazajian:2019oqj}\\
        0.18 & \verb|Planck|+BBN & \cite{Fields:2019pfx,Yeh:2022heq}\\
        0.17 & \verb|ACT|+\verb|Planck|+lensing & \cite{ACT:2025tim} \\
     \hline
     \hline
        0.12 & \verb|COrE| & \cite{CORE:2017oje}\\
        0.12 & \verb|Euclid| & \cite{2011arXiv1110.3193L}\\
        0.12 &  \verb|SO|
        &\cite{SimonsObservatory:2018koc}\\
        \hline
        0.06 & \verb|CMB-S4| & \cite{Abazajian:2019eic,annurev-nucl-102014-021908}\\
        0.06 & \verb|PICO| & \cite{NASAPICO:2019thw}\\
        \hline
        0.014 & \verb|CMB-HD| & \cite{CMB-HD:2022bsz}\\
    \hline
    \end{tabular}
    \caption{Current limits \textit{(upper block)} and forecasts \textit{(lower block)} for the dark radiation abundance parameterized in terms of the shift in the effective number of neutrinos $\Delta N_\text{eff.}$ from various currently ongoing and proposed future experiments.}
    \label{tab:Neff}
\end{table}

In this work we assume that the cold non-thermal Majoron condensate produced via kinetic misalignment is the sole source of the observed dark matter abundance. In section \ref{sec:discuss} we will demonstrate that our preferred parameter  space from cogenesis involves sub-eV scale Majorons. Majorons with keV-scale masses \cite{Irsic:2017ixq,Lopez-Honorez:2017csg,Ballesteros:2020adh,DEramo:2020gpr,Decant:2021mhj} produced via freeze-in \cite{Hall:2009bx} are also viable dark matter candidates, but here we consider only ultralight Majorons, that would be relativistic during BBN or recombination if produced thermally. These thermal Majorons would therefore contribute to the dark radiation abundance during the aforementioned epochs, which is conventionally expressed in terms of the effective number of neutrinos $N_\text{eff.}$, and  faces tight constraints from e.g. \verb|Planck| CMB data together with BAO and the observed light element abundances from primordial nucleosynthesis (BBN) \cite{Planck:2018vyg}
\begin{align}
    N_\text{eff}^\text{BBN} = 2.97^{+0.58}_{-0.54}, \quad N_\text{eff}^\text{Planck+BAO} = 2.99^{+0.34}_{-0.33}.
\end{align}
These measurements  translate to a limit on additional dark radiation 
\begin{align}
    \Delta N_\text{eff.}\equiv  N_\text{eff.}- N_\text{eff.}^\text{SM},
\end{align}
where the Standard Model prediction $N_\text{eff.}^\text{SM}$ lies between 3.043 and 3.044 \cite{Cielo:2023bqp,Drewes:2024wbw}. 
We tabulated existing limits together with projected sensitivities for $\Delta N_\text{eff.}$ from several  ongoing and future experiments in table \ref{tab:Neff}. We do not include the most recent limits from \texttt{SPT} as they seem to indicate $\Delta N_\text{eff.}<0$  \cite{SPT-3G:2025bzu}.
One can determine $\Delta N_\text{eff.}$ from the following relation 
\begin{align}\label{eq:Neff}
    \Delta N_\text{eff.}(T) =   \frac{4}{7}\; g_{*\rho}(T) \left(\frac{10.75}{g_{*S}(T)}\right)^\frac{4}{3}  \frac{\rho_j(T)}{\rho_\text{SM}(T)}\quad \text{with} \quad \rho_\text{SM}(T) = \frac{\pi^2}{30} g_{*\rho}(T) T^4.
\end{align}
For a thermalized  Majoron, that decoupled from the bath at the temperature $T_\text{dec.}$ one would find that \cite{Abazajian:2019oqj}
\begin{align}
    \Delta N_\text{eff.} \simeq 0.027 \left(\frac{106.75}{g_{\rho}(T_\text{dec.})}\right)^\frac{4}{3},
\end{align}
which implies that $T_\text{dec.}$ should be larger than $\SI{100}{\mega\electronvolt}\;(\SI{1}{\mega\electronvolt})$ for $\Delta N_\text{eff.}<0.17\;(0.285)$. However for our scenario we typically find that thermalized Majorons are a problematic subcomponent of too hot dark matter, and hence we focus on the production via freeze-in \cite{Luo:2020fdt} from a thermal bath. 
We approximate the Majoron energy density as \cite{Heeck:2017xbu}
\begin{align}
    \rho_j \simeq 2.5 T\; n_j,
\end{align}
and we evaluate Eq.~\eqref{eq:Neff} at the production temperature $T_\text{prod.}$, that depends on the process at hand. 
Since $\rho_j \sim T n_j \sim T Y s$ and $\rho_\text{SM}\sim T^4\sim T s$ in terms of the dimensionless Majoron yield $Y_j^\text{FI}$ from freeze in and the entropy density $s$, we can write
\begin{align}
    \Delta N_\text{eff.} \simeq 45.2 \frac{Y_j^\text{FI}}{g_S(T_\text{prod.})}.
\end{align}

At temperatures below recombination, say during structure formation, thermal Majorons with masses above the eV-scale would act as a component of hot dark matter, since they were produced while being ultra-relativistic. Limits from structure formation  \cite{Irsic:2017ixq,Lopez-Honorez:2017csg,Ballesteros:2020adh,DEramo:2020gpr,Decant:2021mhj}  enforce that such a too warm component of DM has to be subdominant.
Therefore we require that the thermal abundance is always below 10\% of the abundance from kinetic misalignment  \cite{Boyarsky:2008xj,Peters:2023asu}
\begin{align}\label{eq:hotDM}
    Y_j^\text{FI} < 0.1 Y_\theta = \frac{\SI{0.022}{\electronvolt}}{m_j}.
\end{align}
Typically we find that the out-of-equilibrium conditions lead to stronger constraints on our parameter space, than the current limits on $\Delta N_\text{eff.}$. The previous condition on the hot dark matter abundance usually turns out to be weaker than the requirement of sufficiently long lived dark matter discussed in Eq.~\eqref{eq:tau}. 

\subsubsection{Decays}
We first discuss production from decays, which peak when the temperature falls to the value of the decaying particle's mass. In Ref.~\cite{Biggio:2023gtm} it was shown that the largest decay widths arise when the heavy components of $T$ decay to Majorons. However since these couplings only switch on after the SM has condensed at $T\lesssim v_H$, the authors of Ref.~\cite{Biggio:2023gtm} assume a modified electroweak phase transition that occurs at a larger temperature. We do not make this assumption and focus on the irreducible couplings to the SM like Higgs that lead to the invisible decay widths in Eq.~\eqref{eq:inv1}-\eqref{eq:inv2}. Here the production of Majorons will predominantly occur at $T_\text{prod.}=m_{h_H}$.
The decays $h_{H}\rightarrow jj$ and $h_{H}\rightarrow  Z j$  never reach equilibrium as long as 
\begin{align}\label{eq:non-therm}
    v_T < 
    \begin{cases}
    \SI{5.4}{\tera\electronvolt} \left(\frac{\SI{1}{\tera\electronvolt}}{m_T}\right) \left(\frac{v_\sigma}{\SI{e8}{\giga\electronvolt}}\right),\\
     \SI{84}{\giga\electronvolt} \sqrt{\frac{v_\sigma}{\SI{e8}{\electronvolt}}}.
    \end{cases}
\end{align}
We estimate the resulting Majoron yield following Ref.~\cite{Hall:2009bx}
\begin{align}
    Y_j^\text{FI} \simeq \frac{135}{8\pi^3 g_S(m_{h_H})}\begin{cases}
        \frac{\Gamma\left(h_H \rightarrow j j\right)}{2H(m_{h_H})},\\
         \frac{\Gamma\left(h_H \rightarrow Z j\right)}{H(m_{h_H})},
    \end{cases}
\end{align}
where  the factor of $1/2$ in the first line arises because $h_{H}\rightarrow jj$ produces two Majorons per decay. The corresponding contributions to the dark radiation abundance read
\begin{align}
    \Delta N_\text{eff.} \simeq 
    \begin{cases}
       3\times 10^{-41} \left(\frac{v_T}{\SI{1}{\kilo\electronvolt}}\right)^4 \left(\frac{m_T}{\SI{1}{\tera\electronvolt}}\right)^4  \left(\frac{\SI{e8}{\giga\electronvolt}}{v_\sigma}\right)^4,\\
       10^{-33} \left(\frac{v_T}{\SI{1}{\kilo\electronvolt}}\right)^4 \left(\frac{\SI{e8}{\giga\electronvolt}}{v_\sigma}\right)^2.
    \end{cases}
\end{align}

The radial mode of the lepton number breaking scalar can also source Majorons via the decays
\begin{align}
    \Gamma(h_\sigma\rightarrow  jj) = \frac{1}{32 \pi}\frac{m_S^3}{v_\sigma^2}
\end{align}
which implies \cite{Co:2020jtv}
\begin{align}\label{eq:saxionNeff}
    \Delta N_\text{eff.}\simeq 0.25 \cdot \sqrt{\frac{\SI{10}{\mega\electronvolt}}{m_S}}\cdot \left(\frac{v_\sigma}{10^8\;\text{GeV}}\right).
\end{align}
Here the choice of  mass  $m_S$ was motivated by the parameters required for the Affleck Dine mechanism, see Eq.~\eqref{eq:lambdaAD}. 
The above was derived assuming a thermalized population of $h_\sigma$, which can arise due to thermalization from the coupling to the Higgs
$\lambda_{\sigma H} |\sigma|^2 |H|^2$ \cite{Co:2020dya}. For the other mechanisms that induce the Majoron velocity in section \ref{sec:NOAD}, $m_S$ can be much larger, e.g. $m_S\simeq v_\sigma$, and the dark radiation abundance can be be suppressed far below any current or future limit.

\subsubsection{Scattering}\label{sec:scat}
Majoron production from fermion annihilations like $\overline{f} f \rightarrow Z j$ is typically suppressed due to the small couplings in Eq.~\eqref{eq:cferm}, and due to the dependence on $v_T$ for charged leptons and quarks, this channel only becomes active after the electroweak transition, so that the $Z$ boson production becomes kinematically blocked. Pair production $\overline{f} f \rightarrow j j$ is then the only remaining channel, but this is even more suppressed due to two insertions of the fermion-Majoron-coupling.

In the scalar sector there is an irreducible contribution from the  $\lambda_{\sigma H T}$ coupling first considered in Ref.~\cite{Biggio:2023gtm}, and the scattering rate can be estimated  to be\footnote{Here we work in the unbroken electroweak phase, because the relevant temperature scale is far above the electroweak scale.}
\begin{align}
    \Gamma(H H \rightarrow T  j) \simeq 3\times 10^{-4} \left(\frac{v_T}{v_\sigma}\right)^2 \left(\frac{m_T}{v_H}\right)^4 T.
\end{align}
Since here $\Gamma/H(T) \sim 1/T$ the rate is infrared (IR) dominated, and its contribution is largest at the smallest available temperature. 
Since the triplet is expected to be heavier than the SM Higgs boson, it becomes non-relativistic first, which suppresses the scattering rate exponentially. Therefore we take $T_\text{prod.}\simeq m_T/5.4$ following Ref.~\cite{Biggio:2023gtm}, and we find that this rate does not establish chemical equilibrium as long as 
\begin{align}\label{eq:non-therm2}
    v_T <     \SI{12}{\giga\electronvolt} \left(\frac{\SI{1}{\tera\electronvolt}}{m_T}\right)^\frac{3}{2} \left(\frac{v_\sigma}{\SI{e8}{\giga\electronvolt}}\right).
\end{align}
In the analysis of Ref.~\cite{Biggio:2023gtm}
this condition could not be satisfied because the authors focused on the parameter range with $v_T=\mathcal{O}(\SI{1}{\giga\electronvolt})$, $v_\sigma = \mathcal{O}(\SI{e7}{\giga\electronvolt})$ and $m_T<\SI{1}{\tera\electronvolt}$ in order to maximize the direct detection prospects via scattering with electrons (see the discussion below Eq.~\eqref{eq:tau}). To ameliorate this problem together with the overproduction of Majorons from decays, the authors then assumed an early matter dominated epoch with a subsequent release of entropy. We do not work in the aforementioned region of parameter space and are therefore free from these problems. 

In case the triplet is too heavy to be thermally produced the relevant rate involves a virtual triplet (see Fig.~\ref{fig:SeesawII})
\begin{align}
    \Gamma\left(\sum_i L_i L_i \rightarrow H H  j \right)\simeq 0.2\frac{\sum_i m_{\nu_i}^2}{v_\sigma^2} \frac{T^5}{v_H^4}.
\end{align}
Here we find that $\Gamma/H \sim T^3$ and hence the reaction is UV dominated.
We expect the production temperature to be $T_\text{prod.}=\text{Min}[T_\text{RH},v_\sigma]$, where $T_\text{RH}$ denotes the reheating temperature and we note that Majoron is only present after the lepton number breaking transition at $T=v_\sigma$.\footnote{Here we assume no field excursion in the radial direction of the lepton number breaking scalar.}
This process   never equilibrates for $v_\sigma < T_\text{RH}$ as long as 
\begin{align}
    v_\sigma < \SI{9e12}{\giga\electronvolt} \left(\frac{\SI{2.65e-3}{\electronvolt\squared}}{  \overline{m}_{\nu}^2}\right),
\end{align}
and for $v_\sigma > T_\text{RH}$ as long as 
\begin{align}
    v_\sigma > \SI{3.3e11}{\giga\electronvolt} \left(\frac{\overline{m}_\nu^2 }{\SI{2.65e-3}{\electronvolt\squared}}\right) \left(\frac{T_\text{RH}}{\SI{e12}{\giga\electronvolt}}\right)^\frac{3}{2}.
\end{align}
The yields for all scatterings are estimated following  Ref.~\cite{Elahi:2014fsa}
\begin{align}
    Y_j^\text{FI} \simeq \frac{180}{2(2\pi)^7} \begin{cases}
        \frac{ \Gamma(H H \rightarrow T j)}{H(T) g_s(T)}\Big|_{T=m_T/5.4},\\ 
        \frac{ \Gamma\left(\sum_i L_i L_i \rightarrow H H j \right)}{H(T) g_s(T)}\Big|_{T=\text{Min}[T_\text{RH},v_\sigma]},
    \end{cases}
\end{align}
and the dark radiation abundance turns out to be
\begin{align}
    \Delta N_\text{eff.} \simeq 
    \begin{cases}
        10^{-19} \left(\frac{v_T}{\SI{1}{\kilo\electronvolt}}\right)^2 \left(\frac{m_T}{\SI{1}{\tera\electronvolt}}\right)^3 \left(\frac{\SI{e8}{\giga\electronvolt}}{v_\sigma}\right)^2,\\
        2\times 10^{-10} \left(\frac{ \overline{m}_{\nu}^2}{\SI{2.65e-3}{\electronvolt\squared}}\right)  \left(\frac{v_\sigma}{\SI{e8}{\giga\electronvolt}}\right),
\end{cases}
\end{align}
where we focused on the case $v_\sigma < T_\text{RH}$ to reduce the number of free parameters.

\section{Low scale Leptogenesis (triplet in the  thermal bath)}\label{sec:low}
In this section we consider Leptogenesis with a triplet, that is light enough to be present as a dynamical degree of freedom in the thermal bath. 
For each particle species $X$ we define the following dimensionless abundances and asymmetries 
\begin{align}
    Y_X = \frac{n_X}{s}, \quad \Sigma_X = Y_X + Y_{\overline{X}} \simeq 2  Y_X, \quad \Delta_X =Y_X - Y_{\overline{X}},
\end{align}
and note that the chemical potential is given by 
\begin{align}
    \frac{\Delta_X}{Y_X^\text{eq.}} = 2 \frac{\mu_X}{T},
\end{align}
where the superscript \enquote{eq.} denotes thermal equilibrium.
Throughout this work we resort to the Maxwell-Boltzmann approximation, and only incorporate the difference between bosons and fermions in our treatment of the spectator processes. 
In a $CPT$ violating background from e.g. a Majoron rotation the chemical potentials are shifted via the following prescription \cite{Chun:2023eqc}
\begin{align}
    \mu_X \rightarrow \mu_X + \frac{B-L}{2} \dot{\theta}, 
\end{align}
which in our scenario implies that 
\begin{align}\label{eq:shift}
    \mu_L \rightarrow \mu_L - \frac{\dot{\theta}}{2},\quad 
    \mu_T \rightarrow \mu_T + \dot{\theta}.
\end{align}
The validity of the above shift for the chemical potential of the bosonic triplet $T$  will be explicitly demonstrated in appendix~\ref{sec:KG}.
The chemical potential of the Higgs $\mu_H$ does not get shifted, as this field is not charged under $B-L$.

\subsection{No oscillating radial mode}
\begin{figure}[t!]
    \centering
    \includegraphics[width=0.8\textwidth]{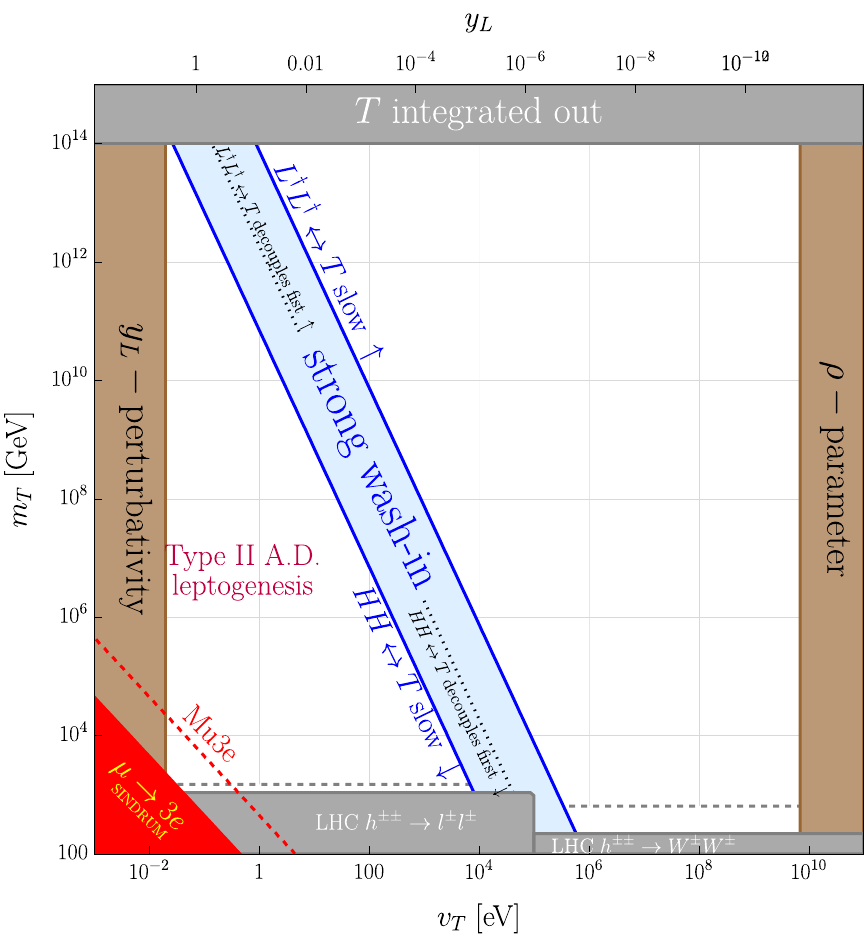}
    \caption{Summary of terrestrial limits on the Type II Seesaw mechanism from section \ref{eq:pheno}. Straight line show current constraints and colored dashed lines refer to projected future limits. 
    In order to show the range of $y_L$ on the upper axis we assumed a fiducial neutrino mass of $\SI{0.05}{\electronvolt}$ for reference. 
    The blue band depicts the \enquote{strong wash-in} regime for Leptogenesis defined in Eq.~\eqref{eq:band}. The \enquote{weak wash-in} regime is situated to the left and to the right of the blue band. In the left part of the plot labeled \enquote{Type II A.D. Leptogenesis} the scenario of Refs.~Refs.~\cite{Barrie:2021mwi,Barrie:2022cub,Han:2023kjg,Kaladharan:2024bop} is operative. For $m_T>\mathcal{O}(\SI{e14}{\giga\electronvolt})$ the triplet would be too heavy to be in the thermal bath at the time the reactions from the lepton number violating Weinberg operator decouple, which will be discussed in section \ref{sec:high}.}
    \label{fig:triplet}
\end{figure}

The Boltzmann equations from Ref.~\cite{Hambye:2005tk} are translated to the notation of Refs.~\cite{Domcke:2020kcp,Chun:2023eqc} to read in a Majoron background
\begin{align}
     \frac{\text{d}}{\text{d}\ln{(T)}}  \left(\frac{\mu_{L}}{T}\right)  &=  \frac{\Gamma_{L}^\text{av.}}{H(T)}  \left(2\frac{\mu_{L}}{T} + \frac{\mu_T}{T}\right)\label{eq:BEL},\\
    \frac{\text{d}}{\text{d} \ln{(T)}} \left(\frac{\mu_H}{T}\right)&=  \frac{\Gamma_H^\text{av.}}{H(T)}  \left(2\frac{\mu_H}{T} - \frac{\mu_T}{T} -\frac{\dot{\theta}}{T}\right),\label{eq:BEH}\\
    \frac{\text{d}}{\text{d} \ln{(T)}} \left(\frac{\mu_T}{T}\right) &= \frac{1}{6}\frac{\Gamma_{L}^\text{av.}}{H(T)}  \left(2\frac{\mu_{L}}{T} + \frac{\mu_T}{T}\right) - \frac{1}{3}\frac{\Gamma_H^\text{av.}}{H(T)}  \left(2\frac{\mu_H}{T} - \frac{\mu_T}{T} -\frac{\dot{\theta}}{T}\right)\label{eq:BET},
\end{align}
and here we defined the  total chemical potential $\mu_L$
\begin{align}
    \mu_{L}=3\mu_{L_i},
\end{align}
in terms of the chemical potential of each generation $\mu_{L_i}$.
We further dropped the contribution from the scattering terms as they are subdominant at $T= m_T$ for small couplings.
Moreover  there is no contribution from $CP$-violating out-of-equilibrium decays, since we work with a single generation of triplets. 
Here the $CP$ conserving tree level decay rates read \cite{Hambye:2005tk}
\begin{align}
     \sum_i \Gamma_{L_i} &\equiv  \Gamma\left(T\rightarrow \sum_i L_i^\dagger L_i^\dagger\right)= \text{Tr}(|y_L|^2) \frac{m_T}{16\pi}= \frac{\sum_i m_{\nu_i}^2}{8\pi v_T^2}m_T,\label{eq:GammaL}\\
    \Gamma_H &\equiv \Gamma\left(T\rightarrow H H\right) = \frac{\lambda_{\sigma H T}^2 v_\sigma^2}{16 \pi m_T}= \frac{v_T^2}{4\pi v_H^4} m_T^3\label{eq:GammaH},
\end{align}
where we ignored the phase space suppression from the SM Higgs mass.
The thermally averaged decay rates in the  Maxwell Boltzmann approximation  are found to be 
\begin{align}
  \Gamma_L^\text{av.} &\equiv \sum_i \frac{\text{K}_1(z)}{\text{K}_2(z)} \frac{\Sigma_T^\text{eq.}}{Y_{L_i}^\text{eq.}} \Gamma_{L_i},\\
    \Gamma_H^\text{av.} &\equiv \frac{\text{K}_1(z)}{\text{K}_2(z)} \frac{\Sigma_T^\text{eq.}}{Y_H^\text{eq.}} \Gamma_H,
\end{align}
in terms of the modified Bessel function of the first (second) kind $\text{K}_1(z)\; \left(\text{K}_2(z)\right)$ and $z\equiv m_T/T$. One can understand the effect of the thermal average from the following expression
\begin{align}\label{eq:taylor}
    \Gamma_i^\text{av.} \simeq \frac{3}{2}\Gamma_i \begin{cases}
        \left(\frac{m_T}{T}\right) \quad &\text{for} \quad T\gg m_T,\\
        \sqrt{\frac{\pi}{2}} \left(\frac{m_T}{T}\right)^\frac{3}{2} e^{-\frac{m_T}{T}} \quad &\text{for} \quad T\ll m_T,
    \end{cases}
    \quad \text{for} \quad i=L,H.
\end{align}

In Eq.~\eqref{eq:BEL} the contribution of $\dot{\theta}$ cancels out between $L$ and $T$  and hence the inverse $L L \rightarrow T^\dagger$ can not produce a chemical potential for either $T$ or $L$. 
The subsequent equation Eq.~\eqref{eq:BEH} on the other hand involves $T$ and $H$, where $\mu_H$ does not get shifted, since $H$ carries no lepton number. Consequently the $\dot{\theta}$ dependence does not cancel and the reaction $HH \rightarrow T$ can produce a chemical potential $\mu_T \sim \dot{\theta}$ that is then transmitted to $\mu_L$ via the decay $T \rightarrow L^\dagger L^\dagger$. This is only possible due to the fact that $T$ is not self conjugate, and can therefore have its own asymmetry described by the Boltzmann equation Eq.~\eqref{eq:BET}. 
An efficient production of lepton asymmetry therefore requires both rates  $\Gamma_H^\text{av.}$ and $\Gamma_L^\text{av.}$  to be fast at $T=m_T$. 
This is not surprising, because \enquote{wash-in} Leptogenesis \cite{Domcke:2020quw} requires the rates that \enquote{wash-out} the asymmetry produced in the conventional out-of-equilibrium decay scenario to be fast, and for scalar triplets it is well known, that \enquote{wash-out} is only efficient, if both rates are in equilibrium \cite{Hambye:2005tk}. One can understand this from the observation that the violation of lepton number requires the presence of both couplings $y_L$ and $\kappa=\lambda_{\sigma H T}v_\sigma/\sqrt{2}$.

\begin{figure}[t!]
    \centering    \includegraphics[width=0.9\textwidth]{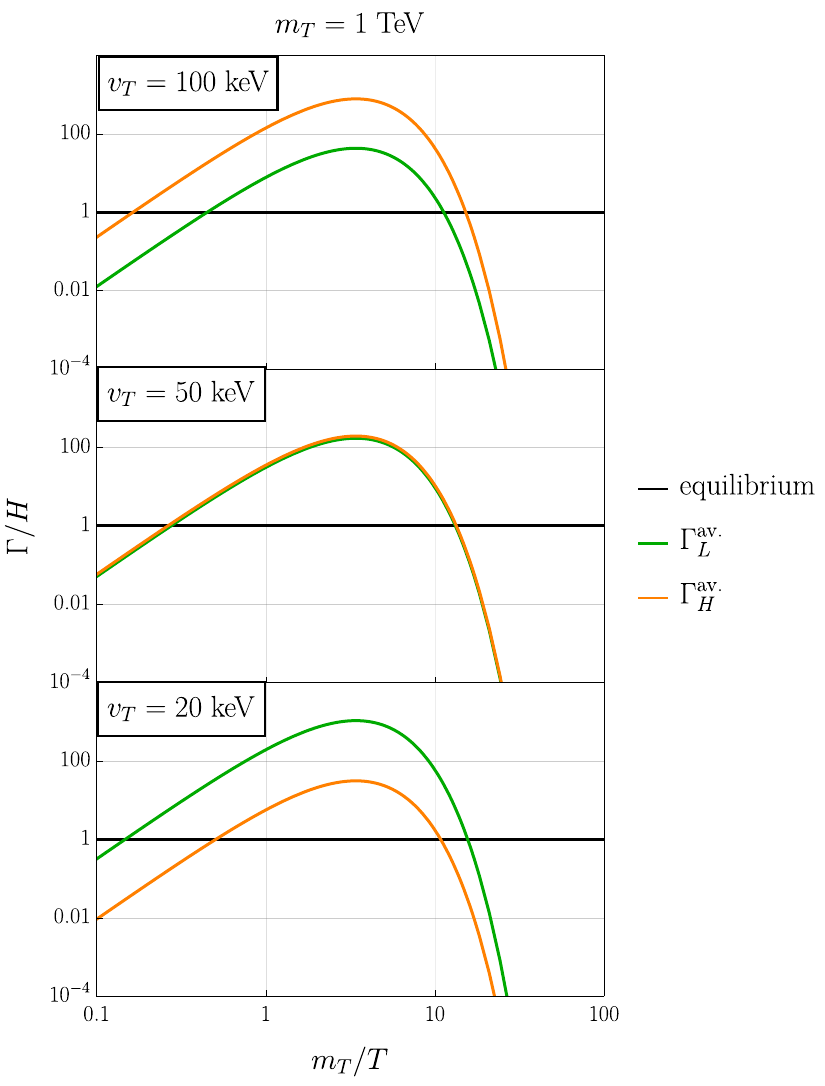}
    \caption{Rates for the processes $L^\dagger L^\dagger \leftrightarrow T$ \textit{(green}) and $HH \leftrightarrow T$ \textit{(orange)} normalized to the Hubble rate as a function of $m_T$ over temperature.
    For constant $m_T$ the values of $v_T$ were varied such that $L^\dagger L^\dagger \leftrightarrow T$ decouples after \textit{(upper panel)}, at the same time as \textit{(middle panel)} or before \textit{(lower panel)}  $HH \leftrightarrow T$.}
    \label{fig:rates}
\end{figure}

The decays and inverse decays reach their maximum around $T=m_T$ (see Fig.~\ref{fig:rates}), and hence for strong washout we impose the  conditions\footnote{At $T=m_T$ the decay rates and thermally averaged decay rates are parametrically of the same order of magnitude, and numerically we obtain essentially the same bound on $v_T$.}
\begin{align}
  K_{L,H}\equiv   \frac{\Gamma_{L,H}^\text{av.}}{H(T)}\Big|_{T=m_T} \geq 1,
\end{align}
which imply that
\begin{align}\label{eq:band}
 \SI{8.5}{\kilo\electronvolt} \sqrt{\frac{\SI{1}{\tera\electronvolt}}{m_T}} <   v_T < \SI{277}{\kilo\electronvolt} \sqrt{\frac{\SI{1}{\tera\electronvolt}}{m_T}}\sqrt{\frac{\overline{m}_\nu^2}{\SI{2.65e-3}{\electronvolt\squared}}}.
\end{align}
We call this the \enquote{strong wash-in} regime, and depicted it as a blue band in Fig.~\ref{fig:triplet}. One can analytically approximate the temperatures at which the processes come into thermal equilibrium 
\begin{align}
    T_L^\text{EQ} &\simeq \SI{2.3}{\tera\electronvolt} \left(\frac{m_T}{\SI{1}{\tera\electronvolt}}\right)^\frac{2}{3} \left(\frac{\SI{100}{\kilo\electronvolt}}{v_T}\right)^\frac{2}{3} \left(\frac{\overline{m}_\nu^2}{\SI{2.65e-3}{\electronvolt\squared}}\right)^\frac{1}{3},\\
    T_H^\text{EQ} &\simeq \SI{6.1}{\tera\electronvolt} \left(\frac{m_T}{\SI{1}{\tera\electronvolt}}\right)^\frac{4}{3} \left(\frac{v_T}{\SI{100}{\kilo\electronvolt}}\right)^\frac{2}{3}.
\end{align}
In order for Leptogenesis to proceed in the strong wash-in regime, we have to additionally  ensure that the rates equilibrate before the critical temperature $T_c$ \cite{Chun:2023eqc}, at which the sphaleron transitions decouple \cite{DOnofrio:2014rug}
\begin{align}
    T_c \simeq \SI{130}{\giga\electronvolt}.
\end{align}
From the conditions $T_{L,H}^\text{EQ} > T_c$ we obtain
\begin{align}\label{eq:freezein}
\SI{308}{\electronvolt} \left(\frac{\SI{1}{\tera\electronvolt}}{m_T}\right)^2<v_T < \SI{7.6}{\mega\electronvolt} \left(\frac{m_T}{\SI{1}{\tera\electronvolt}}\right)\sqrt{\frac{\overline{m}_\nu^2}{\SI{2.65e-3}{\electronvolt\squared}}},
\end{align}
and these inequalities are always satisfied, when we impose the parameter range in Eq.~\eqref{eq:band} for the strong wash-in together with the lower bounds on $m_T$ from Eq.~\eqref{eq:LHC1} and Eq.~\eqref{eq:LHC2}. Also note that since $T_{L,H}^\text{EQ}>m_T$,  equilibration below $T_c$ would imply $m_T <T_c=\SI{130}{\giga\electronvolt}$, which is ruled out by the previously mentioned LHC searches. 

The analytical expressions for the freeze-out temperature  with $1\ll K_{L,H}<10^4$ are estimated to be (see section 6 of Ref.~\cite{Giudice:2003jh})
\begin{align}
    T_L^\text{FO} &\simeq  \frac{m_T}{5\sqrt{\log\left(\frac{9 \sqrt{\frac{5}{2}} M_\text{Pl} \overline{m}_\nu^2}{32\pi^2 \sqrt{g_\rho(T_L^\text{FO})}m_T v_T^2}\right)}},\\
    T_H^\text{FO} &\simeq  \frac{m_T}{5\sqrt{\log\left(\frac{9 \sqrt{\frac{5}{2}} M_\text{Pl} m_T v_T^2}{16\pi^2\sqrt{g_\rho(T_H^\text{FO})}v_H^4}\right)}}.
\end{align}
Here due to weak logarithmic dependence on the model parameters we find that 
\begin{align}\label{eq:TFO}
    T_{L,H}^\text{FO} = \mathcal{O}\left(\frac{m_T}{10}\right).
\end{align}
Note that the above expressions are just used to provide some analytical intuition. All figures will be plotted using the numerically determined freeze-out temperatures. 

The reaction involving $L$ freezes out before the inverse decays to $H$ as long as
\begin{align}\label{eq:focond}
    v_T > \SI{48}{\kilo\electronvolt} \left(\frac{\overline{m}_\nu^2}{\SI{2.65e-3}{\electronvolt\squared}}\right)^\frac{1}{4} \sqrt{\frac{\SI{1}{\tera\electronvolt}}{m_T}},
\end{align}
and a numerical example that demonstrates, which rate decouples first was displayed in Fig.~\ref{fig:rates}.
We also show the regions in the $m_T$ versus $v_T$ parameter plane, in which each of the rates decouples first in Fig.~\ref{fig:triplet}.

The original Boltzmann equations from Ref.~\cite{Hambye:2005tk} satisfy a sum rule
\begin{align}
    2 \Delta_T + \Delta_H - \Delta_{L} =0,
\end{align}
which corresponds to the following condition on the chemical potentials 
\begin{align}\label{eq:Y0}
    6 \mu_T + 2\mu_H - \mu_{L} =0,
\end{align}
and one can check that this is satisfied by  Eqns.~\eqref{eq:BEL}-\eqref{eq:BET}.
This statement is nothing more than hypercharge conservation \cite{Hambye:2005tk} for the triplet and its decay products.\footnote{It is important to note that the Boltzmann equations from Ref.~\cite{Hambye:2005tk} were derived using Maxwell-Boltzmann statistics, which treats bosons and fermions the same. On the other hand the hypercharge constraint in Eqns.~\eqref{eq:Y0}-\eqref{eq:Y} was evaluated using the Fermi-Dirac and Bose-Einstein quantum statistics, due to which the fermionic asymmetries pick up a factor of $1/2$ compared to the bosonic ones. If we were to use Maxwell-Boltzmann statistics for this constraint, we would obtain  $6 \mu_T + 2\mu_H - 2\mu_{L}$ instead of the correct result  $6 \mu_T + 2\mu_H - \mu_{L}$. 
When converting the  $\Delta_i$ to the chemical potentials $\mu_i$ with Maxwell Boltzmann statistics one actually obtains a factor of $Y_{L_i}^\text{eq.}/\Sigma_T^\text{eq.}=1/3$ in front of the first term in Eq.~\eqref{eq:BET}, which would lead to $6 \mu_T + 2\mu_H - 2\mu_{L}$. In order to make this Boltzmann equation compatible with the proper hypercharge conservation constraint derived from quantum statistics, we inserted the missing factor of $1/2$ in Eq.~\eqref{eq:BET} by hand, which explains the factor of $1/6$.} Our analysis treats the lepton number violating processes on the same footing as the other spectator processes \cite{Domcke:2020quw}, which redistribute the lepton asymmetry (see Ref.~\cite{Harvey:1990qw}), and therefore the previously mentioned sum rule is included in the conservation of the total hypercharge of the plasma 
\begin{align}\label{eq:Y}
    3 \left(\mu_{Q_i} + \mu_{u_i} - \mu_{d_i} - \mu_{L_i} -\mu_{e_i}\right) +6\mu_T + 2\mu_H =0,
\end{align}
where we remind the reader that $3\mu_{L_i}=\mu_L$.
We take all Standard Model Yukawa interactions to be fast, which is accurate below about $\SI{e5}{\giga\electronvolt}$ \cite{Nardi:2005hs}, and  include the equilibrated interactions of the triplet 
\begin{align}
\mu_T + 2\mu_{L_i} =0, \quad \mu_T - 2\mu_H + \dot{\theta}=0. 
\end{align}
The sphaleron redistribution coefficient can be read off from 
\begin{align}
    \mu_{B_i} = -\frac{14}{51} \dot{\theta}   = \frac{28}{61} \mu_{(B-L)_i},
\end{align}
and its value of $28/61$ is larger than the sphaleron redistribution coefficient of $28/79$ for models without a thermalized triplet scalar. Note that the value of this coefficient changes for models in which the Majoron is identified with a QCD axion due to its direct coupling to QCD sphalerons, see e.g. \cite{Co:2020jtv}. 
In general the resulting baryon asymmetry can be parameterized as 
\begin{align}\label{eq:MASTER}
    \Delta_B = \frac{7}{153} \text{Min}\left[1,\frac{\Gamma_L^\text{av.}}{H(T)}\right] \text{Min}\left[1,\frac{\Gamma_H^\text{av.}}{H(T)}\right] \left(\frac{T}{v_\sigma}\right)^2 Y_\theta\Big|_{T=\tilde{T}},
\end{align}
with 
\begin{align}\label{eq:Tfo}
    \tilde{T} \equiv 
    \begin{cases}
      \text{Max}\left[T_L^\text{FO},T_H^\text{FO}\right] \quad &\text{for}\; K_L\geq 1 \land K_H \geq 1 \land \text{Max}\left[T_L^\text{FO},T_H^\text{FO}\right] >T_c, \\
        T_L^\text{FO} \quad &\text{for} \; K_L \geq 1 \land K_H <1 \land T_L^\text{FO}>T_c,\\
        T_H^\text{FO} \quad &\text{for} \; K_L <1 \land K_H \geq 1\land T_H^\text{FO}>T_c,\\
        T_c \quad &\text{else}.
    \end{cases}
\end{align}
In the above $\tilde{T}$ is determined by which process decouples first. 
If the rates involving the triplets never thermalize, or only decouple below the sphaleron decoupling at $T_c$, the sphalerons freeze-out first and hence $\tilde{T} = T_c$. Otherwise the condition in Eq.~\eqref{eq:focond} allows one to determine whether $T_L^\text{FO} $ or $T_H^\text{FO}$ is larger. We can deduce from Eq.~\eqref{eq:band}  that either $\Gamma_L^\text{av.}$ or $\Gamma_H^\text{av.}$, but not both, can be slow.  Therefore the asymmetry can only depend on at most a single factor of either $\Gamma_L^\text{av.}/H(T)$ or $\Gamma_H^\text{av.}/H(T)$.\footnote{The only exception to this rule is if $T$ never develops a thermal abundance to begin with, so that both averaged decay rates are rescaled by $\Sigma_T/\Sigma_T^\text{eq.}<1$, where $\Sigma_T$ is the non-thermal abundance. This can suppress even the larger decay rate below the Hubble rate. However since $T$ has gauge couplings thermalization is hard to avoid unless one invokes kinematical suppression via e.g. low-scale reheating with $T_\text{RH}\lesssim m_T$, which will not be considered further in this work.}

Note that spontaneous Leptogenesis in a Majoron background is not affected by the triplet's  gauge interactions, unlike the conventional out-of-equilibrium decay scenario: There the source term for the lepton asymmetry is $\varepsilon_L (\Sigma_T-\Sigma_T^\text{eq.})$, where $\varepsilon_L$ is the amount of $CP$ violation per triplet decay corresponding to the second Sakharov condition, and $\Sigma_T-\Sigma_T^\text{eq.}$ is the deviation from equilibrium encoded in the third Sakharov condition. One expects that the electroweak gauge annihilations $\overline{T} T \rightarrow WW, WY, YY$ keep $\Sigma_T$ closer to $\Sigma_T^\text{eq.}$ reducing the efficiency of the conventional Leptogenesis; however Ref.~\cite{Hambye:2005tk} showed that in the strong washout regime the decays and inverse decays actually are the dominant processes thermalizing the triplet. In spontaneous Leptogenesis the source term is instead given by  $\dot{\theta}$, which breaks $CP$ and $CPT$ spontaneously. Therefore the third Sakharov condition and the corresponding out-of-equilibrium conditions are not required. Furthermore gauge annihilations only act on the \textit{symmetric} component $\Sigma_T$, but we use the inverse decays $HH \leftrightarrow T$ to generate the \textit{asymmetric} component $\Delta_T$, which stores the asymmetry, before releasing it via decays to leptons.

\subsubsection{Strong Wash-In}

\begin{figure}[t!]
    \centering
    \includegraphics[width=0.45\textwidth]{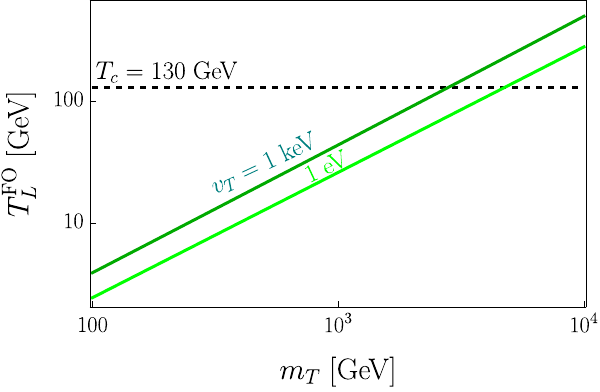}
    \includegraphics[width=0.45\textwidth]{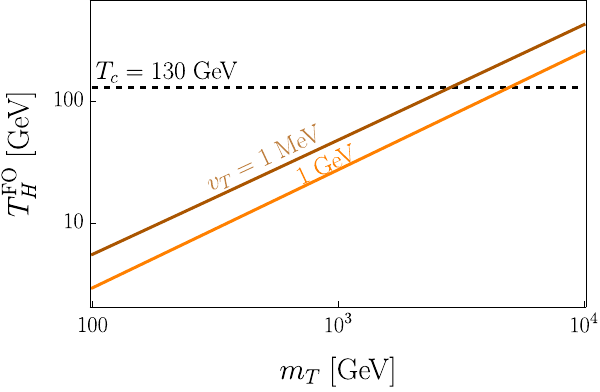}
    \caption{Numerically determined freeze-out temperatures of $\Gamma_L^\text{av.}$ \textit{(left)} and $\Gamma_H^\text{av.}$ \textit{(right)} as a function of $m_T$ for different choices of $v_T=\{ \SI{1}{\electronvolt},\;\SI{1}{\kilo\electronvolt},\;\SI{1}{\mega\electronvolt},\;\SI{1}{\giga\electronvolt}\}$. Note that the choices for $v_T$ not depicted in each respective plot do not lead to successful thermalization.}
    \label{fig:temps}
\end{figure}

In order to develop some intuition for the required parameter space we proceed with analytical estimates:
Saturating the inequality in Eq.~\eqref{eq:focond} means that $\Gamma_L^\text{av.}$ and $\Gamma_H^\text{av.}$ decouple around the same time, and as long as Eq.~\eqref{eq:band} is satisfied we can work in the strong wash-in regime 
\begin{align}
        \Delta_B \simeq  \frac{7}{153}   \left(\frac{\tilde{T}}{v_\sigma}\right)^2 Y_\theta.
\end{align}
Since the freeze-out of the inverse decays typically happens at the temperature given by Eq.~\eqref{eq:TFO},  we deduce that these processes decouple above $T_c$ for $m_T > \mathcal{O}(10) T_c= \SI{1.3}{\tera\electronvolt}$.
It is important to emphasize that this is just a rough  estimate for the sake of analytical understanding. In Fig.~\ref{fig:temps} we plot the numerically determined freeze-out temperature $T_{L,H}^\text{FO}$ for both rates $\Gamma_{L,H}^\text{av.}$,  and find in each case that decoupling actually occurs above $T_c$ for somewhat larger values of $m_T\simeq (3-5)\;\text{TeV}$. 
Our previous estimate allows us to write a simplified approximate expression for the decoupling temperature 
\begin{align}\label{eq:tildeapprox}
    \tilde{T} \simeq \begin{cases}
        \frac{m_T}{\mathcal{O}(10)} \quad & \text{for} \quad m_T > \SI{1.3}{\tera\electronvolt},\\
        T_c \quad &\text{for} \quad  m_T < \SI{1.3}{\tera\electronvolt}.
    \end{cases}
\end{align}
Trading $Y_\theta$ for $m_j$  we find that 
\begin{align}\label{eq:analyt-strong}
    \Delta_B \simeq 8.76\times 10^{-11}  \left(\frac{\SI{e8}{\giga\electronvolt}}{v_\sigma }\right)^2
    \begin{cases}
           \left(\frac{m_T}{\SI{100}{\tera\electronvolt}}\right)^2 \left(\frac{\SI{1}{\electronvolt}}{m_j}\right) \quad & \text{for} \quad m_T > \SI{1.3}{\tera\electronvolt},\\
           \left(\frac{\SI{0.2}{\milli\electronvolt}}{m_j}\right) \quad &\text{for} \quad  m_T < \SI{1.3}{\tera\electronvolt}.
    \end{cases}
\end{align}
Here we need $v_T\simeq\SI{4.8}{\kilo\electronvolt}$ for $m_T=\SI{100}{\tera\electronvolt}$ from Eq.~\eqref{eq:focond}.

\subsubsection{Weak Wash-In}
Next we show an analytical expression for the case when one of the rates is suppressed compared to the other
\begin{align}
    \Delta_B \simeq \frac{7}{153} \left(\frac{\tilde{T}}{v_\sigma}\right)^2 Y_\theta
    \begin{cases}
    \frac{\Gamma_L^\text{av.}}{H}\Big|_{\tilde{T}} \quad \text{for} \quad  v_T>\SI{277}{\kilo\electronvolt} \sqrt{\frac{\SI{1}{\tera\electronvolt}}{m_T}}\sqrt{\frac{\overline{m}_\nu^2}{\SI{3e-3}{\electronvolt\squared}}},\\
    \frac{\Gamma_H^\text{av.}}{H}\Big|_{\tilde{T}} \quad \text{for} \quad v_T <  \SI{8.5}{\kilo\electronvolt} \sqrt{\frac{\SI{1}{\tera\electronvolt}}{m_T}},
    \end{cases}
\end{align}
where $\tilde{T}$ is still given by Eq.~\eqref{eq:tildeapprox}.
In case $\Gamma_L^\text{av.}$ never thermalizes  we obtain
\begin{align}\label{eq:Lslow}
      \Delta_B \simeq 8.76\times 10^{-11} \left(\frac{\SI{e8}{\giga\electronvolt}}{v_\sigma}\right)^2
      \sqrt{\frac{\overline{m}_\nu^2}{\SI{2.65e-3}{\electronvolt\squared}}} 
      \begin{cases}
      \left(\frac{m_T}{\SI{10}{\tera\electronvolt}}\right) 
      \left(\frac{\SI{100}{\kilo\electronvolt}}{v_T}\right)^2
      \left(\frac{\SI{2.7}{\milli\electronvolt}}{m_j}\right),\\
      \left(\frac{m_T}{\SI{1}{\tera\electronvolt}}\right)^\frac{5}{2} 
      \left(\frac{\SI{500}{\kilo\electronvolt}}{v_T}\right)^2
      \left(\frac{\SI{71}{\micro\electronvolt}}{m_j}\right).
      \end{cases}
\end{align}
If instead $\Gamma_H^\text{av.}$ never thermalizes  we find 
\begin{align} \label{eq:Hslow}
    \Delta_B \simeq 8.76\times 10^{-11} \left(\frac{\SI{e8}{\giga\electronvolt}}{v_\sigma}\right)^2
    \begin{cases}
    \left(\frac{m_T}{\SI{10}{\tera\electronvolt}}\right)^3 \left(\frac{v_T}{\SI{1}{\kilo\electronvolt}}\right)^2
    \left(\frac{\SI{0.5}{\milli\electronvolt}}{m_j}\right),\\
    \left(\frac{m_T}{\SI{1}{\tera\electronvolt}}\right)^\frac{9}{2} \left(\frac{v_T}{\SI{5}{\kilo\electronvolt}}\right)^2 
    \left(\frac{\SI{81}{\micro\electronvolt}}{m_j}\right).
    \end{cases}
\end{align}
One should be careful with the above expressions for $m_T <\SI{1.3}{\tera\electronvolt}$ as they non-linearly depend on $m_T$ due to the Boltzmann suppression $e^{-m_T/T_c}$ (see Eq.~\eqref{eq:taylor}).

\subsection{Constraints}
Overall it is important to emphasize, that our Leptogenesis scenario can in principle work with most  triplet masses above the collider limits in  Eqns.~\eqref{eq:LHC1}-\eqref{eq:LHC2}. More specifically we can accommodate TeV-scale triplets in reach of present and  next generation collider experiments.

The backreaction of the fermionic asymmetry generation on the Majoron rotation is negligible as long as $T\ll v_\sigma$ \cite{Co:2019wyp}, so the condensate persists until later times and can source the dark matter abundance. Here this constraint implies  $\tilde{T} \ll v_\sigma$ so that 
\begin{align}
    v_\sigma \gg T_c \quad \text{for} \quad  m_T < \SI{1.3}{\tera\electronvolt},
\end{align}
as well as 
\begin{align}\label{eq:BR2}
    m_T \ll 10 v_\sigma \quad \text{for} \quad  m_T > \SI{1.3}{\tera\electronvolt}.
\end{align}

Furthermore we have to ensure that the rolling Majoron only gets trapped in its potential, after baryogenesis has completed at the temperature $\tilde{T}$. Therefore we demand that 
\begin{align}
    \tilde{T} > T_\text{trap},
\end{align}
where the trapping temperature is defined in Eq.~\eqref{eq:trap}. For the relevant parameter space we find that this condition holds as long as $m_j<\mathcal{O}(\SI{10}{\kilo\electronvolt})$, which is always satisfied.

In order for our assumption of radiation domination to be valid, we have to demand the kinetic energy of the Majoron rotation is subdominant to the energy density of the thermal plasma at least at the time of asymmetry production 
\begin{align}\label{eq:kination}
    \frac{\dot{\theta}^2 v_\sigma^2}{2} < \frac{\pi^2}{30} g_{\rho}(\tilde{T}) \tilde{T}^4,
\end{align}
which corresponds to an upper limit on $Y_\theta$.\footnote{Note that an era of kination domination requires a previous epoch of matter domination as shown in Ref.~\cite{Gouttenoire:2021jhk}. Since the radial mode in a quartic potential redshifts as radiation and its thermalization also produces radiation, which redshifts slower $(\sim 1/a^4)$ than the Majoron's kinetic energy $(\sim 1/a^6)$, this would necessitate an additional ingredient to realize the intermediate matter domination. Overall this region of parameter space corresponds to a cosmic history that is more involved than simple radiation domination.}
Here we follow the treatment of Ref.~\cite{Chun:2023eqc}, where $Y_\theta=\dot{\theta}v_\sigma^2/s$ is fixed to reproduce the baryon asymmetry. Using this together with the analytical result from the strong wash-in regime in Eq.~\eqref{eq:analyt-strong} we find that 
\begin{align}
    v_\sigma < 
    \begin{cases}
            \SI{3e11}{\giga\electronvolt}\left(\frac{m_T}{\SI{10}{\tera\electronvolt}}\right) \quad & \text{for} \quad m_T > \SI{1.3}{\tera\electronvolt},\\
        \SI{1.2e10}{\giga\electronvolt} \quad &\text{for} \quad  m_T < \SI{1.3}{\tera\electronvolt},
    \end{cases}
\end{align}
where we used $g_{\rho}(\tilde{T}) = 106.75 = g_s (\tilde{T})$. 
For the more suppressed asymmetry production from weak wash-in with a slow $\Gamma_L^\text{inv.}$ discussed in Eq.~\eqref{eq:Lslow} we find  that
\begin{align}
    v_\sigma < 
      \left(\frac{\overline{m}_\nu^2}{\SI{2.65e-3}{\electronvolt\squared}}\right)
    \begin{cases}
    \SI{2.2e10}{\giga\electronvolt}
    \left(\frac{\SI{100}{\kilo\electronvolt}}{v_T}\right)^2 \quad & \text{for} \quad m_T > \SI{1.3}{\tera\electronvolt},\\
    \SI{4.5e9}{\giga\electronvolt}
    \left(\frac{m_T}{\SI{1}{\tera\electronvolt}}\right)^\frac{5}{2} \left(\frac{\SI{500}{\kilo\electronvolt}}{v_T}\right)^2 \quad & \text{for} \quad m_T < \SI{1.3}{\tera\electronvolt},
    \end{cases}
\end{align}
and for the case with a slow $\Gamma_H^\text{av.}$ from Eq.~\eqref{eq:Hslow}
we deduce 
\begin{align}
    v_\sigma <
    \begin{cases}
        \SI{4e9}{\giga\electronvolt} 
        \left(\frac{m_T}{\SI{10}{\tera\electronvolt}}\right)^2 
        \left(\frac{v_T}{\SI{1}{\kilo\electronvolt}}\right)^2\quad & \text{for} \quad m_T > \SI{1.3}{\tera\electronvolt},\\
        \SI{5e9}{\giga\electronvolt} \left(\frac{m_T}{\SI{1}{\tera\electronvolt}}\right)^\frac{9}{2} \left(\frac{v_T}{\SI{5}{\kilo\electronvolt}}\right)^2 \quad & \text{for} \quad m_T < \SI{1.3}{\tera\electronvolt}.
    \end{cases}
\end{align}

Our calculation assumed that $\theta$ is a slowly varying background field. In order for $\theta$ to remain approximately constant over the time scale of thermal processes $1/T$ we impose that \cite{Co:2020jtv}
\begin{align}\label{eq:large}
  |\dot{\theta}| < \tilde{T},
\end{align}
and proceed in a similar vein to the previous limit.
For all cases we find that the $v_\sigma$ dependence divides out. In the strong wash-in regime with $Y_\theta$ fixed via the observed baryon asymmetry the adiabaticity condition is always satisfied. In the weak wash-in regime with  a slow $\Gamma_L^\text{av.}$ discussed in Eq.~\eqref{eq:Lslow} we obtain an upper limit on $v_T$ of 
\begin{align}
    v_T < \sqrt{\frac{\overline{m}_\nu^2}{\SI{2.65e-3}{\electronvolt\squared}}} 
    \begin{cases}
        \SI{163}{\mega\electronvolt}\sqrt{\frac{\SI{10}{\tera\electronvolt}}{m_T}} \quad & \text{for} \quad m_T > \SI{1.3}{\tera\electronvolt},\\
        \SI{1}{\giga\electronvolt} \left(\frac{m_T}{\SI{1}{\tera\electronvolt}}\right)^\frac{5}{4}\quad & \text{for} \quad m_T < \SI{1.3}{\tera\electronvolt},
    \end{cases}\label{eq:adiab1}
\end{align}
 and for  the case with a slow $\Gamma_H^\text{av.}$ from Eq.~\eqref{eq:Hslow} we find a lower limit on $v_T$ of 
 \begin{align}
     v_T > 
     \begin{cases}
        \SI{1.5}{\electronvolt} \sqrt{\frac{\SI{10}{\tera\electronvolt}}{m_T}} \quad & \text{for} \quad m_T > \SI{1.3}{\tera\electronvolt},\\
         \SI{2.3}{\electronvolt} \left(\frac{\SI{1}{\tera\electronvolt}}{m_T}\right)^\frac{9}{4}\quad & \text{for} \quad m_T < \SI{1.3}{\tera\electronvolt}.
     \end{cases}\label{eq:adiab2}
\end{align}
By comparing these bounds with the definition of the weak wash-in regime in Eq.~\eqref{eq:band}, one can deduce that they do not impose significant constraints on our parameter space.

The authors of Ref.~\cite{Harvey:1990qw} argued that one should take
\begin{align}
    |\dot{\theta}|<m_T,
\end{align}
to avoid Bose condensation of the triplet scalar. From Eq.~\eqref{eq:large} we see, that  $|\dot{\theta}|$ is bounded from above by $\tilde{T}$, whose order of magnitude can be found in Eq.~\eqref{eq:tildeapprox}, and this evinces that Bose condensation of the triplet does not occur. 

\subsection{Chiral hypermagnetic instability}
The impact of a rotating pNGB on the overproduction of baryon asymmetry via the chiral hypermagnetic instability was analyzed in Ref.~\cite{Co:2022kul}. We gieve a brief summary of their arguments here: 
The coupling of the Majoron to the leptons can induce a chemical potential in hypercharge of 
\begin{align}
    \mu_{\text{Y}5}= \sum_i s_i g_i Q_\text{Y}[\psi_i]^2 \mu_i \equiv c_5 \dot{\theta}, 
\end{align}
where $g_i$ is the number of degrees of freedom for each multiplet and $s_i$ equals $+1\;(-1)$ for left-(right-)chiral fermions.
If all SM fermions are in equilibrium we find that
\begin{align}
    \frac{ \mu_{\text{Y}5}}{T}= 51.4 Y_\theta \left(\frac{T}{v_\sigma}\right)^2.
\end{align}
For the fastest growing helicity mode one finds a rate of growth of  \cite{Co:2022kul}
\begin{align}
    \Gamma_\text{CPI} = \frac{\alpha_\text{Y}^2 \mu_{\text{Y}5}^2}{2\pi^2 \sigma_\text{Y}},
\end{align}
in terms of the hypercharge finestructure constant $\alpha_\text{Y}\simeq 0.01$, as well as the thermal conductivity in the SM plasma before the electroweak phase transition  $\sigma_\text{Y}\simeq 54 T$ \cite{Arnold:2000dr}.
For the Affleck-Dine mechanism the authors of Ref.~\cite{Co:2022kul} require that 
\begin{align}\label{eq:hyper}
    \frac{ \Gamma_\text{CPI}}{H}\Big|_{T_S} < c_\text{CPI} = 10,
\end{align}
because for $T<T_S\;(T>T_S)$ one can deduce from the time dependence of $\dot{\theta}\sim T^3\;(T)$ that $\Gamma_\text{CPI}/H\sim \dot{\theta}^2/T^3\sim T^3 \;(1/T)$ during radiation domination. We generalize this result to other mechanism for the Majoron's initial velocity by considering only $\dot{\theta}\sim T^3$. Thus the rate is UV dominated and should be evaluated at the temperature $T_\text{kin}$ at which the Majoron rotation is initiated. To be model-independent we take $T_\text{kin}\simeq \tilde{T}$, where the temperature of baryogenesis $\tilde{T}$ that was defined in Eq.~\eqref{eq:Tfo}. Furthermore we again fix $Y_\theta$ by reproducing the observed baryon asymmetry, as we did in the discussion below Eq.~\eqref{eq:kination}.

In the strong wash-in regime we find from Eq.~\eqref{eq:analyt-strong} that Eq.~\eqref{eq:hyper} is always satisfied for $\tilde{T}=T_c\; (m_T<\SI{1.3}{\tera\electronvolt})$ 
and for $\tilde{T}=m_T/10\; (m_T>\SI{1.3}{\tera\electronvolt})$ 
that the absence of the instability implies
\begin{align}
    m_T>\SI{647}{\kilo\electronvolt},
\end{align}
which is also always satisfied. 
In the weak wash-in regime with a slow $\Gamma_L^\text{av.}$ we obtain from Eq.~\eqref{eq:Lslow} that
\begin{align}\label{eq:CPL1}
    v_T < \sqrt{\frac{\overline{m}_\nu^2}{\SI{2.65e-3}{\electronvolt\squared}}} 
    \begin{cases}
    \SI{1.5}{\mega\electronvolt}  \left(\frac{\SI{10}{\tera\electronvolt}}{m_T}\right)\quad & \text{for}\quad m_T > \SI{1.3}{\tera\electronvolt},\\\SI{3.6}{\mega\electronvolt} \left(\frac{M_T}{\SI{1}{\tera\electronvolt}}\right) \quad & \text{for}  \quad m_T <\SI{1.3}{\tera\electronvolt}.
    \end{cases}
\end{align}
If instead $\Gamma_H^\text{av.}$ is slow we obtain from Eq.~\eqref{eq:Hslow} that 
\begin{align}\label{eq:CPL2}
    v_T> \begin{cases}
    \SI{76}{\electronvolt}\left(\frac{\SI{10}{\tera\electronvolt}}{m_T}\right)^\frac{3}{4}
        \quad & \text{for}\quad m_T > \SI{1.3}{\tera\electronvolt},\\
        \SI{205}{\electronvolt}\left(\frac{\SI{1}{\tera\electronvolt}}{m_T}\right)^9\quad & \text{for}\quad m_T < \SI{1.3}{\tera\electronvolt}.
    \end{cases}
\end{align}

\subsection{Oscillating radial mode}
So far we have assumed that the vev of $\sigma$ is stabilized at its present day value $v_\sigma$ during Leptogenesis. For a Majoron rotation from the  Affleck-Dine mechanism in section~\ref{sec:AD} we have to ensure that the  dynamics at the beginning of the radial oscillations, when the radial mode takes a large field value  $S_i \gg v_\sigma$, do not  affect these results. Processes like $S HH \leftrightarrow T$ could modify the triplet asymmetry production at early time and its rate scales as 
\begin{align}
    \Gamma(S HH \leftrightarrow T) \simeq \frac{\lambda_{\sigma H T}^2}{16\pi} \frac{S^2}{T},
\end{align}
where for a quartic potential with $S\sim 1/a$ we find $\Gamma(S HH \leftrightarrow  T)/H(T) \sim 1/T$ during radiation domination, implying that this rate is IR dominated. The large field value never affects Leptogenesis, if the reaction involving it only equilibrates after $h_\sigma$ has settled to its true minimum at $v_\sigma$, which for a quartic potential occurs at 
\begin{align}
    T_\text{min}= \frac{v_\sigma}{S_i}T_\text{osc.},
\end{align}
where $T_\text{osc.}$ was defined in Eq.~\eqref{eq:Tosc}.
This implies the condition
\begin{align}\label{eq:seq}
    v_T< \SI{11}{\mega\electronvolt} \left(\frac{\SI{1}{\tera\electronvolt}}{m_T}\right)^2 \left(\frac{v_\sigma}{\SI{e9}{\giga\electronvolt}}\right)^\frac{3}{2} \left(\frac{T_\text{osc.}}{\SI{e15}{\giga\electronvolt}}\right)^\frac{3}{2} \left(\frac{M_\text{Pl.}}{S_i}\right).
\end{align}
Here we assume that the production temperature of the baryon asymmetry  $\tilde{T}$ defined in Eq.~\eqref{eq:tildeapprox} is below $ T_\text{min}$, which implies
\begin{align}
    \tilde{T} < \SI{e5}{\giga\electronvolt} \left(\frac{v_\sigma}{\SI{e9}{\giga\electronvolt}}\right) \left(\frac{M_\text{Pl.}}{S_i}\right) \left(\frac{T_\text{osc.}}{\SI{e15}{\giga\electronvolt}}\right).
\end{align}
If $m_T>\SI{1.3}{\tera\electronvolt}$ we can deduce from Eq.~\eqref{eq:tildeapprox} that
\begin{align}
    m_T < \SI{e6}{\giga\electronvolt} \left(\frac{v_\sigma}{\SI{e9}{\giga\electronvolt}}\right) \left(\frac{M_\text{Pl.}}{S_i}\right) \left(\frac{T_\text{osc.}}{\SI{e15}{\giga\electronvolt}}\right).
\end{align}
Thus we can sequester the generation of the rotation in the Affleck Dine scenario at the large temperature $T_\text{osc.}$ from the generation of the baryon asymmetry at the much lower temperature $\tilde{T}$.

\section{High scale Leptogenesis (triplet not in the thermal bath)}\label{sec:high}
Here we comment on the features of the scenario where the triplet is heavy enough to be integrated out at all times.
In case the triplet is not part of the thermal bath, we can work in the effective operator limit making use of the Weinberg operator. Consequently one would expect that Leptogenesis in this regime is not sensitive to the UV completion of the Weinberg operator in a similar manner to the scenarios discussed in Refs.~\cite{Kusenko:2014uta,Ibe:2015nfa,Daido:2015gqa,Co:2020jtv,Chao:2023ojl,Chun:2023eqc,Datta:2024xhg,Wada:2024cbe}.
The Boltzmann equation for the lepton chemical potential reads
\begin{align}
     \frac{\text{d}}{\text{d}\ln{(T)}} \left(\frac{\mu_L}{T}\right)=  \frac{\Gamma_W}{H(T)}\left(\frac{\mu_L}{T}-\frac{\dot{\theta}}{T}\right),
\end{align}
but here we will not discuss Leptogenesis further as it has been covered in a wide array of literature before \cite{Kusenko:2014uta,Ibe:2015nfa,Daido:2015gqa,Co:2020jtv,Chao:2023ojl,Chun:2023eqc,Datta:2024xhg,Wada:2024cbe}. Instead we comment on the implications for the Type II Seesaw. 
Here the rate for the  process  $LL \leftrightarrow HH$ and all processes related by crossing symmetry is 
\begin{align}
    \Gamma_W\equiv \Gamma\left(LL \leftrightarrow HH\right) = \frac{1}{4\pi^3} \frac{\sum_i m_{\nu_i}^2}{v_H^4}T^3,
\end{align}
which freezes out at 
\begin{align}\label{eq:TWFO}
    T_W^{\text{FO}} \simeq \SI{2.41e14}{\giga\electronvolt} \left(\frac{\SI{2.65e-3}{\electronvolt\squared}}{\overline{m}_\nu^2}\right).
\end{align}
For this estimate to be self-consistent we have to impose that the triplet is at least heavier than the freeze-out temperature 
\begin{align}\label{eq:conv}
    m_T > \mathcal{O}(\SI{e14}{\giga\electronvolt}),
\end{align}
which was depicted in Fig.~\ref{fig:triplet}.
Such heavy triplet masses require large values of $v_\sigma$  to maintain the smallness of $v_T$ in Eq.~\eqref{eq:vT} for perturbative values of  $\lambda_{\sigma H T}<\sqrt{4\pi}$. We can use this to find a lower limit on the lepton number breaking scale of 
\begin{align}\label{eq:sigmaLOWER}
    v_\sigma > \frac{v_T}{\sqrt{4\pi}} \frac{m_T^2}{v_H^2} \simeq \SI{5e13}{\giga\electronvolt} \left(\frac{v_T}{\SI{1}{\electronvolt}}\right) \left(\frac{m_T}{\SI{e14}{\giga\electronvolt}}\right)^2.
\end{align}

A distinction occurs when the radial mode of $\sigma$ develops a large initial field value as in the Affleck-Dine mechanism discussed in section~\ref{sec:AD}.
Because the Weinberg operator for the Type II Seesaw is sensitive to an insertion of this vev and hence its dynamics (see Fig.~\ref{fig:SeesawII} and Eq.~\eqref{eq:SeeswII}), the rate for processes like $S LL\leftrightarrow HH$ changes to (see e.g. Ref.~\cite{Berbig:2023uzs,Co:2024oek})
\begin{align}
\Gamma_{SW} \equiv \Gamma\left(S LL \leftrightarrow HH\right) = \frac{1}{4\pi^3} \frac{\sum_i m_{\nu_i}^2}{v_H^4}T^3 \left(\frac{S}{v_\sigma}\right)^2 = \Gamma_{W}  \left(\frac{S}{v_\sigma}\right)^2.
\end{align}
For concreteness we assume a quartic potential for $\sigma$ so that during radiation domination 
\begin{align}
    S(T)= S_i \frac{T}{T_\text{osc.}}.
\end{align}
Assuming that $S$ is already moving this process freezes out at
\begin{align}\label{eq:TSWFO}
   T_{SW}^\text{FO}= \SI{e8}{\giga\electronvolt} \left(\frac{\SI{2.65e-3}{\electronvolt\squared}}{\overline{m}_\nu^2}\right) \left(\frac{v_\sigma}{\SI{e10}{\giga\electronvolt}}\right)^\frac{2}{3}\left(\frac{M_\text{Pl.}}{S_i}\right)^\frac{2}{3} \left(\frac{T_\text{osc.}}{\SI{e14}{\giga\electronvolt}}\right)^\frac{2}{3},
\end{align}
and once can see that the $v_\sigma/S_i$ suppression lowers the freeze-out temperature. This could in principle allow for a smaller $m_T$, if we only require it to be absent from the plasma below $ T_{SW}^\text{FO}$. Note however that the scenario with large $S_i$ faces strong constraints from dark radiation due to  irreducible contribution from decays of the radial mode in Eq.~\eqref{eq:saxionNeff}.
Requiring $\Delta N_\text{eff.}<0.17$ implies
\begin{align}
    v_\sigma < \SI{2.5e12}{\giga\electronvolt} \left(\frac{T_\text{osc.}}{\SI{6.5e15}{\giga\electronvolt}}\right)^2 \left(\frac{M_\text{Pl.}}{S_i}\right),
\end{align}
where we used the largest possible $T_\text{osc.}$ from Eq.~\eqref{eq:ToscLimit}. Note that Leptogenesis in this regime is independent of $v_T$, but as we argued in Eq.~\eqref{eq:vTLow} $v_T$ can not be too far below the eV-scale. If we use this information in Eq.~\eqref{eq:sigmaLOWER}, we find that 
\begin{align}
    v_\sigma > \SI{2.5e12}{\giga\electronvolt} \left(\frac{v_T}{\SI{0.02}{\electronvolt}}\right) \left(\frac{m_T}{\SI{1.6e14}{\giga\electronvolt}}\right)^2.
\end{align}
On the one hand this example demonstrates that we can not lower the triplet mass much below the conventional upper limit in Eq.~\eqref{eq:conv}, and on the other hand we see that at best there is only a narrow allowed parameter window. The available parameter space is even further reduced once we include the other reactions producing Majorons from section \ref{sec:scat}. Together with the fact that  high scale scenario is hard to test experimentally these arguments illustrate, why we de not consider the high-scale scenario in more detail.

\section{Discussion}\label{sec:discuss}

\begin{figure}[t!]
    \centering
    \includegraphics[width=0.55\linewidth]{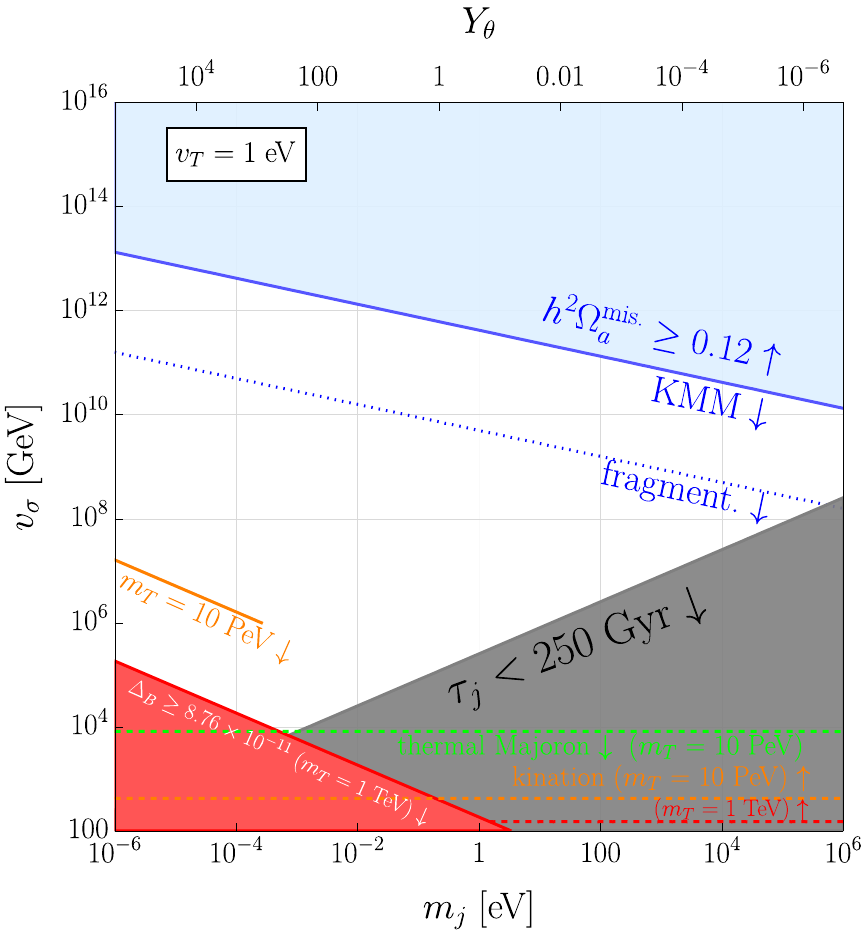}
    \includegraphics[width=0.55\linewidth]{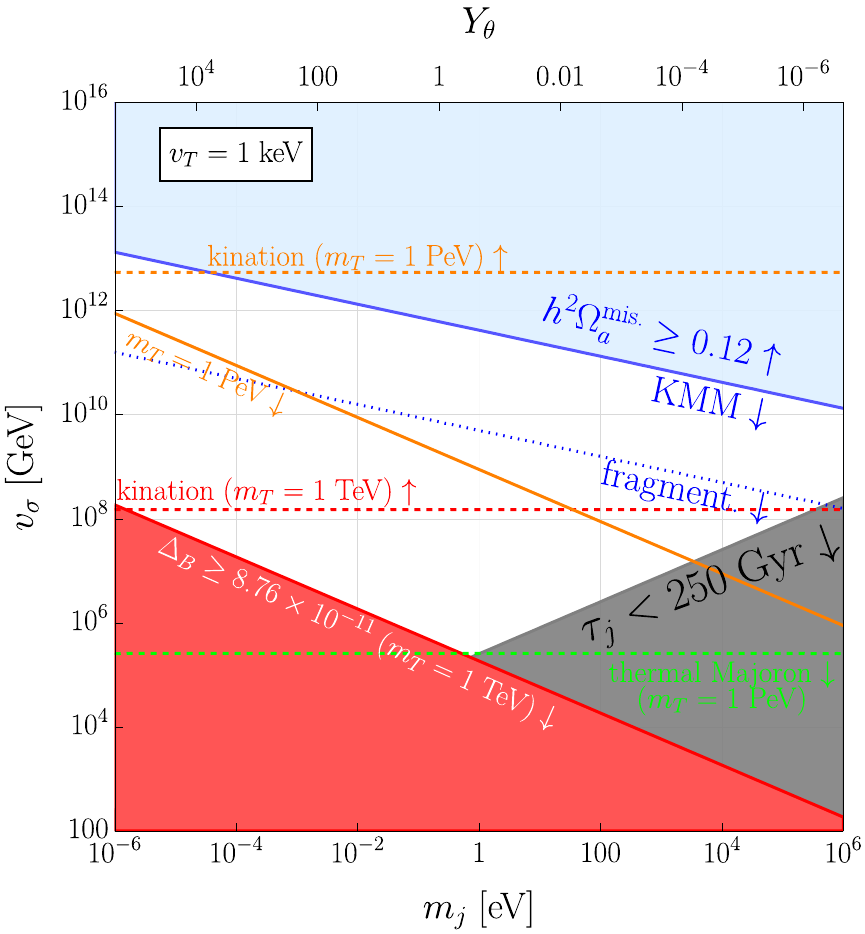}
    \caption{Parameter space for the cogenesis of the baryon asymmetry and dark matter relic
abundance via kinetic misalignment in the plane spanned by the lepton number breaking scale $v_\sigma$ and the Majoron mass $m_j$ for $v_T=\SI{1}{\electronvolt}$ \textit{(upper panel)} and $v_T=\SI{1}{\kilo\electronvolt}$ \textit{(lower panel)}. An explanation of all colored lines and regions is given section~\ref{sec:discuss}.}
    \label{fig:9-6}
\end{figure}

\begin{figure}[t!]
    \centering
    \includegraphics[width=0.55\textwidth]{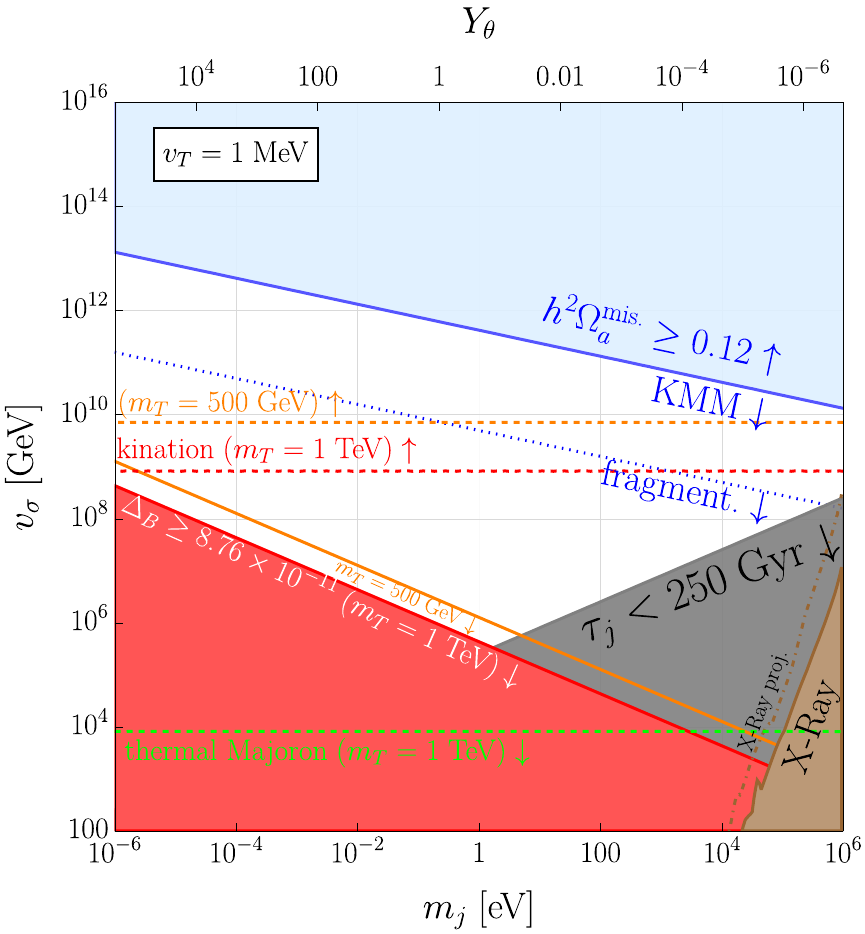}
    \includegraphics[width=0.55\textwidth]{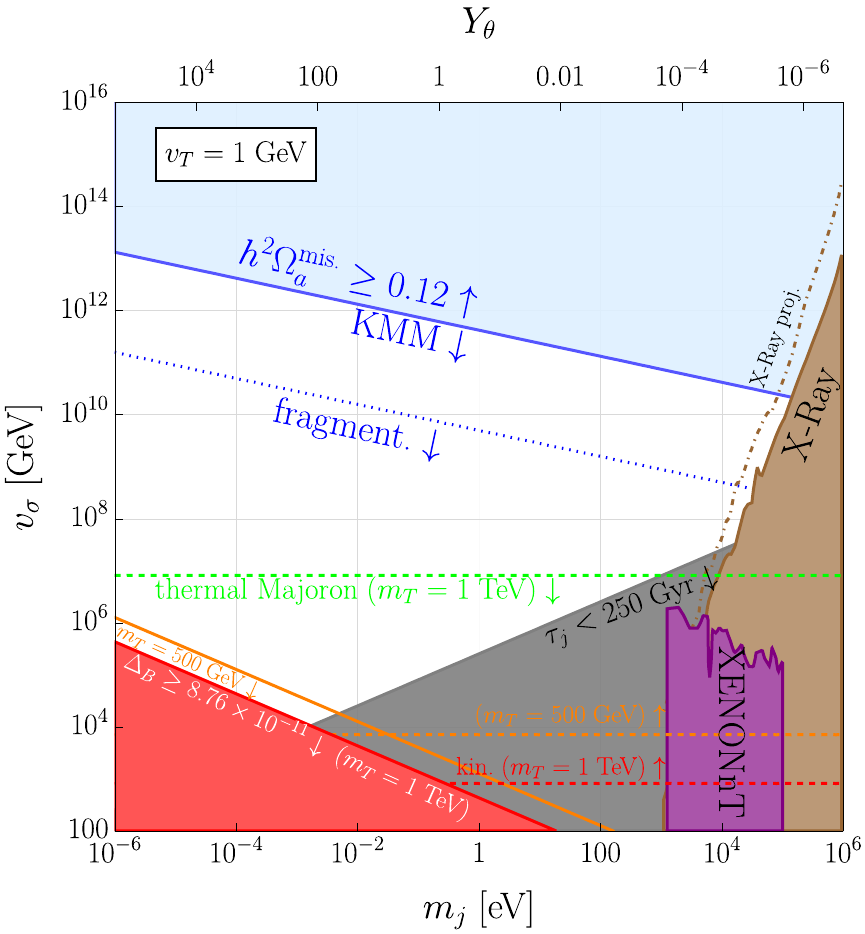}
    \caption{Parameter space for the cogenesis of the baryon asymmetry and dark matter relic
abundance via kinetic misalignment in the plane spanned by the lepton number breaking scale $v_\sigma$ and the Majoron mass $m_j$ for $v_T=\SI{1}{\mega\electronvolt}$ \textit{(upper panel)} and $v_T=\SI{1}{\giga\electronvolt}$ \textit{(lower panel)}. An explanation of all colored lines and regions is given section~\ref{sec:discuss}.}
    \label{fig:3-0}
\end{figure}

\begin{figure}[t!]
    \centering
    \includegraphics[width=0.45\textwidth]{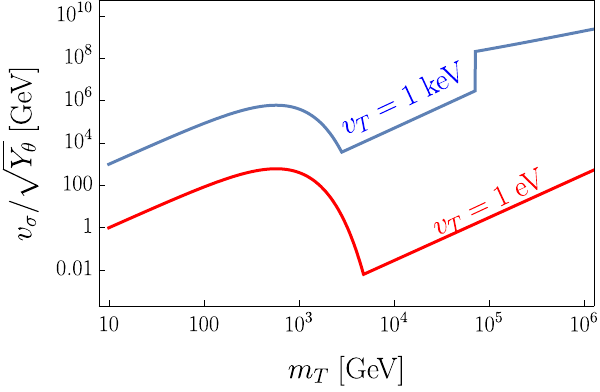}
    \includegraphics[width=0.45\textwidth]{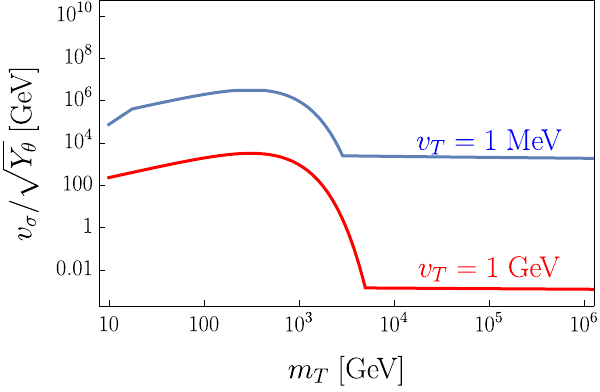}
    \caption{The required value of $v_\sigma/\sqrt{Y_\theta}$ for successful Leptogenesis  from Eq.~\eqref{eq:MASTER} as a function of $m_T$ for different values of $v_T=\{ \SI{1}{\electronvolt},\;\SI{1}{\kilo\electronvolt},\;\SI{1}{\mega\electronvolt},\;\SI{1}{\giga\electronvolt}\}$. In the \textit{left} figure $\Gamma^\text{av.}_H$ is out of equilibrium, and in the \textit{right} figure $\Gamma_L^\text{av.}$ is out of equilirbium. For a fixed $m_T$ the  \textit{left} plot can be mapped to the parameter space in Fig.~\ref{fig:9-6}, and the \textit{right} plot can be translated to the parameter space in  Fig.~\ref{fig:3-0}.}
    \label{fig:vsig}
\end{figure}

\begin{figure}[t!]
    \centering
    \includegraphics[width=0.55\linewidth]{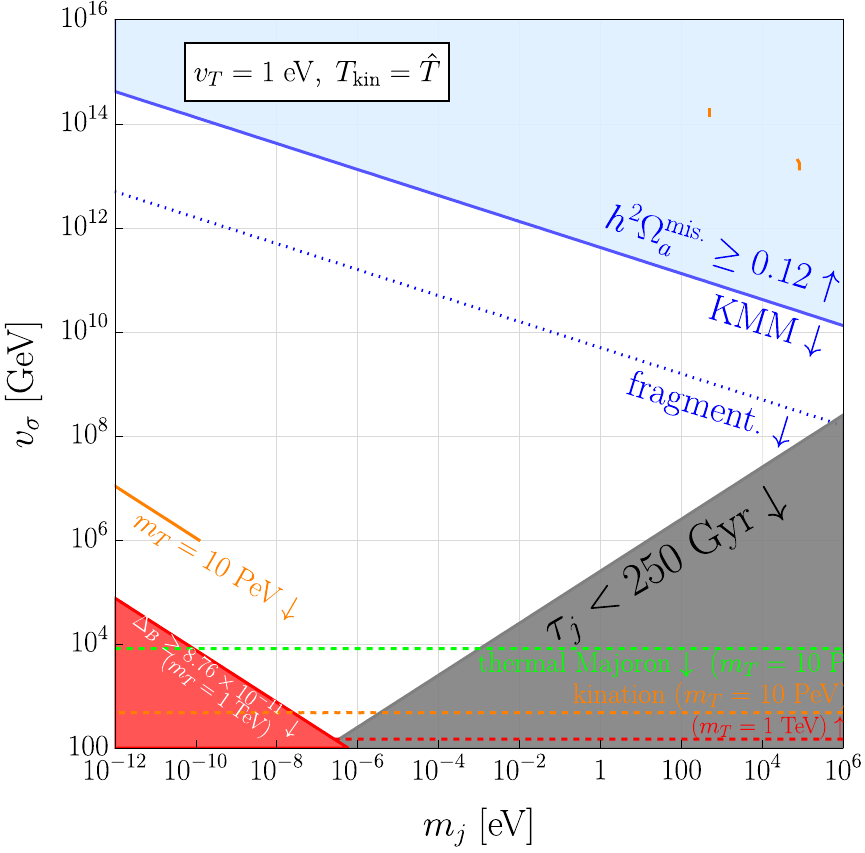}
    \includegraphics[width=0.55\linewidth]{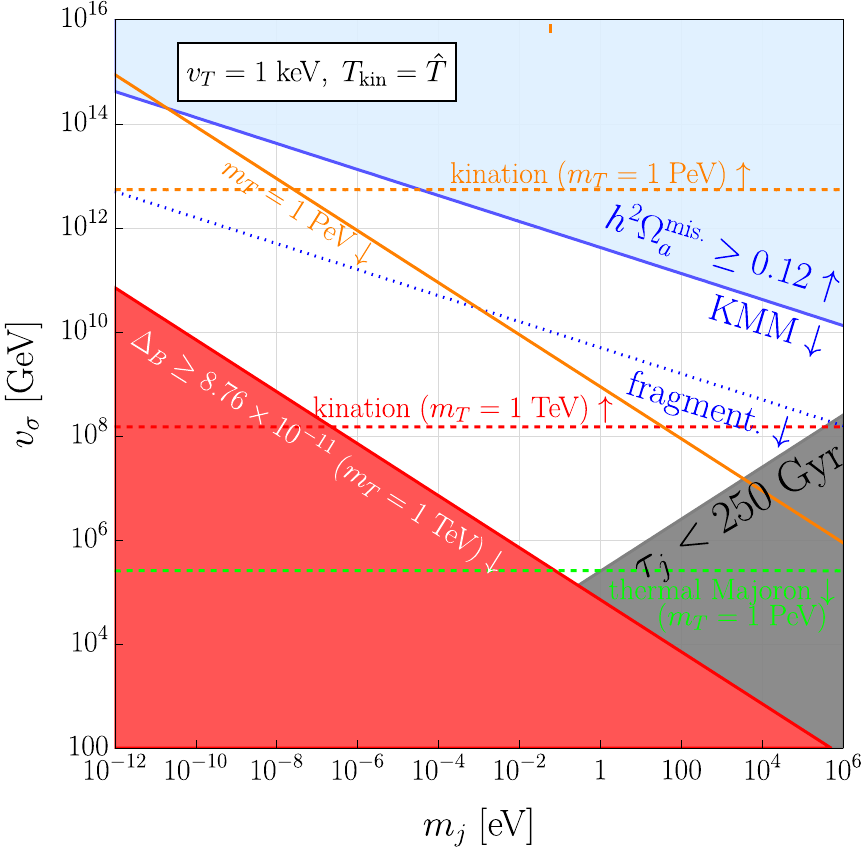}
    \caption{Parameter space for the cogenesis of the baryon asymmetry and dark matter relic
abundance via kinetic misalignment in the plane spanned by the lepton number breaking scale $v_\sigma$ and the Majoron mass $m_j$ for $v_T=\SI{1}{\electronvolt}$ \textit{(upper panel)} and $v_T=\SI{1}{\kilo\electronvolt}$ \textit{(lower panel)}. For the acoustic misalignment we take $T_\text{kin}=\tilde{T}$ and consider a scale-invariant scalar power spectrum with $\mathcal{P}_{\mathcal{R}}(k)=2\times 10^{-9}$.}
    \label{fig:A9-6}
\end{figure}

\begin{figure}[t!]
    \centering
    \includegraphics[width=0.55\textwidth]{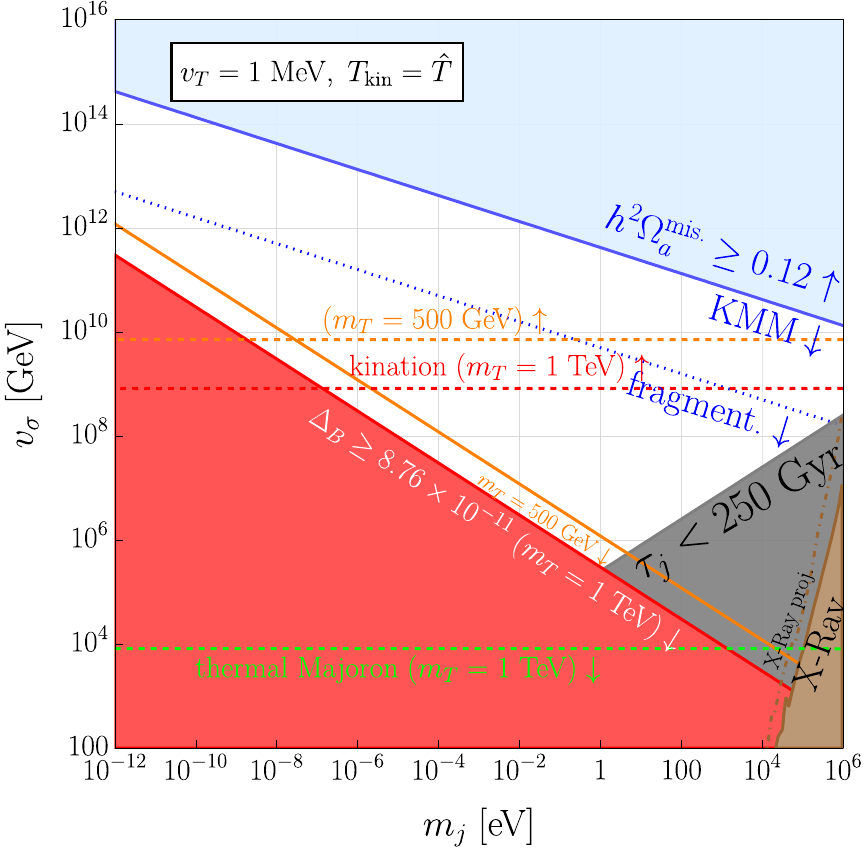}
    \includegraphics[width=0.55\textwidth]{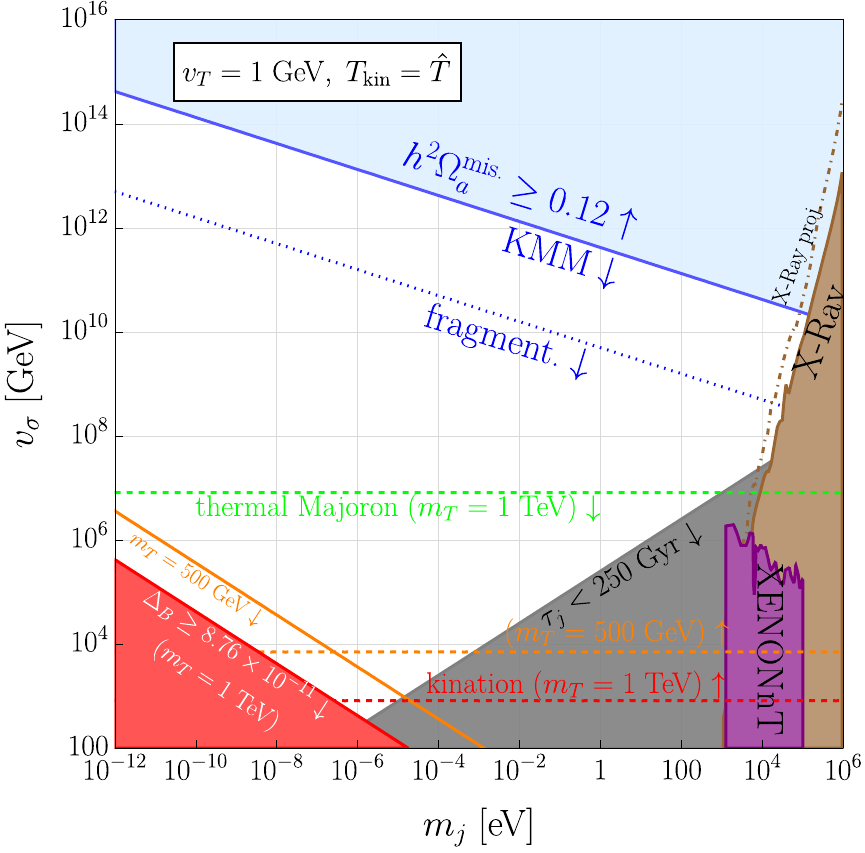}
    \caption{Parameter space for the cogenesis of the baryon asymmetry and dark matter relic
abundance via the sum of kinetic acoustic  misalignment in the plane spanned by the lepton number breaking scale $v_\sigma$ and the Majoron mass $m_j$ for $v_T=\SI{1}{\mega\electronvolt}$ \textit{(upper panel)} and $v_T=\SI{1}{\giga\electronvolt}$ \textit{(lower panel)}. For the acoustic misalignment we take $T_\text{kin}=\tilde{T}$ and consider a scale-invariant scalar power spectrum with $\mathcal{P}_{\mathcal{R}}(k)=2\times 10^{-9}$.}
    \label{fig:A3-0}
\end{figure}
  
\subsection{Parameter Space}
We display our collection of predictions and constraints in the $v_\sigma$ versus $m_j$ parameter plane in Figs.~\ref{fig:9-6}-\ref{fig:3-0}, where we varied $v_T$ in steps of three orders of magnitude between $\SI{1}{\electronvolt}$ and $\SI{1}{\giga\electronvolt}$. 

Let us  discuss the implications for dark matter:
We will first focus on the case of pure kinetic misalignment and later on discuss the implications from also including the acoustic misalignment contribution defined in Rq.~\eqref{eq:acoustic}.
Here we also display the value of the Majoron yield $Y_\theta$ that is required for a given $m_j$ to explain the dark matter relic abundance via kinetic misalignment, see Eq.~\eqref{eq:theta-param}.
For each point  one can then check, if a given model from sections~\ref{sec:AD}-\ref{sec:NOAD} can realize the required $Y_\theta$. For the Affleck-Dine scenario one can then also check, whether the constraint from the sequestering in Eq.~\eqref{eq:seq} is satisfied. 
In all plots  kinetic misalignment dominates over the conventional misalignment (see Eq.~\eqref{eq:KMMoperative}) in the region below the blue line labeled \enquote{KMM $\downarrow$}. On this blue line the relic abundance is saturated by the usual misalignment contribution from Eq.~\eqref{eq:mis}, where we chose the post-inflationary lepton number breaking scenario with an patch-averaged initial misalignment angle of  $\sqrt{\braket{\theta_i^2}}=2.1$ \cite{GrillidiCortona:2015jxo}. For smaller $\theta_i$ from the pre-inflationary scenario the relic abundance is fixed for larger $v_\sigma$. Below the dotted blue line labeled \enquote{fragment. $\downarrow$} the Majoron condensate fragments into higher momentum excitations as described in the discussion surrounding Eq.~\eqref{eq:fragment} from Ref.~\cite{Eroncel:2022vjg}.
We checked that the entire depicted parameter space is compatible with the recently derived limits on the free-streaming length and isocurvature 
from the Lyman-$\alpha$ forest \cite{Harigaya:2025pox}.

In the black area the dark matter lifetime from decays to neutrinos (Eq.~\eqref{eq:dec-nu}) or photons (Eq.~\eqref{eq:dec-gamma}) is below $250\;\text{Gyr}$. We find that the photon decay mode requires larger $v_T$ of the order $\mathcal{O}(\SI{1}{\giga\electronvolt})$ to be relevant at all, which is close to the upper bound from electroweak precision observables in Eq.~\eqref{eq:EWPT}. Existing limits from X-Ray and gamma ray searches were displayed as brown regions and projected future limits as brown dot-dashed lines (consult Ref.~\cite{Panci:2022wlc} and references therein for details). From both panels in Fig~\ref{fig:3-0} it can be deduced that the aforementioned limits only become relevant for $v_T>\mathcal{O}(\SI{1}{\mega\electronvolt})$, and they can dominate over the lifetime constraint for large $m_j$ if $v_T=\mathcal{O}(\SI{1}{\giga\electronvolt})$. 

The direct detection limit from Majorons scattering off electrons obtained by the \verb|XENONnT| \cite{XENON:2022ltv} collaboration for an exposure of $1.16\;\text{ton}\times\text{year}$ was drawn in purple. For a larger exposure the exclusion limit would move upwards to larger $v_\sigma$. The lower panel in Fig.~\ref{fig:3-0} reveals, that this limit also  only matters for  $v_T=\mathcal{O}(\SI{1}{\giga\electronvolt})$ together with  $m_j=\mathcal{O}(\SI{10}{\kilo\electronvolt})$, $v_\sigma = \mathcal{O}(\SI{e6}{\giga\electronvolt})$. The authors of Ref.~\cite{Biggio:2023gtm} attempted to populate this parameter region via the freeze-in production of Majorons, which required an earlier electroweak phase transition together with an early matter dominated epoch. 

Concerning dark radiation and hot dark matter our results indicate that the strongest limits come from the non-thermalization conditions in Eq.~\eqref{eq:non-therm} and Eq.~\eqref{eq:non-therm2}, which was drawn as a green line labeled \enquote{thermal Majoron $\downarrow$}. 
By comparing both equations one finds that the limit from scattering is stronger. From the lower panel in Fig.~\ref{fig:3-0} it is clear that most of the parameter space for  $v_T=\mathcal{O}(\SI{1}{\giga\electronvolt})$ would be ruled out by the Majoron thermalization. This finding mirrors the conclusions reached in  Ref.~\cite{Biggio:2023gtm}, and was one of the reasons, why the authors of that reference considered early matter domination. Depending on the combination of $v_T$ and $m_T$ this constraint rules out $v_\sigma$ below $\mathcal{O}(\SI{e4}{\giga\electronvolt})$-$\mathcal{O}(\SI{e7}{\giga\electronvolt})$. We verified that the constraints from the dark radiation abundance encoded in $\Delta N_\text{eff.}$ and the limit on the hot dark matter abundance in Eq.~\eqref{eq:hotDM} were subleading.

Next we discuss what conclusions can be drawn for Leptogenesis:
For a given $m_T$ we showcase the line on which the observed value of the baryon asymmetry $\Delta_B = 8.76\times 10^{-11}$ \cite{Planck:2018vyg,Schoneberg:2024ifp} is reproduced by Eq.~\eqref{eq:MASTER}. Below this line the asymmetry is overproduced for the same $m_T$.  In all plots we use the color red for $m_T=\SI{1}{\tera\electronvolt}$ and the color orange for a second representative choice of $m_T$, which depends on $v_T$. Above the dashed red and orange line there would be an epoch of kination during the asymmetry production (see Eq.~\eqref{eq:kination} and the paragraph below), which is not ruled out by data, but would change the details of our calculation via the change in the redshifting behavior. Here we do not use the analytical approximations from section~\ref{sec:low}, but employ numerical results instead.

For $v_T=\SI{1}{\electronvolt}$ shown in the upper panel of Fig.~\ref{fig:9-6} we are in the weak wash-in regime with a slow $\Gamma_H^\text{av.}$, described by Eq.~\eqref{eq:Hslow} for both $m_T=\SI{1}{\tera\electronvolt}$ and $m_T=\SI{10}{\peta\electronvolt}$.
Here both choices of $m_T$ are not viable because the kination constraint implies $v_\sigma<\mathcal{O}((100-1000)\;\text{GeV})$ and cuts away almost the entire parameter space. Note that we truncated the orange line for Leptogenesis with $m_T=\SI{10}{\peta\electronvolt}$ to comply with the backreaction constraint in Eq.~\ref{eq:BR2}.

If we consider $v_T=\SI{1}{\kilo\electronvolt}$ as displayed in the lower panel of Fig.~\ref{fig:9-6}, we still find ourselves in the weak wash-in regime with a slow $\Gamma_H^\text{av.}$, described by Eq.~\eqref{eq:Hslow} for both $m_T=\SI{1}{\tera\electronvolt}$ and $m_T=\SI{1}{\peta\electronvolt}$. Here the Leptogenesis  lines lie at larger values of $v_\sigma$ compared to the previous panel, since the rate $\Gamma^\text{av.}_H$ is larger for a larger $v_T$ as can be seen in Eq.~\eqref{eq:GammaH}. For $m_T=\SI{1}{\tera\electronvolt}$ the viable window for dark matter and baryon asymmetry cogenesis is $\mathcal{O}(\SI{e5}{\giga\electronvolt})<v_\sigma< \mathcal{O}(\SI{e8}{\giga\electronvolt})$ and $\mathcal{O}(\SI{1}{\electronvolt}) > m_j >\mathcal{O}(\SI{1}{\micro\electronvolt})$ due to the DM lifetime and kination. Here the Majoron fragments in the entire viable parameter space. For  $m_T=\SI{1}{\peta\electronvolt}$ the viable window is $\mathcal{O}(\SI{e7}{\giga\electronvolt})<v_\sigma<\mathcal{O}(\SI{e10}{\giga\electronvolt})$, where the lower limit comes from the lifetime and the  upper limit comes from the overabundance of conventional misalignment Majorons. For smaller $\theta_i$ this limit could be lifted to $\mathcal{O}(\SI{e13}{\giga\electronvolt})$, when the kination bound becomes relevant. 

For both  $v_T=\SI{1}{\electronvolt},\;\SI{1}{\kilo\electronvolt}$ we find that the lines for the baryon asymmetry and kination constraint move to larger $v_\sigma$ for both $\SI{300}{\giga\electronvolt}\lesssim m_T\lesssim\SI{1}{\tera\electronvolt}$  
and $m_T>(10^5-10^6)\;\text{GeV}$, which can be deduced from the left panel of Fig.~\ref{fig:vsig}, and which motivated our choices of $m_T=(1-10)\;\text{PeV}$.

We continue with the choice $v_T=\SI{1}{\mega\electronvolt}$ that was presented in the upper panel of Fig.~\ref{fig:3-0}. Leptogenesis for  both $m_T=\SI{1}{\tera\electronvolt}$ and $m_T=\SI{500}{\giga\electronvolt}$ occurs in the weak wash-in regime with slow $\Gamma_L^\text{av.}$ described in Eq.~\eqref{eq:Lslow}. 
For $m_T=\SI{1}{\tera\electronvolt}$  ($\SI{500}{\giga\electronvolt}$) cogenesis works for $\mathcal{O}(\SI{e5}{\giga\electronvolt})<v_\sigma< \mathcal{O}(\SI{e9}{\giga\electronvolt})$ and $\mathcal{O}(\SI{1}{\electronvolt}) > m_j >\mathcal{O}(\SI{1}{\micro\electronvolt})$
($\mathcal{O}(\SI{e6}{\giga\electronvolt})<v_\sigma< \mathcal{O}(\SI{e10}{\giga\electronvolt})$ and $\mathcal{O}(\SI{10}{\electronvolt}) > m_j >\mathcal{O}(\SI{0.1}{\micro\electronvolt})$).
Here even the parameter space in which the Majoron does not fragment is reachable .

The largest considered value of $v_T=\SI{1}{\giga\electronvolt}$ was motivated by the constraint from the electroweak rho-parameter in Eq.~\eqref{eq:EWPT}, and the corresponding parameter space is visible in the lower panel of Fig.~\ref{fig:3-0}. One can observe that the whole parameter space for Leptogenesis  with  $m_T=\SI{1}{\tera\electronvolt}$ and $m_T=\SI{500}{\giga\electronvolt}$ is ruled out by the aforementioned constraint from Majoron thermalization in green and the bound for avoiding Majoron kination.

The choice of $m_T=\SI{500}{\giga\electronvolt}$ in both panels of Fig.~\ref{fig:3-0} can be understood via the right panel of Fig.~\ref{fig:vsig}: For constant $v_T$ the lines for the baryon asymmetry and kination constraint move to larger $v_\sigma$ for $\SI{50}{\giga\electronvolt}\lesssim m_T\lesssim\SI{1}{\tera\electronvolt}$. This time the choice $m_T\gg \SI{1}{\tera\electronvolt}$ does not allow one to raise $v_\sigma$: Both curves on the right of Fig.~\ref{fig:vsig} decrease with growing $m_T$, and eventually become flat for $m_T>\mathcal{O}(\SI{e4}{\giga\electronvolt})$, unlike the curves in the left panel of Fig.~\ref{fig:vsig}, which start to grow again significantly for  $m_T>\mathcal{O}(\SI{e4}{\giga\electronvolt})$.
The lines of constant baryon asymmetry with the same choice of $m_T$ are found at larger values of $v_\sigma$ for $v_T=\SI{1}{\mega\electronvolt}$ compared to $v_T=\SI{1}{\giga\electronvolt}$, because the rate $\Gamma_L^\text{av.}$ in Eq.~\eqref{eq:GammaL} is inversely proportional to $v_T^2$. 

We conclude our analysis based on kinetic misalignment only by noting that the viable range for cogenesis with $m_T=\mathcal{O}(\SI{1}{\tera\electronvolt})$ is given by
\begin{align}
 \mathcal{O}(\SI{1}{\kilo\electronvolt}) <   v_T < \mathcal{O}(\SI{1}{\mega\electronvolt}),
\end{align}
and this range includes \textit{both} the strong and weak wash-in regimes.  This band for $v_T$ is compatible with the constraints from the adiabaticity of $\dot{\theta}$ in Eqns.~\eqref{eq:adiab1}-\eqref{eq:adiab2} and the absence of the chiral hypermagnetic instability in Eqns.~\eqref{eq:CPL1}-\eqref{eq:CPL2}.

Next we include also the acoustic misalignment contribution from Eq.~\eqref{eq:acoustic} together with the kinetic misalignment. We assume a scale invariant scalar power spectrum with an amplitude of $2\times 10^{-9}$ and for a model-independent estimate compatible with baryogenesis we take $T_\text{kin}$ to be equal to $\tilde{T}$, defined in Eq.~\eqref{eq:Tfo}. Note that $T_\text{kin}$ can in principle be much larger, but this can only be analyzed for a given mechanism initiating the Majoron's kination behavior.  From the upper panel in Fig.~\ref{fig:A9-6} and the lower panel in Fig.~\ref{fig:A3-0}, we can deduce that the acoustic misalignment shifts the parameter space for $v_T = \SI{1}{\electronvolt},\; \SI{1}{\giga\electronvolt}$ to much lower $m_j$ than before. Conversely the lower panel in Fig.~\ref{fig:A9-6} and the upper panel in Fig.~\ref{fig:A3-0} demonstrate that this change is far less pronounced for $v_T = \SI{1}{\kilo\electronvolt},\; \SI{1}{\mega\electronvolt}$ and the preferred values of $m_j$ are affected only by order one factors. The underlying reason is that acoustic misalignment becomes more important for smaller $v_\sigma$ and with only kinetic misalignment we found that $v_T = \SI{1}{\electronvolt},\; \SI{1}{\giga\electronvolt}$ required $v_\sigma$ around two orders of magnitude below than what was needed for $v_T = \SI{1}{\electronvolt},\; \SI{1}{\giga\electronvolt}$. Therefore we conclude that the inclusion of acoustic misalignment does not significantly shift our preferred parameter space.

\subsection{Model discrimination}
The previous discussion revealed that while cogenesis is viable over a wide range of parameters, the required values of $v_T$ and $m_j$ are too tiny, to ever give a signal for direct or indirect DM detection.
Since the Majoron does not thermalize, we also do not expect cosmological signatures except a   freeze-in suppressed contribution to $\Delta N_\text{eff.}$, which is rather generic from a beyond the SM viewpoint, and can not be used to single out our scenario. Our preferred values  of $v_\sigma$ are also far too large to lead to observable rates for Majoron emission in neutrino-less double beta decay \cite{Kharusi:2021jez,CUPID-0:2022yws}. For the $v_T$ we consider we find no appreciable rates for invisible Higgs decays to Majorons (see Eq.~\eqref{eq:vTInv1} and Eq.~\eqref{eq:vTInv2}). Because the couplings or the Type II Seesaw Majoron to charged leptons and quarks are flavor diagonal, tree level lepton flavor violation or rare meson decays with Majoron emission are also unavailable. The Majoron-photon coupling also offers no chance of detectability, since is suppressed by either the Majoron-pion mixing or by small fermion masses, since lepton number is not anomalous with respect to electromagnetism.\footnote{This can be avoided in models that identify the Majoron with the QCD axion such as e.g. Refs.~\cite{Ballesteros:2016euj,Ballesteros:2016xej}, or via additional model building \cite{Liang:2024vnd}.} Therefore the Majoron itself does not offer ways of directly probing this model.

If the Majoron rotation is excited via the Affleck-Dine mechanism with a quartic potential for $\sigma$, the radial mode of $\sigma$ will typically have masses of $\mathcal{O}(\text{MeV})$. Thermalization of the radial mode via the Higgs portal coupling $\lambda_{\sigma H}$ could lead to observable mixing with the Higgs \cite{Co:2020dya,Co:2020jtv}. However, while potentially observable, this avenue is not specific to our model, and can arise in wide array of beyond the SM contexts.

This means that we have to resort to  the phenomenology of the triplet scalar. For our band of $\mathcal{O}(\SI{1}{\kilo\electronvolt}) <   v_T < \mathcal{O}(\SI{1}{\mega\electronvolt})$ we can see from Fig.~\ref{fig:triplet} that lepton flavor violation from decays is unlikely to be observable, and we do not expect a more refined analysis of nuclear conversion processes to drastically change this conclusion. In this range, or in the narrower strong wash-in regime from Eq.~\eqref{eq:band} with $\SI{8.5}{\kilo\electronvolt}<v_T<\SI{277}{\kilo\electronvolt}$ for $m_T=\SI{1}{\tera\electronvolt}$, we find that the doubly charged scalar $h^{\pm\pm}$ can decay to both $l^\pm l^\pm$ or $W^\pm W^\pm$ with branching fractions depending on the precise value of $v_T$ (see Eq.~\eqref{eq:vTW}).

The Inflationary Affleck Dine Type II Seesaw Leptogenesis of Refs.~\cite{Barrie:2021mwi,Barrie:2022cub,Han:2023kjg,Kaladharan:2024bop} can also operate with TeV-scale triplets.
This particular model needs $\Gamma_L^\text{av.}$ to be fast in order to transmit the asymmetry stored in the inflaton to the lepton sector, while having a slow  $\Gamma_H^\text{av.}$, to avoid washout of the lepton asymmetry into the SM higgs doublet, which implies 
\begin{align}
    v_T < \SI{8.5}{\kilo\electronvolt} \sqrt{\frac{\SI{1}{\tera\electronvolt}}{m_T}}.
\end{align}
This limit corresponds to the weak wash-in regime, as can be observed in Fig.~\ref{fig:triplet}. For this model we expect a TeV-scale triplet to preferably decay to leptons. Such a  smaller $v_T$ can also potentially lead to observable rates in experiments probing lepton flavor violation \cite{Barrie:2022ake,Han:2025ifi}. 

The last mechanism for low scale Type II Seesaw Leptogenesis involves a resonant enhancement from at least two quasi-degenerate triplets \cite{Pilaftsis:2003gt,Bechinger:2009qk}. Assuming that this also occurs in the strong wash-out regime like its high scale counterpart in Ref.~\cite{Hambye:2005tk}, the range of vevs is the same as in Eq.~\eqref{eq:band}. Such a setup could be potentially distinguished from our spontaneous Leptogenesis scenario by the presence of a compressed spectrum with two almost degenerate doubly-charged scalars. We expect that the corresponding pair of doubly charged Higgses could appear experimentally as two overlapping resonances, which could at least in principle be discriminated from a single narrower resonance.

\section{Conclusion}\label{sec:conclusion}
In this work we have demonstrated that spontaneous wash-in Type II Seesaw Leptogenesis can successfully  occur in a Majoron background. Our approach addresses the observed neutrino mass scale, the origin of the asymmetry between matter and anti-matter, as well as the nature of dark matter in the form of the Majoron.

\begin{itemize}
    \item \textbf{Majoron Dark Matter and Cogenesis:}
    Long lived enough Majoron dark matter from kinetic misalignment \cite{Co:2019jts,Chang:2019tvx,Co:2020dya} and Leptogenesis  
    can be simultaneously realized for a lepton number breaking scale of around $\mathcal{O}(\SI{e5}{\giga\electronvolt})<v_\sigma< \mathcal{O}(\SI{e8}{\giga\electronvolt})$ and a Majoron mass between about  $\mathcal{O}(\SI{1}{\electronvolt}) > m_j >\mathcal{O}(\SI{1}{\micro\electronvolt})$. This range of parameters does not change if we also include the recently discovered acoustic misalignment mechanism \cite{Bodas:2025eca,Eroncel:2025qlk}. 
    In this set-up the Majoron is essentially invisible from an experimental point of view,as all of its couplings are suppressed by the large $v_\sigma$.
    
    \item \textbf{Triplet Scalar phenomenology:} This set up requires only a single electroweak triplet, whose lightest possible mass is \textit{not} limited by Leptogenesis  but  rather \textit{only} by collider constraints to lie above (220-1080) GeV \cite{ATLAS:2018ceg,Novak:2024igo}. In other words our mechanism is fully compatible with a triplet mass of 1 TeV in reach of dedicated searches at current or future colliders. 
    In this mass range we require a triplet vev of $\mathcal{O}(\SI{1}{\kilo\electronvolt}) <   v_T < \mathcal{O}(\SI{1}{\mega\electronvolt})$ for Leptogenesis  due to limits from the absence of kination, the absence of the chiral hypermagnetic instability \cite{Co:2022kul} and thermalized Majorons. 
    For a TeV-scale triplet the aforementioned range of $v_\sigma$ implies, that the small neutrino mass scale arises from a combination of tiny values of $\lambda_{\sigma H T}$ together with a suppressed Yukawa coupling $y_L$ (see Eqns.~\eqref{eq:m1}-\eqref{eq:m2}).
    We predict no observable lepton flavor violation from decays in next generation experiments, while the treatment of nuclear conversion requires a dedicated analysis.  Here the doubly charged scalar has decay modes to both  $l^\pm l^\pm$ and $W^\pm W^\pm$, and which decay dominates depends on the precise value of $v_T$.

    \item \textbf{Comparison to the rotating triplet model of Refs.~\cite{Barrie:2021mwi,Barrie:2022cub,Han:2023kjg,Kaladharan:2024bop}:}
    Another recent approach to non-thermal Leptogenesis  in the Type II Seesaw was based on the Affleck-Dine mechanism, and took the inflaton to be a linear combination of the SM like Higgs and the neutral component of the triplet \cite{Barrie:2021mwi,Barrie:2022cub,Han:2023kjg,Kaladharan:2024bop}. This model features a cubic and a quintic \cite{Barrie:2021mwi,Barrie:2022cub,Han:2023kjg}  (quartic \cite{Kaladharan:2024bop}) term in the scalar potential, and one requires that the cubic term is sufficiently small to avoid the classical washout \cite{Cheung:2012im} of the asymmetry via scalar field dynamics. Consequently  $v_T$ is sourced from these two separate terms.  One  advantage of our realization is that  it works with a  single phase dependent term in the scalar potential. Moreover the aforementioned  model is complimentary to ours, as its minimal field content provides no dark matter candidate, whereas our singlet-double-triplet Majoron model comes with dark matter \enquote{out of the box}, but inflation was not considered so far. In our model it is straightforward to promote the radial mode of the lepton number breaking scalar $\sigma$ to the inflaton \cite{Ballesteros:2016euj,Ballesteros:2016xej}, either with or without considering the admixture of the SM like Higgs. The coherent motion of the Majoron can then arise from its slow roll dynamics during inflation (see e.g. Refs.~\cite{Lee:2023dtw,Lee:2024bij}). In this case one has to carefully treat preheating along the lines of Refs.~\cite{Ballesteros:2016euj,Ballesteros:2016xej}, in order to avoid non-thermal symmetry restoration or the production of topological defects \cite{Kofman:1995fi}.

     \item \textbf{Model discrimination:}
     If a triplet scalar were to be discovered at a collider, the  model of Refs.~\cite{Barrie:2021mwi,Barrie:2022cub,Han:2023kjg,Kaladharan:2024bop} predicts that the doubly charged scalar predominantly decays to $l^\pm l^\pm$. This could be used to distinguish it from our model, which \textit{can} have decays to $W^\pm W^\pm$ as the dominant mode for $v_T>\SI{40}{\kilo\electronvolt}$. Another way to discriminate both scenarios would be an observation of lepton flavor violating decays, which can arise in the aforementioned model but are not expected in out preferred parameter space.  Low scale resonant Leptogenesis  \cite{Pilaftsis:2003gt,Bechinger:2009qk} could be differentiated from both non-thermal scenarios by a precise measurement of the doubly charged scalar resonance, since resonant Leptogenesis predicts two almost degenerate triplets with overlapping resonances.
\end{itemize}

\section{Acknowledgments}
The author would like to thank Juan Herrero-García  for pointing out that the labels for $h^{\pm\pm}\rightarrow l^\pm l^\pm$ and $h^{\pm\pm}\rightarrow W^\pm W^\pm$ in the previous iteration of figure \ref{fig:triplet} were accidentally switched.
MB is supported by \enquote{Consolidación Investigadora Grant CNS2022-135592}, and also funded by \enquote{European Union NextGenerationEU/PRTR}, as well as the Generalitat Valenciana APOSTD/2025 Grant No. CIAPOS/2024/148.

\appendix

\section{Effective chemical potential for bosons in a pNGB background}\label{sec:KG}

To the best of our knowledge the  effective chemical potential prescription in a cosmological pNGB background (here for the Majoron case) \cite{Chun:2023eqc}
\begin{align}\label{eq:muX}
    \mu_X \rightarrow \mu_X + \frac{B-L}{2} \dot{\theta}, 
\end{align}
has not been derived in the literature for the case of $X$ being a boson.
Ref.~\cite{Arbuzova:2016qfh} showed that this prescription is valid for Dirac fermions, and in Ref.~\cite{Chun:2023eqc} it was demonstrated that it also holds for Majorana fermions, as long as the temperature is far above the Majorana mass. 

We begin our discussion by removing the Majoron phase
\begin{align}
    \theta \equiv \frac{j}{v_\sigma}
\end{align}
from the scalar interaction
\begin{align}
    \lambda_{\sigma H T} \sigma H^\dagger \varepsilon T H^*, 
\end{align}
by a field redefinition
\begin{align}
    T\rightarrow e^{i \theta} T,
\end{align}
and we suppress the weak interaction gauge indices throughout this section. 
This redefinition shifts the Lagrangian (here we do not show the  gauge interactions)
\begin{align}
    \left(\partial_\mu T\right)\left(\partial^\mu T^* \right) - m_T^2 |T|^2 
\end{align}
to 
\begin{align}
    \partial_\mu T \partial^\mu T^* + \left(\partial_\mu \theta \partial^\mu \theta -m_T^2\right)  |T|^2 + i \partial_\mu \theta \left(\left(\partial^\mu T\right)T^* - T (\partial^\mu T^*)\right).
\end{align}
The second interaction is the coupling of $\partial_\mu \theta$ to the scalar lepton number current $j_L^\mu$,
\begin{align}
    j_L^\mu =  \left(\partial^\mu T\right)T^* - T \partial^\mu T^*,
\end{align}
which is analogous to the fermionic case. However for bosons there also exists a quadratic interaction  $\partial_\mu \theta \partial^\mu \theta |T|^2$, that is not present for fermions.
This term only affects the normalization of the Majoron's kinetic term $ \partial_\mu \theta \partial^\mu \theta v_\sigma^2$, which is negligible here due to the fact that $v_\sigma \gg v_T$.

Next we will demonstrate, that the interaction of $T$ with $\theta$ shifts the energy levels of particle and anti-particles. 
We continue by the deriving the equation of motion for the triplet 
\begin{align}
    \left[\Box +m_T^2 -\partial_\mu \theta \partial^\mu \theta -i \Box \theta - 2 i \partial_\mu \theta \partial^\mu \right]T=0.
\end{align}
In the following we take $v_\sigma$ to be constant, so that $\partial_\mu \theta = (\partial_\mu j)/v_\sigma$.
In cosmology we are usually interested in spatially homogeneous, time-dependent backgrounds so that $\partial_i \theta=0$ and $\dot{\theta}=\partial_t \theta \neq 0$, which simplifies the equation of motion to 
\begin{align}
        \left[\left(\partial_t - i \dot{\theta}\right)^2 -\partial_i^2 + m_T^2  -i \ddot{\theta} \right]T=0.
\end{align}
For spontaneous Baryogenesis the background field $\theta$ only changes adiabatically \cite{Cohen:1987vi,Cohen:1988kt,Arbuzova:2016qfh}, so that we have $\dot{\theta}\simeq \text{const.}$, and we can take $\ddot{\theta}\simeq 0$. The dispersion relation is obtained by applying the equation of motion to a plane wave solution $\sim e^{-i(E t -\vec{p} \vec{x})}$ for the triplet field and we find 
\begin{align}
    E_{\pm} = \pm\sqrt{|\vec{p}|^2+m_T^2} -\dot{\theta}.
\end{align}
As usual we identify the positive energy solutions $E_+$ with particles $T$ of positive energy $E_T=E_+$, and the negative energy solutions $E_-$ with \textit{anti-}particles of positive energy $E_{\overline{T}}=-E_-$. We further define 
\begin{align}
    E_0 \equiv \sqrt{|\vec{p}|^2+m_T^2},
\end{align}
and obtain the shifted energies 
\begin{align}
    E_T = E_0 - \dot{\theta},\quad E_{\overline{T}}= E_0 + \dot{\theta}.
\end{align}
The Bose-Einstein distribution function for $T$ and $\overline{T}$ read 
\begin{align}
 f_T &= \frac{1}{(2\pi)^3}   \frac{1}{e^{\frac{-(E_T-\mu_T)}{T}}-1},\\
  f_{\overline{T}} &= \frac{1}{(2\pi)^3}   \frac{1}{e^{\frac{-(E_{\overline{T}}+\mu_T)}{T}}-1},
\end{align}
and the triplet asymmetry turns out to be
\begin{align}
    \Delta n_T \equiv n_T - n_{\overline{T}} =  (\mu_T +\dot{\theta}) T^2.
\end{align}
From this is evident that $\dot{\theta}\neq0$  can effectively  be accounted for by a shift of the triplet's chemical potential 
\begin{align}
    \mu_T \rightarrow \mu_T + \dot{\theta},
\end{align}
which we used in Eq.~\eqref{eq:shift} of the main text, and which is equivalent to Eq.~\eqref{eq:muX}, since the triplet carries a lepton number of $Q_L[T]=-2$.
Bose condensation of the triplet scalar  can be avoided as long as \cite{Harvey:1990qw}
\begin{align}
    |\dot{\theta}| < m_T.
\end{align}

\section{Triplet rotation?}\label{sec:tripletRot}
In the main text we considered the lepton number breaking field $\sigma$ to be realized as a rotating condensate in the post-inflationary universe, and used the triplet essentially as a mediator to communicate the effect of $\dot{\theta}\neq0 $ to the leptons. One may wonder, if our scenario could also involve a rotation of the triplet itself similar to the inflationary set-up considered in Refs.~\cite{Barrie:2021mwi,Barrie:2022cub,Han:2023kjg}:

In appendix A of Ref.~\cite{Co:2020jtv} it was suggested that a coupling of the form $\sigma H_u H_d$, where $H_{u,d}$ denote Higgs doublets with opposite hypercharges, could both source a large  field value and initiate a rotation of $H_{u,d}$ from the oscillation of the radial mode in $\sigma$. The large field value of the Higgses can arise, because the aforementioned term is essentially a $\sigma$-dependent mass term for $\phi^2\equiv H_u H_d$. However our model features the term $ \lambda_{\sigma H T} \sigma H^t \varepsilon T H$, which can not be interpreted as a mass term for $T$ or $H$, so we do not expect it to induce large field values. If either field  $H$ or $T$  forms a coherent condensate ($\braket{H^0},\braket{T^0} \gg \text{temperature}$) via other dynamics, one could use the $ \lambda_{\sigma H T}$ term to transfer the Majoron rotation to the angular components of $H^0$ and $T^0$ in the same vein as in Ref.~\cite{Domcke:2022wpb}. Then the rotating $T$ could source a lepton asymmetry directly via the decay $T\rightarrow L^\dagger L^\dagger$. A potential caveat is that both $\braket{H^0},\braket{T^0} \neq0$ break electroweak symmetry and hence the $B+L$ violating sphaleron transitions, that converts the lepton asymmetry into one for baryons, are expected to be out of equilibrium. This could be circumvented, if the decays happen late enough so that thermal corrections have time to restore electroweak symmetry.

We can avoid the formation of a triplet condensate during inflation by either giving it a positive Hubble-dependent mass squared \cite{Redi:2022llj}, or by impeding the development of quantum fluctuations via a large coupling to another field, that is heavy during inflation \cite{Enqvist:2011pt}.

Another potential problem is that this charge transfer reduces $Y_\theta$ by a fraction $f$ and $f Y_\theta$ is depleted by the decays of $T$ into leptons. One then has to ensure, that $(1-f)Y_\theta$ is still large enough to source the dark matter relic abundance, see Eq.~\eqref{eq:abund}. Due to these complications we do not consider the direct rotation of $T$ further.

\bibliographystyle{JHEP}
\bibliography{typeIIMajoron}

\end{document}